\documentclass[natbib,smallextended]{svjour3}
\journalname{Astronomy Astrophysics Review}

\smartqed  

\usepackage{graphicx}
\usepackage{xcolor}
\usepackage{txfonts}
\usepackage{fancyhdr}
\usepackage{lastpage}
\usepackage[colorlinks=true,linkcolor=blue,citecolor=blue,urlcolor=teal]{hyperref}
\usepackage{natbib}

\DeclareSymbolFont{mymath}{T1}{ybv}{m}{it}
\SetSymbolFont{mymath}{normal}{T1}{ybv}{m}{it}
\DeclareSymbolFontAlphabet{\mathnormal}{mymath}
\DeclareMathSymbol{\vel}{\mathalpha}{mymath}{`v}

\begin{document}

\title{The evolution of CNO elements in galaxies}




\author{Donatella Romano}
\authorrunning{D. Romano}
\institute{Donatella Romano \at
  INAF, Osservatorio di Astrofisica e Scienza dello Spazio\\
  Via Gobetti 93/3, Bologna, 40129, Italy\\
  \email{donatella.romano@inaf.it}\\
  ORCID: 0000-0002-0845-6171 \\
}

\maketitle


\abstract{After hydrogen and helium, oxygen, carbon, and nitrogen -- hereinafter, the CNO elements -- are the most abundant species in the universe. They are observed in all kinds of astrophysical environments, from the smallest to the largest scales, and are at the basis of all known forms of life, hence, the constituents of any biomarker. As such, their study proves crucial in several areas of contemporary astrophysics, extending to astrobiology. In this review, I will summarize current knowledge about CNO element evolution in galaxies, starting from our home, the Milky Way. After a brief recap of CNO synthesis in stars, I will present the comparison between chemical evolution model predictions and observations of CNO isotopic abundances and abundance ratios in stars and in gaseous matter. Such a comparison permits to constrain the modes and time scales of the assembly of galaxies and their stellar populations, as well as stellar evolution and nucleosynthesis theories. I will stress that chemical evolution models must be carefully calibrated against the wealth of abundance data available for the Milky Way before they can be applied to the interpretation of observational datasets for other systems. In this vein, I will also discuss the usefulness of some key CNO isotopic ratios as probes of the prevailing, galaxy-wide stellar initial mass function in galaxies where more direct estimates from starlight are unfeasible.}




\pagestyle{fancy}
\fancyhf{}
\fancyhead[OL]{The evolution of CNO elements in galaxies}
\fancyhead[EL]{Page \thepage\ of \pageref{LastPage}}
\fancyhead[ER]{D. Romano}
\fancyhead[OR]{Page \thepage\ of \pageref{LastPage}}

\renewcommand{\headrulewidth}{1.1pt}

\raggedbottom

\setcounter{tocdepth}{3} 
\tableofcontents

\section{Introduction and historical perspective}
\label{sec:intro}

In the late fifties, \cite{burb57} wrote their monumental paper titled \emph{``Synthesis of the Elements in Stars"}, hereby laying the foundations of modern stellar nucleosynthesis theory. In that paper, the authors discuss the sequence of processes that lead to the synthesis of all isotopes known in nature and highlight the importance of parallel advances in nuclear physics and astronomy. These, in turn, rest on the constant interplay among observations, experiments and theory. Progresses in nucleosynthesis theory were spurred by \citeauthor{burb57}'s thoughts and recommendations and are recorded in a series of review papers and textbooks \citep[e.g.,][]{trur84,trim91,page97,wall97,clay03,nomo13}. It is beyond the scope of the present work to give a full account of all possible nucleosynthesis paths activating in stars. The aim being to review studies of carbon, nitrogen, and oxygen (CNO) evolution in galaxies, it will suffice to recall that: (i) virtually all $^{14}$N observed in the universe originates in hydrostatic hydrogen burning in stars; (ii) the synthesis of $^{13}$C, $^{15}$N, and $^{17}$O is a consequence of hydrogen burning in stars via both the cold and the hot CNO cycles \citep[see][for a review]{wies10}; (iii) $^{12}$C, $^{16}$O, and $^{18}$O have their origin in helium-burning phases \citep{salp52,clay03}. Different isotopes are produced in different amounts and on different time scales by stars of various initial masses and chemical compositions \citep{tins79}. Stellar duplicity also plays a non-negligible role \citep[see, e.g.,][for a discussion of the effects of binary interactions on carbon production in massive stars]{farm21}. Therefore, in order to follow properly the evolution of the abundances and abundance ratios of different CNO elements in the interstellar medium (ISM), one has to rely on detailed chemical evolution models, relaxing the instantaneous recycling approximation (IRA) and considering the occurrence of binary systems \citep[][and references therein]{matt21}.

Oxygen, carbon, and nitrogen follow hydrogen and helium as the most abundant elements in the universe and may aggregate to form the basic bricks of life. Their spectral features are observed in planetary and stellar atmospheres as well as in gaseous matter in a variety of environments \citep[see, e.g., reviews by][]{wils94,tine13,maio19,rand21}. Presolar dust grains isolated in meteorites feature many CNO compounds \citep{zinn14,nitt16}.

Because of their ubiquity and heterogeneous stellar production sites, the CNO elements have been the subject of extensive study. In their pioneering work, \cite{audo75} and \cite{vigr76} used simple Galactic evolutionary models to link the lower than solar $^{12}$C/$^{13}$C ratio inferred from molecular data of the local ISM to delayed production of $^{13}$C from low-mass stars, including novae. They further noted that the hot CNO cycle occurring in the fast explosive phases of novae may be responsible for the production of $^{15}$N \citep[see also][]{dear78} and that the minor CNO isotopes may be not only produced, but also destroyed in stellar interiors \citep[see also][]{salb55}, meaning that \emph{negative yields}\footnote{Following \cite{maed92}, the element yield $p_j$ of a star of initial mass $m$ is the newly formed mass of element $j$ that is ejected in the ISM during the whole stellar lifetime normalized to the initial mass of the star. Nowadays, the term yield is more often employed to denote the unnormalized ejected mass, $m_j^{\rm{new}} = m p_j$, in solar mass units \citep[][among others]{kara07,cris11,limo18,vent20,vent21,cinq22}. The yield may be positive or negative, depending on whether element $j$ is produced or destroyed in the star.} must be implemented in chemical evolution codes.

In the eighties, the first numerical models relaxing IRA as well as higher quality molecular data became available. Taking advantage of this, \cite{tosi82} analysed the behavior of several CNO isotopic ratios with the galactocentric distance. She concluded that the $^{12}$C/$^{13}$C and $^{16}$O/$^{17}$O ratios behave as typical primary-to-secondary\footnote{We remind the reader that a primary element is synthesized in stellar interiors starting from H and He, while a secondary one needs some metal seeds already present at star's birth to be produced. By virtue of this, the ratio of two primary elements stays almost constant with time/metallicity, while the ratio of a primary to a secondary element is expected to decrease as the metallicity increases.} element ratios, $^{14}$N has both a primary and a secondary component, and $^{15}$N is a regular secondary element. She also pointed out that some unknown mechanism is inhibiting the ejection of $^{18}$O in the ISM over the last few Gyr of Galactic evolution.

The problem of the dual nature of the main N isotope has been a longstanding one. \cite{trur71} first noticed that some primary $^{14}$N production from intermediate-mass stars with initial mass in the range 2--4 M$_\odot$, resulting from convective mixing of material from the helium-burning region to the hydrogen-burning layer is needed, being the sole secondary production insufficient to account for the solar system abundance of this element. The [N/Fe]\footnote{In the standard spectroscopic notation, [X/Y] represents the logarithmic ratio of quantity X/Y with respect to the Sun, i.e., $\log(N_{\mathrm{X}}/N_{\mathrm{Y}}) - \log(N_{\mathrm{X}}/N_{\mathrm{Y}})_\odot$, where $N_{\mathrm{X}}$ and $N_{\mathrm{Y}}$ are the abundances by number of elements X and Y, respectively.}~$\simeq$~0.0 ratios measured in metal-poor Galactic halo dwarfs by \cite{sned74} then suggested that this mechanism is at work also in massive stars \citep[see also][]{barb83}. Observations of nitrogen abundances in H\,II regions of our own as well as nearby spiral and irregular galaxies, comprising low-metallicity dwarf galaxies \citep[e.g.,][]{smit75,peim78,lequ79,garn90}, reinforced the idea that $^{14}$N must have a substantial primary component, though the exact stellar mass range appointed to its primary production remained debated \citep[e.g.,][]{edmu78,matt85,matt86}: hot bottom burning was readily identified as the process leading to primary $^{14}$N (and $^{13}$C) production in massive asymptotic giant branch (AGB) stars \citep{renz81}, but we had to wait the early 2000s to have stellar rotation pinned down as a mechanism able to trigger substantial primary $^{14}$N production in low-metallicity massive stars \citep{maed00,meya02}.

Coming back to early Galactic chemical evolution (GCE) studies, it is worth mentioning the work of \cite{dant91} and \cite{matt91}. These authors implemented a self-consistent calculation of the nova rate  in their GCE model and concluded that novae can represent a non negligible source of $^{13}$C and $^{15}$N in the Galaxy\footnote{This result was confirmed a dozen years later by \cite{roma03}, who adopted detailed nucleosynthesis prescriptions from a grid of hydrodynamical nova models computed by \cite{jose98}.}. In particular, novae would not overproduce $^{13}$C and $^{15}$N as long as these elements are synthesized proportionally to $^7$Li and no major contribution to their primary component comes from intermediate-mass stars. A couple of years later, \cite{henk93} modeled the evolution of CNO isotopes in the nuclear regions of active galaxies and first suggested that a stellar initial mass function (IMF) skewed in favour of high-mass stars could be responsible for the large $^{18}$O/$^{17}$O and small $^{16}$O/$^{18}$O ratios observed in starbursts. Indeed, following earlier detections of $^{13}$CO in five galaxies by \cite{encr79}, observations of carbon monoxide isotopes at millimeter wavelengths were adding up for extragalactic sources, pointing to physical processes in mergers and starbursts sensibly different than those acting in secularly evolving systems \citep{aalt91,caso92}. The idea of an IMF varying with the environment has been given further support in recent years, as we will see in Sect.~\ref{sec:beyond}.

The problem of the evolution of C and O isotopes in the Galaxy was then reassessed by \cite{pran96}, who confirmed the puzzling behaviour of $^{18}$O pointed out in previous studies and the $^{17}$O overproduction on Galactic scales, possibly due to the low $^{17}$O(p, $\gamma$)$^{18}$F and $^{17}$O(p, $\alpha$)$^{14}$N reaction rates adopted in stellar nucleosynthesis calculations. Later on, \cite{roma03}, \cite{koba11}, and \cite{nitt12} still found the need for a reduction of the stellar yields of $^{17}$O to explain its solar system abundance. An improved direct measurement of the 64.5 keV resonance strength in the $^{17}$O(p, $\alpha$)$^{14}$N reaction performed at the Laboratory for Underground Nuclear Astrophysics (LUNA), a unique facility located deep underground in the Gran Sasso National Laboratory, Italy, has recently increased by a factor of two the rate of this reaction at temperatures relevant to shell hydrogen burning in red giant branch (RGB) and AGB stars \citep{brun16}. By adopting the augmented rate, \cite{stra17} have obtained $^{17}$O yields from 15 to 40 per cent smaller for stars with initial mass in the range 2--20 M$_\odot$, thus alleviating the discrepancies between GCE model predictions and $^{17}$O measurements.

When interpreting the CNO abundance data by means of classic GCE models, one relies in essence on a generalization of the time-delay model \citep[][see also a discussion in \citealt{vinc16}]{tins79}. The time-delay model in its narrower sense explains the [O/Fe] versus [Fe/H] pattern traced by stars in the solar vicinity as due to prompt injection of oxygen from short-lived core-collapse supernovae (CCSNe) and delayed iron pollution from type Ia supernovae (SNeIa), the fundamental point being that elements produced via primary nucleosynthetic processes behave as secondary elements from the point of view of chemical evolution when manufactured by low-mass stars with long lifetimes \citep[][see also \citealt{matt89}]{tins79}. On the other hand, primary elements forged in the outer layers of massive stars may exhibit a pseudo-secondary character because of metallicity dependence driven by stellar mass loss \citep{maed92}. It is to this that no unanimous verdict has been reached yet on whether most of $^{12}$C we observe today in the solar neighbourhood comes from low-mass \citep{tins79,chia03,gavi05,matt10} or massive stars \citep{pran94,henr00,aker04,roma20}: in fact, depending on the adopted stellar yields, different authors reach opposite conclusions. Possible variations in the stellar IMF complicate further the picture \citep{matt10,roma20}. Alternative interpretations foresee inhomogeneous chemical enrichment in the framework of hydrodynamical cosmological simulations coupled to alterations of CNO synthesis due to failed SNe at high metallicities \citep{vin18a,vin18b}.

Clearly, our capability of constructing well-sound chemical evolution models that explain self-consistently CNO abundance measurements in various galaxies rests mostly on the robustness of the assumed stellar yields \citep[see][]{roma10,cote17}. These must be computed for as dense a grid of initial stellar masses and metallicities as possible. However, until recently, grids of stellar yields dense enough were woefully lacking in the literature. In particular, yields for very metal-poor and/or super-solar metallicity stars were often missing, essentially due to the paucity, or even absence, of corresponding data to calibrate the uncertain parameters of stellar evolution at the lowest/highest metallicities. I will come back to this issue in Sect.~\ref{sec:nuc}. Because of this dearth of stellar yield data, uncertain interpolations of the yields in the very metal-poor domain have often been used in chemical evolution models for dwarf spheroidal (dSphs), dwarf irregular (dIrrs), blue compact dwarf (BCDs), and ultrafaint dwarf galaxies (UFDs). Similarly, when dealing with systems forming high fractions of metal-rich stars, such as elliptical galaxies and bulges of spirals, chemical evolution modelers have rather arbitrarily extrapolated solar-metallicity yields to the super-solar metallicity regime.

As regards nearby low-mass galaxies, dSphs and UFDs are currently devoid of gas; their CNO abundances are derived from giant stars, with the obvious drawback that the measured abundances of C and N need corrections for stellar evolutionary effects in order to obtain information on the ISM composition at the time the stars were born \citep[e.g.,][and references therein]{ishi14,kirb15,lard16}. On the other hand, dIrrs and BCDs contain large fractions of gas; thus, their present-day CNO abundances can be obtained quite easily from ultraviolet (UV) and optical nebular emission lines \citep[see][for a review]{anni22}. Models aimed at explaining abundance measurements in star-forming dwarf galaxies considered self-enrichment of H\,II regions following gasping or bursting star formation \citep[][and references therein]{tosi91,marc95,tols09}, often taking into account the development of galactic winds \citep[][among others]{matt83,pily93,cari95,pily99}. In particular, differential outflows powered by SN explosions and preferentially enriched in SN ejecta were invoked to explain the observed low metallicities and relatively high N/O ratios \citep{marc94,recc01,roma06}. Galactic winds flowing from intensely star-forming regions can significantly impact the structure and subsequent evolution of the host galaxies \citep{veil05}. Current observational evidence seems to point to a weak feedback in low-mass galaxies \citep{mcqu19,manc20,xuyi21} with preferential loss of metals \citep{mcqu15}, in agreement with the results of hydrodynamical simulations that resolve the cooling radius of the expanding interstellar bubbles \citep[][and references therein]{rom19a}.

As for metal-rich systems, distinction must be drawn between those that can be resolved in stars, such as the inner regions of the Milky Way, and those with ensemble properties from light-averaged spectra, like massive early-type galaxies at both low and high redshifts. Recently, \cite{bens21} have measured C and O abundances of 70 microlensed dwarf, turnoff and subgiant stars residing in the Galactic bulge, thus providing the first statistically significant sample of bulge stars with C abundances not altered by stellar evolutionary processes. The almost total lack of C abundances for unevolved bulge stars until 2021 \citep[see][and references therein]{roma20} was related to the inherent difficulties in obtaining high-resolution spectra of stars that have apparent magnitudes as faint as $V =$ 18--21 at the distance of the bulge. However, during a microlensing event the brightness of the target increases, making possible to get spectra suitable for detailed abundance analyses with reasonable exposure times on large telescopes \citep{bens21}. Nitrogen abundances for a large number of bulge stars are provided by the APOGEE-2 survey \citep[see][and references therein]{kisk21} but only for giants, which requires uncertain evolutionary corrections \citep[][and references therein]{gris21}. Light-averaged element abundance ratios for 3802 early-type galaxies from the Sloan Digital Sky Survey \citep[SDSS,][]{york00} including [C/Fe], [N/Fe], and [O/Fe], are presented by \cite{joha12}. In the most massive systems, the [C/O] ratio is slightly higher than the solar value, namely, [C/O]$\sim$ 0.05, while [N/O] is sub-solar. More recently, near-infrared (NIR) and millimeter telescopes have allowed to unravel the properties of distant galaxies, encompassing the determination of the N/O versus O/H and N/O versus stellar mass relations as well as measurements of CNO isotope ratios from observations of the gas phase \cite[e.g.,][see also \citealt{maio19}, and references therein]{zhan18,brow19,hayd22}.

Overall, providing a coherent picture of the evolution of all seven stable CNO isotopes in galaxies of different morphological type remains one of the open quests for modern astrophysics. This review aims at giving an overview of recent advances in complementary research fields that will hopefully result in an improved understanding of the evolution of CNO elements in galaxies in the next few years. Section~\ref{sec:nuc} discusses progresses in stellar evolution and nucleosynthesis theory, without neglecting to mention the many uncertainties that still plague stellar yield calculations. Section~\ref{sec:models} deals with advances in chemical evolution modeling driven by improvements in the underlying theory as well as instrumental developments that provide more precise and accurate data. In closing the paper, a discussion (Sect.~\ref{sec:bridge}) is followed by a list of future challenges and novel perspectives that are opening up (Sect.~\ref{sec:next}).

\section{Synthesis of CNO elements in stars}
\label{sec:nuc}

We have now a firm grasp of the thermonuclear reactions that trigger the synthesis of the CNO elements in stars. We know that $^{12}$C is synthesised during He burning through the so-called triple-$\alpha$ capture process, whose rate is dominated by capture into the 7.65~MeV Hoyle state in $^{12}$C \citep{hoyl54,free14}, and that $^{16}$O is simultaneously produced at the expenses of $^{12}$C via fusion with another $\alpha$ particle (He nucleus). The cold CNO cycle \citep{weiz38,beth39} takes place in the H-burning zones of main sequence\footnote{Main sequence stars less massive than about 1.5~M$_\odot$ burn hydrogen in the core through the $pp$ chains; the CN(O) cycle becomes dominant at higher masses.} and giant branch stars, leading to the production of $^{14}$N, $^{13}$C, and $^{17}$O through its different branches; stellar convection can then bring the products of CNO cycle burning to the surface \citep{fowl55,burb57,dear74}. The activation of the hot CNO cycle at temperatures exceeding $10^8$~K explains the formation of $^{15}$N and $^{17}$O in H-rich zones \citep[][and references therein]{audo73,wies10}. Finally, most of $^{18}$O produced early in He burning through the $^{14}$N($\alpha$, $\gamma$)$^{18}$F($\beta^+$)$^{18}$O reaction chain is destroyed by $^{18}$O($\alpha$, $\gamma$)$^{22}$Ne reactions occurring at higher temperatures in stellar cores. However, some $^{18}$O may survive in the He shell of massive stars, especially if fresh $^{14}$N is brought to the shell by rotational mixing \citep{hirs07,limo18}. Even if these nucleosynthetic paths may seem crystal clear, we are, sadly, far from having a satisfactory account of CNO production in stars.

The relative rates of the triple-$\alpha$ and $^{12}$C($\alpha$, $\gamma$)$^{16}$O reactions as functions of temperature and density determine the carbon-to-oxygen abundance ratio at the end of He burning, thus affecting the abundances of later nucleosynthesis products (in massive stars) and the composition of the stellar remnants. Unfortunately, the $^{12}$C($\alpha$, $\gamma$)$^{16}$O cross section in the Gamow window corresponding to temperatures for hydrostatic He-burning in stars is very small, of the order of a few $10^{-17}$ barn, well below current direct measurement sensitivity \citep{kunz02}. Laboratory measurements are thus performed at higher energies and the results extrapolated to the astrophysically relevant energy range, which makes the stellar rate of the $^{12}$C($\alpha$, $\gamma$)$^{16}$O reaction one of the most important unsettled rates in nuclear astrophysics \citep[see, e.g.,][]{buch06,alio22}. As for the triple-$\alpha$ reaction, a new measurement of the radiative width of the Hoyle state (after more than 40 years!) by \cite{kibe20} has resulted in a 34\% increase in the reaction rate; its consequences on full stellar nucleosynthesis calculations still need to be explored \citep[see, however,][]{fiel18}. Regarding H burning through the CNO cycles, it is worth mentioning the results obtained over 30 years by the LUNA Collaboration as well as ongoing experiments for many key reactions such as, for instance, $^{14}$N(p, $\gamma$)$^{15}$O, $^{15}$N(p, $\gamma$)$^{16}$O, $^{17}$O($p$, $\alpha$)$^{14}$N, $^{17}$O($p$, $\gamma$)$^{18}$F, $^{18}$O($p$, $\alpha$)$^{15}$N, $^{18}$O($p$, $\gamma$)$^{19}$F, $^{12}$C($p$, $\gamma$)$^{13}$N, and $^{13}$C($p$, $\gamma$)$^{14}$N \citep[e.g.,][and references therein]{anan22}.

Updated rates for these reactions impact the production of CNO elements in stars \citep{herw06,fiel18} and, thus, our comprehension of the origin of stardust grains \citep{palm11,luga17}, mixing processes in giant stars \citep[][and references therein]{stra17}, stellar age dating \citep{imbr04,sala15,casa19}, and chemical evolution of galaxies \citep{roma03,koba11,roma17,rom19b}.

\subsection{Single stars}
\label{sec:single}

Even if the rates of all the reactions involved in stellar nucleosynthesis calculations were known with the necessary accuracy and precision, we would still be struggling with several major problems. First, integration of nuclear reaction networks is a difficult task that requires special care in the choice of the integration method to avoid numerical instability outbreaks \citep[see, e.g., discussions in][]{long14,yagu22}.

Second, different treatments of convection\footnote{Large-scale turbulent motions of the stellar fluid.} in one-dimensional (1D) stellar evolution codes -- including overshoot of convective eddies\footnote{Convective motions do not stop at the point where the acceleration imparted to the fluid elements is zero (which sets the boundaries of the convective region): because of inertia, in fact, the velocity of the convective cells may be non-null even if the acceleration vanishes. This causes material spillover from unstable to stable, stratified regions, which is called overshooting or, sometimes, convective boundary mixing \citep[see, e.g.,][]{roxb65,cast71,sala17}.} -- rely on parametric recipes to approximate physical phenomena that would require a hydrodynamical non-local description, which leads to several shortcomings \citep{renz87,kupk17,arne19}. Other mixing processes have been identified and incorporated in the computations \citep[see the review by][]{sala17}, among which semiconvection \citep{schw58,lang85,merr95}, thermohaline (or fingering) convection \citep{cha07a,eggl08,cant10,deni10}, atomic diffusion \citep[][and references therein]{mich70,mich15}, magnetic buoyancy \citep{buss07,deni09}, meridional circulation and shear-induced turbulence\footnote{The latter processes refer to instabilities driven by stellar rotation.} \citep[][and references therein]{maed09}. Besides taking into account the contribution of these and additional processes generally not included in stellar models -- such as, for instance, gravity waves or internal magnetic fields -- it would be desirable to have their couplings and interactions properly modelled. For example, \cite{cha07b} have shown that stellar magnetic fields may inhibit thermohaline mixing, while rotation might contrast the effects of atomic diffusion \citep[see, e.g.,][]{deal20} in low-mass stars\footnote{According to \cite{deal20}, neglecting rotation and atomic diffusion when estimating the ages of F-type stars may result in errors as large as 25\%.}. Stellar mass loss (see next paragraph) can suppress microscopic diffusion in the outer stellar layers as well \citep[e.g.,][]{vauc95} and \cite{maed13} have clearly shown that some well-known instabilities must be treated jointly in rotating stars, as they interact with each other.

Mass loss prescriptions deeply impact the outcome of stellar evolution models, including the predicted yields \citep[e.g.,][]{chio78,maed92,stan07,beas21}. Stars lose mass at widely varying rates during their lifespan, with a strong dependence of the wind strenght on stellar mass, metallicity and molecular opacities \citep[see, e.g.,][]{mari02,vink05,ven10}. At present, mass loss rates can not be deduced from first principles in 1D stellar models that, hence, rely on empirical relations or theoretical predictions. Only time-averaged rates are implemented, although it is clear that both episodic and eruptive events can occur in real stars. Low- and intermediate-mass stars (objects with initial masses between about 1 and 9 times the mass of our Sun) lose a large fraction of their mass before leaving the AGB. Mass loss rates $\dot{M}_{\mathrm{AGB}} \sim 10^{-8}$--$10^{-5}$~M$_\odot$ yr$^{-1}$ are quite typical of AGB stars; rates as high as $10^{-4}$~M$_\odot$ yr$^{-1}$ are observed at the end of the AGB phase\footnote{For comparison, the mass loss rate of the Sun is $\dot{M}_\odot \sim 2 \times 10^{-14}$~M$_\odot$ yr$^{-1}$, taking into account the contributions from both quasi-steady solar wind and coronal mass ejections \citep[e.g.,][]{mish19}.}. Mass loss rates measured in massive stars span about 8 orders of magnitude. Stars with initial mass in the range 10--25~M$_\odot$ become red supergiants (RSGs), while stars above 40~M$_\odot$ become Wolf-Rayet (WR) stars, with some overlapping in-between. At variance with low- and intermediate-mass stars, for which wind mass loss is relatively unimportant until the most advanced phases of the evolution, massive stars above $\sim$20~M$_\odot$ expel significant fractions of their envelopes already on the main sequence; mass loss in the post-main sequence determines the final fate of the star, i.e., the kind of SN explosion \citep[see][]{smit14}. I refer the reader to the excellent reviews by \cite{hoef18}, \cite{deci21}, and \cite{vink22} -- focusing, respectively, on low- and intermediate-mass stars, AGB and RSG stars, and stars more massive than 25~M$_\odot$ -- for exhaustive reports on mass-loss rate measurements in different wavelength bands, recent advances in the understanding of the mechanisms underpinning and sustaining the outflows, the role of wind clumping and porosity, and any remaining issues in assessing the dependence of the mass-loss rates on the fundamental stellar parameters.

In the following, I review studies of stellar evolution and nucleosynthesis, focusing mostly on those that have produced grids of yields adequate for use in GCE studies. Moreover, I concentrate on CNO element production.

\subsubsection{Low- and intermediate-mass stars}
\label{sec:lims}

The evolution and nucleosynthesis of single low- and intermediate-mass stars, including super-AGB stars, are nicely reviewed in \cite{kara14} and \cite{dohe17}. In this section, I summarize briefly the evolutionary paths of these stars in relation to CNO element production. In addition, I report on the latest theoretical developments.

Stars with zero-age main-sequence masses in the range $\sim$1--9~M$_\odot$ have lifetimes varying from about 12 Gyr to $<$30~Myr, and enrich the ISM with significant quantities of $^{12}$C, $^{13}$C, $^{14}$N, and $^{17}$O (only to mention the elements that are the focus of this review). They are, furthermore, efficient sinks of $^{15}$N and $^{18}$O. The outer stellar layers are expelled by pulsation-enhanced dust-driven winds when the stars are climbing the giant branches.

During the ascent of the RGB, when the star is burning H in a shell surrounding the contracting He core, material previously exposed to partial H burning in the stellar interior is conveyed to the surface (first dredge-up episode). Low-mass stars ($m \le 3$~M$_\odot$, with the limiting mass partly dependent on the initial chemical composition) may experience non-standard (also referred to as extra) mixing processing in the upper part of the RGB that further modifies the chemical composition of the envelope \citep[see][for the impact of such processes on C and N atmospheric abundances]{stan09,laga19}. At this stage, the stellar envelope is enriched in $^{13}$C, $^{14}$N, and $^{17}$O, while the abundances of $^{12}$C, $^{16}$O, and $^{18}$O decrease. The RGB phase ends when He ignites in the core\footnote{Massive intermediate-mass stars at low metallicities ignite He earlier in cores larger than their higher metallicity counterparts and, thus, avoid the RGB phase and first dredge-up event \citep{gira96}.}. Following core He exhaustion, the star develops a complex structure, with a degenerate CO core surrounded by a thermally unstable He-burning shell, separated from an outer H-burning shell by a He-rich region that contains the ashes of H burning. On top of the H-burning shell, a thin radiative layer\footnote{In massive AGB stars, the radiative layer is absent and the bottom of the convective envelope overlaps with the H-burning shell.} is overlaid by a deep convective envelope. When He firstly ignites in the shell following the contraction of the core, a substantial amount of energy is released, which leads to an inwards expansion of the envelope. In stars more massive than about 4~M$_\odot$, the H-burning layer is extinguished and the convective envelope can penetrate in the inner regions: this is the second dredge-up episode that brings to the surface the products of complete H burning, leading to a further increase in $^{14}$N and decrease in $^{12}$C, $^{13}$C, and $^{15}$N. In the most massive intermediate-mass stars C ignites off center in the core in conditions of partial degeneracy, burning in an inward-propagating flame that eventually produces a CO-Ne or ONe core \citep{garc94,farm15}. Objects climbing the AGB with ONe cores are called super-AGB stars\footnote{In super-AGB stars, the second dredge-up event occurs at different stages of the C-burning phase, depending on the initial mass of the star. Moreover, in the most massive super-AGB stars, the so-called corrosive second dredge up is observed: the bottom of the convective envelope penetrates more deeply, enriching the surface layers with $^{12}$C, $^{18}$O, and $^{16}$O \citep[][and references therein]{dohe17}.}.

Now, the thermally-pulsing AGB evolution begins, in which long periods of quiescent H-burning in shell, known as interpulse phases, are interspersed with shell He flashes/pulses powering effective convective motions that homogenize the intershell region; as the pulse fades, convection in the intershell retreats. The energy produced by the pulse makes the envelope to expand (while the star is overall contracting), which switches off the H-burning shell. The base of the outer convective envelope can thus reach regions previously mixed by intershell convection and carry to the surface matter enriched in He-burning products (third dredge up): the star becomes enriched in $^{12}$C. In stars with mass above 3--5~M$_\odot$ (the exact threshold value depending on the metallicity and on the details of convection modelling) the profiles of the thermodynamic variables in the regions above the degenerate core are so steep that the convective envelope is partially overlapped with the H-burning shell. The innermost regions of the external mantle of these stars are therefore exposed to H-burning nucleosynthesis, a phenomenon usually referred to as hot bottom burning (HBB): most $^{12}$C is converted to $^{14}$N and $^{13}$C, moreover, large amounts of $^{17}$O are produced and quickly mixed into the envelope \citep{renz81,wood83}. The comparison between model predictions and abundances observed in AGB stars seems to suggest that some extra mixing is required in low-mass, low-metallicity AGBs \citep{boot95}, but the nature of the process driving the mixing is unclear at present. Regarding solar-metallicity AGB stars, \cite{kara10} have demonstrated that the inclusion of extra mixing on the RGB impacts the subsequent evolution, leading to theoretical C and N abundances consistent with those measured in Galactic carbon-rich stars.

\paragraph{Yields from low- and intermediate-mass stars}

Low- and intermediate-mass stars pollute the ISM through wind mass ejection. The net yield of element $j$ of a star of initial mass $m$, that is, the amount of newly-produced element $j$ ejected in the ISM during the star lifespan, is
\begin{equation}
m_j^{\rm{new}} = m p_j = \int_0^{\tau(m)}[X_j(t) - X_j^{\rm{init}}] \, \dot{M}(t) \, {\rm d} t,
\label{eq:1}
\end{equation}
where $\tau(m)$ is the stellar lifetime, $X_j(t)$ is the abundance of element $j$ in the ejecta at any time, $X_j^{\rm{init}}$ is the abundance of element $j$ in the gas out of which the star is born, and $\dot{M}(t)$ is the mass loss rate. A negative yield simply means that element $j$ is destroyed in the stellar interior rather than produced. The total amount of mass ejected in the form of element $j$ is obtained by adding to $m_j^{\rm{new}}$ the mass originally in the form of element $j$ that is ejected without any nuclear processing in the star:
\begin{equation}
m_j^{\rm{tot}} = m_j^{\rm{old}} + m_j^{\rm{new}} = (m-m_{\rm{remn}}) \, X_j^{\rm{init}} + m p_j,
\label{eq:2}
\end{equation}
where $m_{\rm{remn}}$ is the mass of the remnant; $m_j^{\rm{tot}}$ is, obviously, always positive (it can be zero in practice for deuterium and $^6$Li). Allowing for episodic/eruptive mass loss events rather than for a smooth mass loss rate during a star's evolution could lead to large variations in the predicted yields, especially if the eruptive events coincide with peaks in the surface abundances. Therefore, while many uncertainties (in nuclear reaction rates, molecular opacities, treatment of mixing processes, boundaries of mixed regions, etcetera) affect the calculation of the yields, perhaps the largest uncertainties arise from the assumed mass loss recipes.

\cite{renz81} computed parameterized synthetic evolutionary models for stars with initial mass in the range 1--8~M$_\odot$ and initial metallities $Z =$ 0.004 and 0.02, from the main sequence to the planetary nebula ejection (or carbon ignition in the core), and produced the first grid of metallicity-dependent yields for low- and intermediate-mass stars. The effects of dredge ups, HBB and mass loss were included in the computations and, interestingly, the primary and secondary $^{13}$C and $^{14}$N components provided separately. Since then, improved treatments of the AGB phase and exploration of wider ranges of parameters have made available finer grids of stellar yields suitable for use in GCE studies \citep[e.g.,][]{vand97,mari01,izza04,karb10}. The first yields for super-AGB stars spanning a useful range of initial metallicities, $Z =$~0.0001--0.04, have appeared more recently \citep{sies10}.

However, chemical evolution modellers have the annoying habit of asking always for more, with more refined and extended grids of yields usually being on top of the list (closely followed by `more data'). Indeed, in order to avoid pernicious interpolations and extrapolations \citep[see][their Sect.~2.3]{roma10}, yields for use in GCE codes should cover, as homogeneously and densely as possible, the mass and metallicity ranges of stars that contribute to the chemical enrichment of a variety of stellar systems, both `here and now' and `there and then'. In recent years, there has been a considerable improvement in this direction. In the last decade, Ventura and collaborators \citep{vent13,vent14,vent18,vent20,vent21} have put forward a homogeneous yield set for low- and intermediate-mass stars, including super-AGB stars, covering a large grid of initial masses (from about 1 to 8~M$_\odot$) and metallicities ($Z = 3 \times 10^{-7}$, $3 \times 10^{-5}$, $3 \times 10^{-4}$, 0.001, 0.004, 0.008, 0.014, 0.03, and 0.04). All the models are computed with the stellar evolutionary code \textsc{aton} \citep[][and references therein]{vent98}; the yields are currently provided for light elements up to Fe and are being complemented with the ones for $s$-process elements using a newly developed post-processing code coupled to the \textsc{aton} models \citep{yagu22}.

\begin{figure}[t]
\centering
\includegraphics[width=0.381\textwidth]{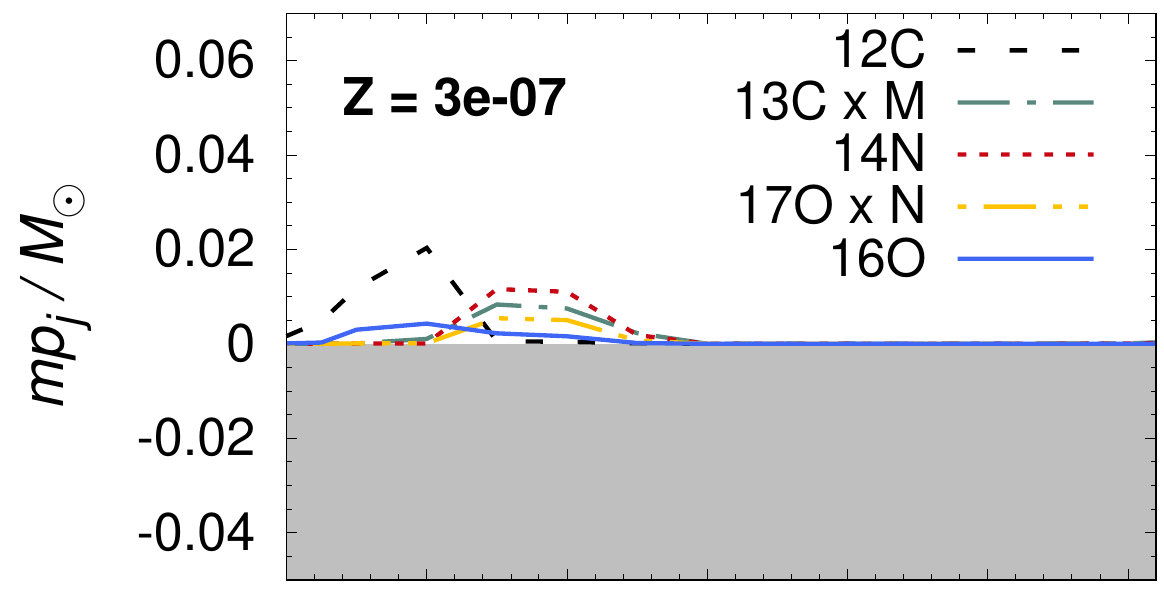}
\includegraphics[width=0.3176\textwidth]{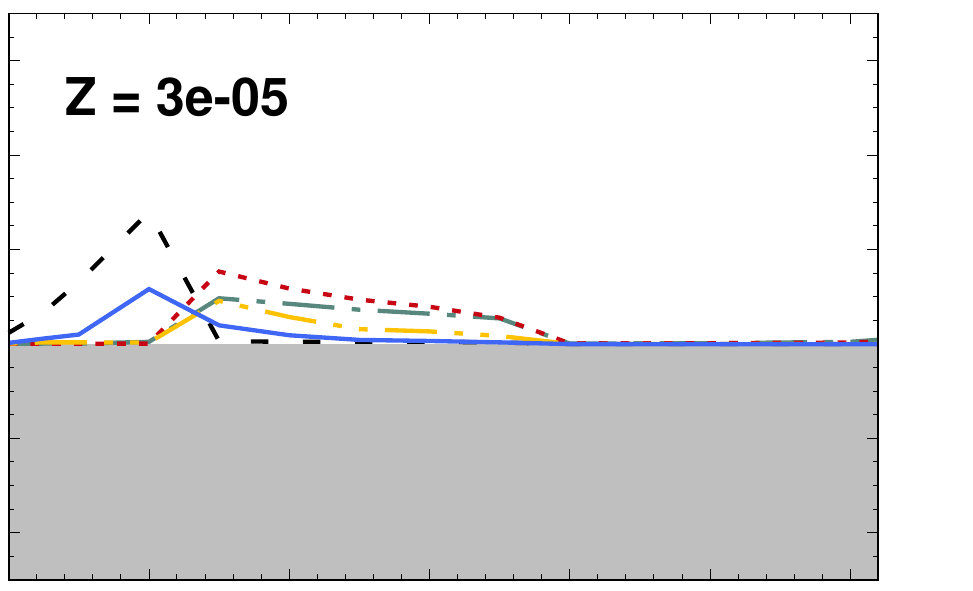} \hspace{-0.42cm}
\includegraphics[width=0.3176\textwidth]{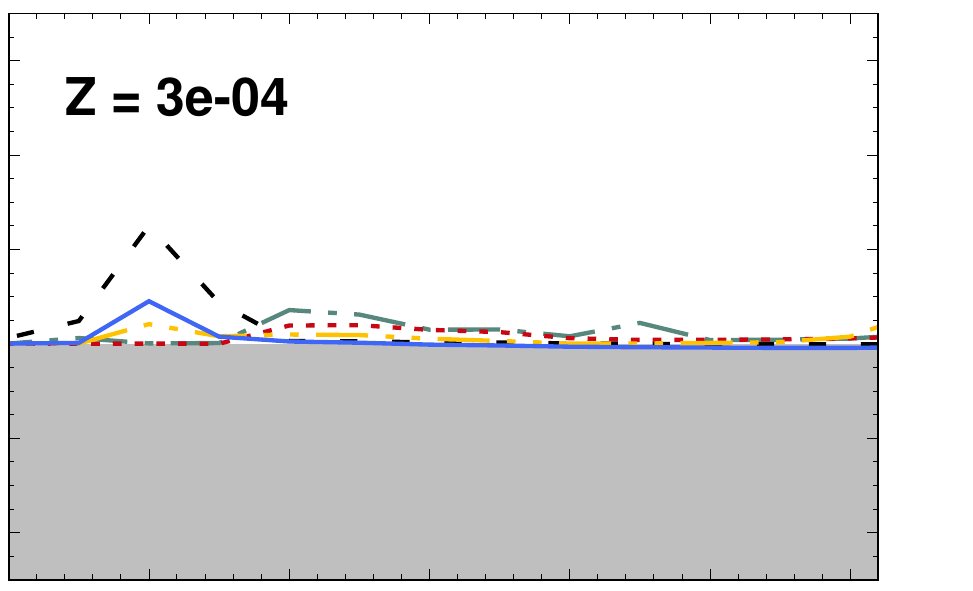}
\includegraphics[width=0.381\textwidth]{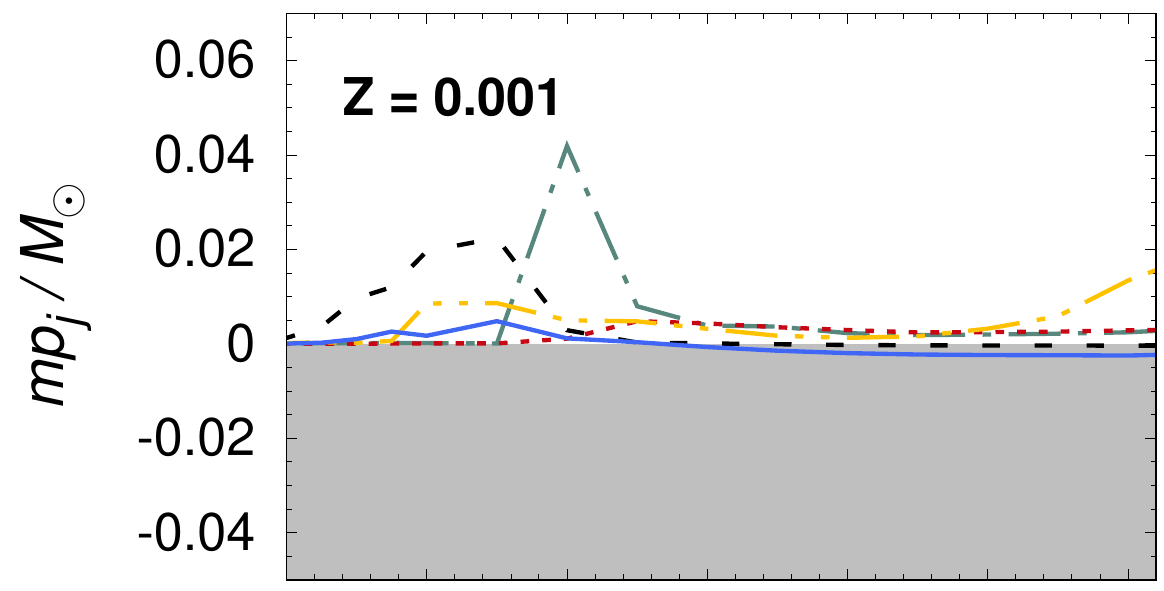}
\includegraphics[width=0.3176\textwidth]{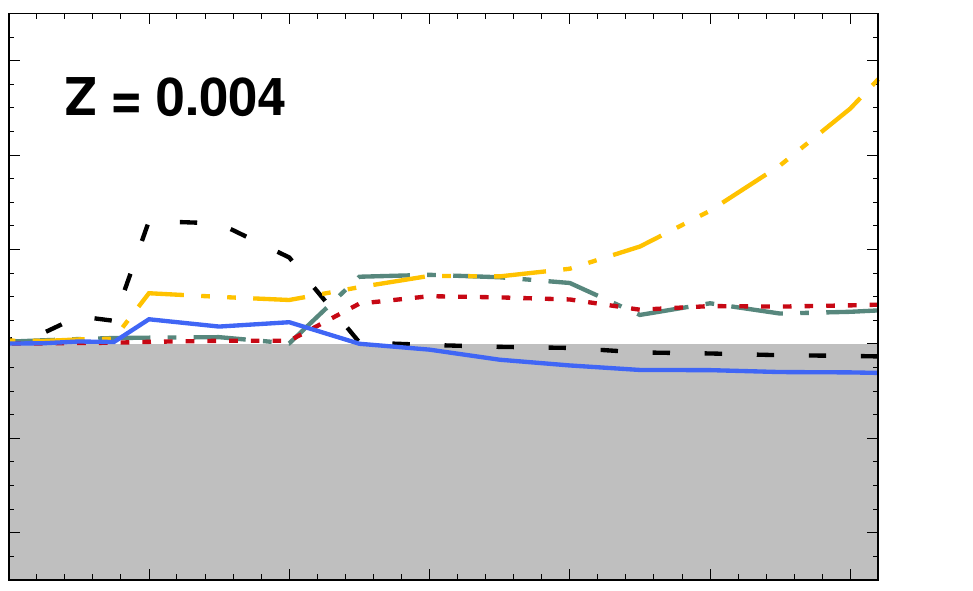} \hspace{-0.42cm}
\includegraphics[width=0.3176\textwidth]{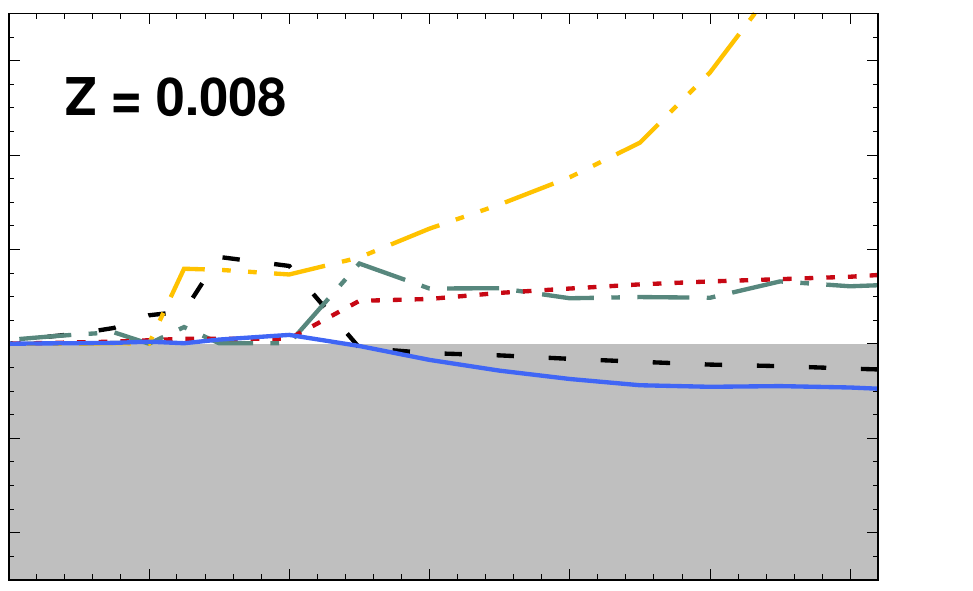}
\includegraphics[width=0.381\textwidth]{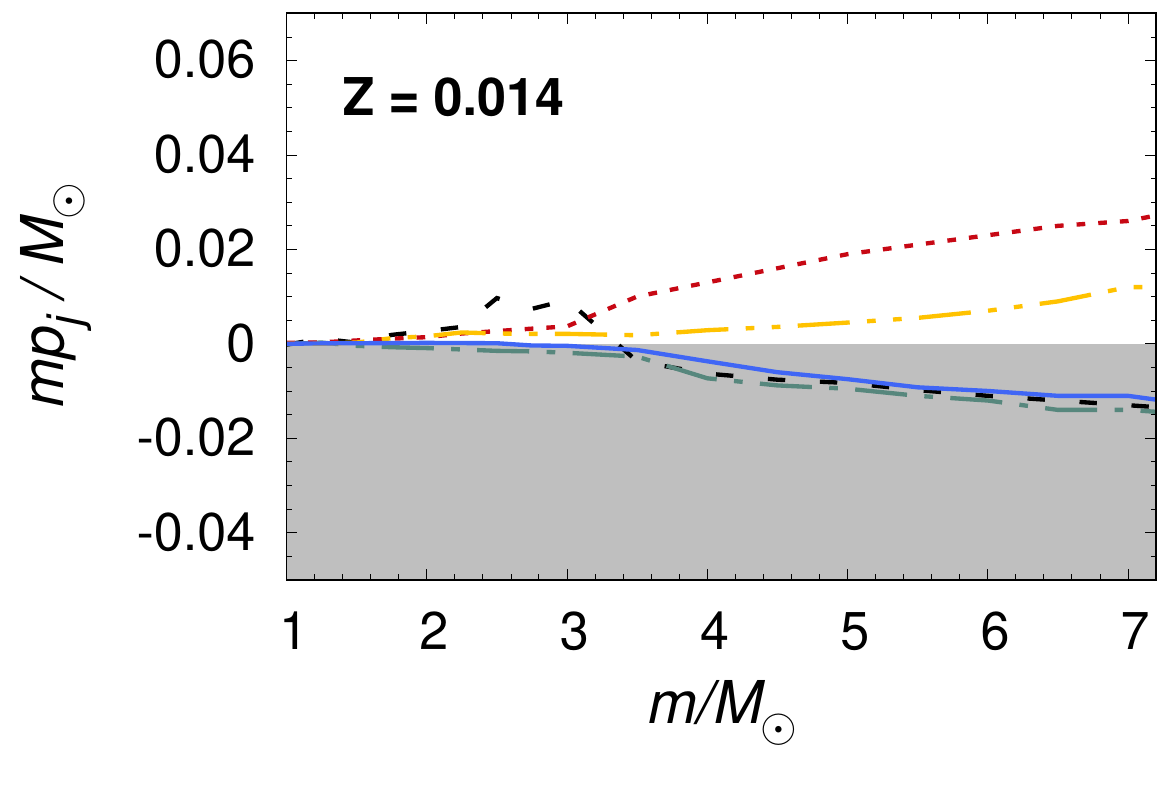}
\includegraphics[width=0.3176\textwidth]{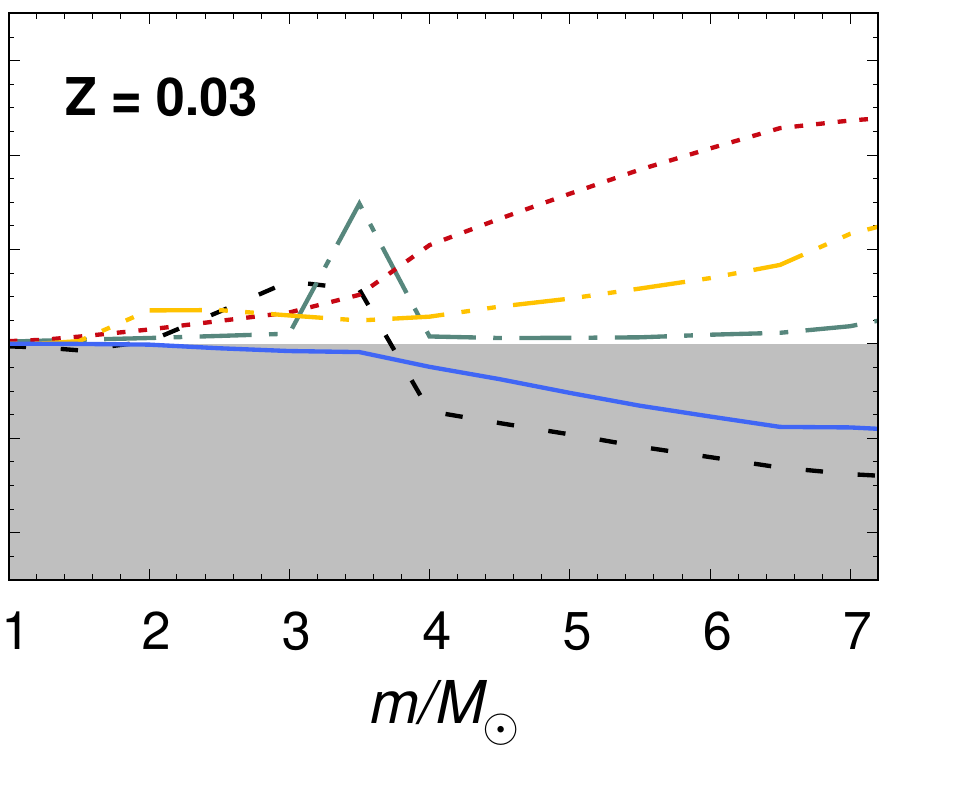} \hspace{-0.42cm}
\includegraphics[width=0.3176\textwidth]{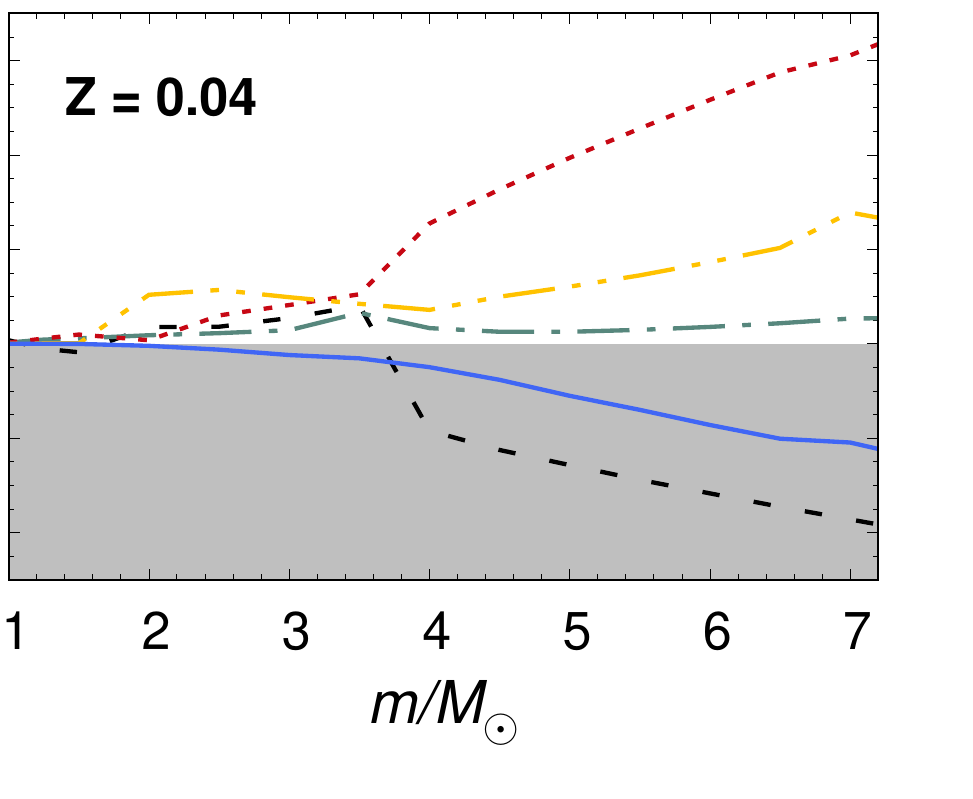}
\caption{ Stellar yields of $^{12}$C (dashed black lines), $^{13}$C (dash-dotted green lines), $^{14}$N (dotted red lines), $^{16}$O (solid blue lines), and $^{17}$O (dash-dot-dot yellow lines) for a homogeneous set of low- and intermediate-mass stellar models \citep{vent13,vent14,vent18,vent20,vent21} computed by means of the \textsc{aton} numerical code \citep[][and references therein]{vent98}. The yields of $^{13}$C and $^{17}$O are multiplied by magnification factors ($\mathrm{M} = 100$, $\mathrm{N} = 1000$  for $Z < Z_\odot$ and $\mathrm{M} = 10$, $\mathrm{N} = 100$ for $Z \ge Z_\odot$) to make them clearly visible. The yields do not include the amount of mass ejected without nuclear processing; hence, negative yields indicate that a given element is destroyed inside the star rather than produced.}
\label{fig:LIMS}
\end{figure}

Figure~\ref{fig:LIMS} shows the full grid of yields for the stable CNO isotopes from the work of Ventura and collaborators. The yields of $^{15}$N and $^{18}$O -- mostly negative, especially at high metallicities, owing to the more efficient CNO cycling -- are not shown, in order to avoid overcrowding. It is immediately seen that low-mass stars are the main $^{12}$C producers, which is readily explained by the fact that they experience only the third dredge up, while their higher-mass counterparts are exposed to HBB. In the low-mass domain\footnote{Notice how this changes with metallicity, being limited to $m \le 2$~M$_\odot$ in the ultra metal-poor regime and spreading to $m \le 3.5$~M$_\odot$ at super-solar metallicities.}, the $^{12}$C yield increases with increasing the stellar mass, owing to the larger number of third dredge-up episodes experienced at higher masses. When HBB turns on, the $^{12}$C yields turn negative. The  $^{16}$O yield follows a similar qualitative behaviour. As regards $^{14}$N, the (scarce) production in the low-mass domain is due to the occurrence of the first dredge up, while significant production takes place when HBB is activated. It is worth emphasizing that the effects of HBB are more pronounced at low metallicities and that the minimum mass for HBB increases with $Z$. Overall, the largest $^{14}$N yields are found at high masses and high metallicities ($Z \ge 0.008$), where secondary production dominates \citep[see a similar figure in][where the primary and secondary contributions to $^{14}$N synthesis are shown separately]{vinc16}. Likewise $^{14}$N, $^{13}$C is brought to the surface by the first dredge up and its surface abundance thereby increases in stars of all masses; however, unlike $^{14}$N, its surface abundance may decrease as a consequence of the second dredge up. A further enhancement can result from HBB if creation via proton capture on $^{12}$C prevails over destruction through the reaction $^{13}$C($p$, $\gamma$)$^{14}$N. The spikes in the predicted $^{13}$C yield may depend on some fine tuning of the HBB process, as also noticed by \cite{mari01}. The only oxygen isotope produced in noticeable amounts by the models is $^{17}$O. For it to be produced, it is necessary that the reaction chain $^{16}$O(p, $\gamma$)$^{17}$F($\beta^+$ $\nu$)$^{17}$O overwhelms the sink reactions $^{17}$O(p, $\gamma$)$^{18}$F and $^{17}$O(p, $\alpha$)$^{14}$N. Interestingly, the astrophysical rate of the latter reaction has been revised upwards lately \citep{brun16}. We refer the reader to the original papers by \citet{vent13,vent14,vent18,vent20,vent21} for a more exhaustive discussion of the dependencies of the yields on the model parameters.

Most recently, \cite{cinq22} have applied the post-processing nucleosynthesis code \textsc{monsoon} developed at Monash University \citep{cann93,luga12} to stellar evolutionary models computed by \cite{kara22} and published a grid of yields for stars with initial mass in the range 1--8~M$_\odot$ and initial metallicities $0.04 \le Z \le 0.10$. It is worth noticing that there are no other yields in the literature for metallicities exceeding twice the solar value. This study shows that the efficiency of the third dredge up is strongly reduced as the metallicity increases. Furthermore, metal-rich stars have higher opacities, are less dense and, thus, reach lower temperatures at the base of their convective envelopes, which allows for only partial hydrogen burning, if any. Nevertheless, irrespective of the temperature at the base of the envelope, the metal-rich intermediate-mass stars produce large amounts of $^{13}$C, $^{14}$N, and $^{17}$O, while efficiently destroying $^{15}$N, $^{16}$O, and $^{18}$O, which suggests that the nucleosynthetic output is determined by a combination of the effects of the first and second dredge ups.

Given the importance of the adopted mass loss prescriptions in determining how and when the AGB phase ends and, hence, in setting the stellar yields, it should be noted that the mass loss rates implemented in the super-solar metallicity models of \cite{cinq22} and \cite{vent20} -- which are taken, respectively, from \cite{vassi93} and \cite{bloe95} -- have been calibrated against metallicity in a limited range. In fact, \cite{vassi93} have evolved models with $0.89 \le m/$M$_\odot \le 5$ and $0.001 \le Z \le 0.016$, while \cite{bloe95} considered stars with initial masses between 1 and 7~M$_\odot$ and initial composition $(X, Y, Z) = (0.739, 0.24, 0.021)$. It is, thus, unclear if the formulae employed still hold in the case of very metal-rich stars.

The yield set by \cite{cinq22} complements previous grids computed with the same post-processing tool for stars of different initial chemical composition presented in \citet[][$0.9 \le m/$M$_\odot \le 6$ and $Z = 1 \times 10^{-4}$]{luga12}, \citet[][$1 \le m/$M$_\odot \le 7$ and $Z = 1 \times 10^{-3}$]{fish14}, \citet[][$1 \le m/$M$_\odot \le 7$ and $Z = 0.0028$]{kara18}, and \citet[][$1 \le m/$M$_\odot \le 8$ and $Z =$~0.007, 0.014, and 0.03]{kara16}. Coupling with the results of computations by \cite{doha14,dohb14}, who investigated five metallicities from $Z = 1 \times 10^{-4}$ to 0.02, allows to extend the mass grid up to $\sim$9~M$_\odot$ (the exact value of the limiting mass depending on metallicity) and offers a homogeneous grid of yields suitable for chemical evolution modelling applications\footnote{Consistent yields for extremely metal-poor stars are published in \cite{gilp21}, but do not cover the 1--3~M$_\odot$ mass range.}.

Other yield sets well-suited for chemical evolution studies can be retrieved by querying the Frascati Raphson Newton Evolutionary Code (FRANEC) Repository of Updated Isotopic Tables and Yields (FRUITY) database\footnote{\url{http://fruity.oa-teramo.inaf.it/}}. The FRUITY web interface allows to download yield tables of all elements from H to Bi for stars with initial masses in the range 1.3--6~M$_\odot$, for a fine grid of initial metallicities ($Z = 2 \times 10^{-5}$, $5 \times 10^{-5}$, $1 \times 10^{-4}$, $3 \times 10^{-4}$, 0.001, 0.002, 0.003, 0.006, 0.008, 0.01, 0.014, and 0.02). Moreover, it is possible to explore the effects of different stellar rotational velocities ($\upsilon_{\mathrm{rot}} =$ 0, 10, 30, or 60 km~s$^{-1}$) and standard versus extended $^{13}$C pockets on the final yields. The relevant stellar models are thoroughly discussed in the accompanying series of papers \citep[see][and references therein]{cris09,cris11,cris15}.

Last but not the least, the NuGrid stellar data set ought to be mentioned. Stellar models are computed with the \textsc{mesa} (Modules for Experiments in Stellar Astrophysics) stellar evolution code \citep{paxt11} and nucleosynthesis is calculated in post-processing using the NuGrid \textsc{mppnp} code \citep[see][]{pign16}. At present, the public database\footnote{\url{https://nugrid.github.io/content/data}} includes complete yields from H to Bi for $m =$~1, 1.65, 2, 3, 4, 5, 6, 7, 12, 15, 20, and 25~M$_\odot$ and $Z = 1 \times 10^{-4}, 0.001, 0.006, 0.01,$ and 0.02 \citep[for a detailed description of the models and yields, see][]{pign16,ritt18,batt19,batt21}. The use of the same stellar evolution and post-processing codes as well as the adoption of a common nuclear reaction network ensure a high degree of internal consistency. Relevant to the subject of this review, in low-mass and massive models at low metallicities H-ingestion in the He shell leads to substantial $^{15}$N production \cite[see][]{pign15}. Although this database has the clear advantage of covering homogeneously different stellar mass realms, it has a major shortcoming in that it does not provide stellar yields for stars above 25~M$_\odot$.

\subsubsection{Notes on the fate of the most massive AGB stars}
\label{sec:ecsn}

\begin{figure}
\centering
\includegraphics[width=0.55\textwidth]{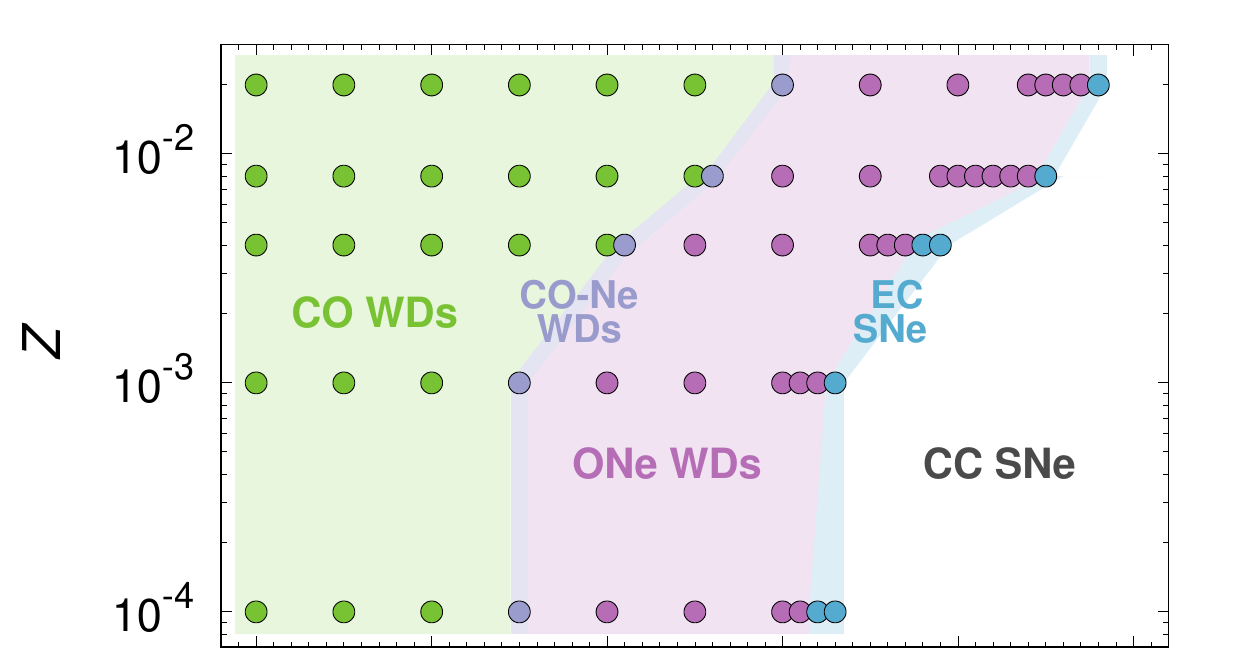}
\includegraphics[width=0.55\textwidth]{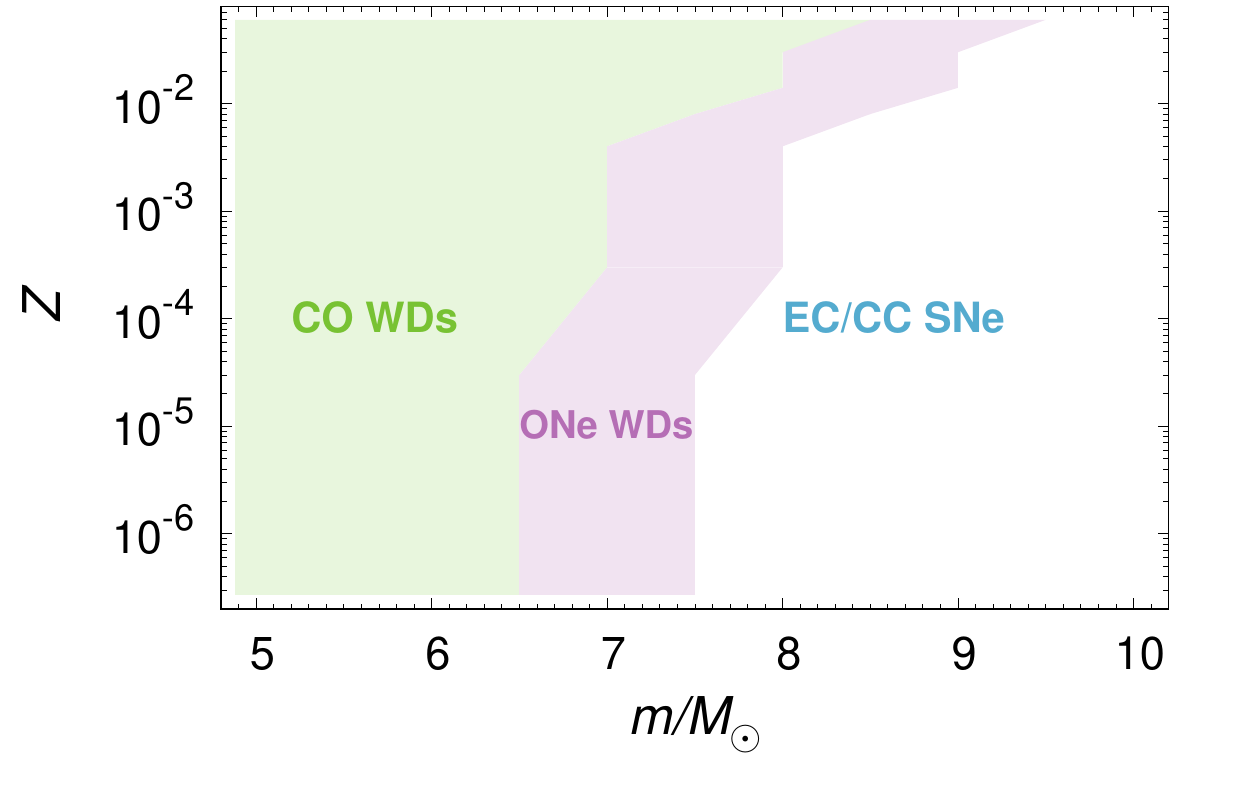}
\caption{ \emph{Top panel:} final fates of intermediate-mass stars as a function of metallicity after \cite{dohe15}. Each circle represents a model star and the different colours tell if the remnant is a CO WD, a CO-Ne WD, an ONe WD, or an ECSN. \emph{Bottom panel:} same as the top panel, but for the stellar models computed by \citet[][and P.~Ventura, private communication]{vent13,vent14,vent18,vent20,vent21}.}
\label{fig:remn}
\end{figure}

Figure~\ref{fig:remn} illustrates the endpoints of the evolution of intermediate-mass stars with zero-age main-sequence masses between 5 and 10~M$_\odot$ and initial metallicities in the interval $Z =$~0.0001--0.02 as reported by \citet[][\emph{top panel}]{dohe15} and in the range $Z = 3 \times 10^{-7}$--0.04 after \citet[][and P.~Ventura, private communication, \emph{bottom panel}]{vent13,vent14,vent18,vent20,vent21}. The two sets of models are in reasonable qualitative agreement; both show that the minimum masses for the onset of carbon burning, formation of neutron stars, and explosion as SN increase with metallicity. However, in the \textsc{aton} models the ONe white dwarf (WD) progenitors are confined to a narrower range of initial masses.

Broadly speaking, stars in a narrow initial mass range\footnote{The exact interval depends on the metallicity and on the treatment of semiconvective mixing during He core burning and overshooting from the convective core during central H and He burnings \citep{poel08,sies10,vent11,farm15}.} climb the AGB with ONe cores after igniting carbon off-center under partially degenerate conditions (see previous section). Simulations can now follow the evolution of this class of objects all the way through the entire thermally pulsing super-AGB phase. This is a rather demanding computational task, mostly because of the extremely short time step, of the order of $10^{-3}$--$10^{-2}$ yr, dictated by H burning during the interpulse phases \citep{sies10}, and the large number of thermal pulses expected \citep{dohe17}. The evolution of super-AGB stars resembles that of their lower-mass counterparts, however, owing to the massive ONe cores, they develop weaker thermal instabilities and may reach very high temperatures at the base of the convective envelope \citep[up to $140 \times 10^7$~K, depending on the efficiency of convective energy transport,][]{sies10}. This leads to very efficient HBB. Therefore, super-AGB stars may become powerful $^{13}$C, $^{14}$N, and $^{17}$O forges, with yields heavily dependent on the temperature at the base of the convective envelope and on the rate of consumption of the mantle \citep[hence, on the treatment of convection, see][]{sies10,vent11,dohe17}.

 The fate of the most massive intermediate-mass stars is highly uncertain, as the velocity of erosion of the convective mantle and the rate of core growth deeply influence the outcome of the models \citep{poel08,vent11,woos15,leun20}. Stars with ONe core masses in excess of $\sim$1.35~M$_\odot$ \citep{schw16} ignite neon and end their lives as electron-capture SNe (ECSNe). Objects with smaller ONe cores die quietly, leaving cooling ONe WDs as remnants \citep[e.g.,][]{leun20}. Primordial (metal-free) stars in this mass interval could grow their degenerate cores up to the Chandrasekhar limit owing to the inefficient mass loss and explode as type 1.5 SNe \citep{gilp07,poel08}.
In summary, the competition between core growth (influenced by the third dredge-up efficiency) and envelope removal by mass loss (influenced by dredge ups that change the envelope opacity and, hence, the strength of the stellar winds) dictates whether a star dies as an ONe WD or an ECSN. \cite{poel08} estimate that from about 3 to 20\% of all SNe might be ECSNe. \cite{dohe15} and \cite{jone19} both point to low percentages, 1--5\%. Since ECSNe are characterized by explosion energies and nickel production factors an order of magnitude lower than those of typical Fe CCSNe \citep{kita06,wana09,hira21,zhas21}, a reliable estimate of their expected rate would be highly welcome as it would improve the modelling of the energetic and chemical feedback in galaxies.


\subsubsection{High-mass stars}
\label{sec:high}

In the next paragraphs, I briefly recap the main facts regarding the evolution of massive stars that evolve in isolation. The section closes with some considerations about CNO element production from single stars that enter the main sequence with masses in excess of 10~$M_\odot$.

It is common wisdom by now that OB stars with initial masses between 8--12 and 20--25~M$_\odot$ evolve through the RSG phase, which is characterized by the presence of an extended convective envelope, and explode as type II-P SNe, leaving neutron stars as remnants and dominating the rate of explosions in the nearby universe \citep[][and references therein]{smar09,lang12,chie13}. The evolutionary routes of stars above 20--25~M$_\odot$, instead, are still shrouded in mistery and there is no unanimous consensus on their final destiny \citep[e.g.,][]{lang12}. Stars with initial masses in the range 40--120~M$_\odot$ possibly evolve through the WR phase sheding most of their H-rich envelopes and explode as type Ibc SNe, leaving behind either neutron stars or black holes \citep[see][]{hege03}. Of particular interest are the evolution and fate of the most massive objects ($m > 80$~M$_\odot$ and up to 350~M$_\odot$) that may die as pair instability SNe \citep[PISNe, also called pair creation SNe,][]{lang07,yuso13,leun19}. For a star to die as a PISN, it is necessary that it is left with a massive H-rich envelope at the pre-SN stage. Since the higher the metallicity, the more efficient the erosion of the outer layers by mass loss, there is a metallicity threshold above which essentially no PISNe are expected \citep[but see][and references therein]{geor17}. \cite{lang07} estimate that at redshift $z = 5$ one should see one PISN per 100 SNe, while this fraction drops to 1/1000 in the local universe. As we will see in Sect.~\ref{sec:models}, including PISN nucleosynthesis in chemical evolution models might have important consequences for our understanding of the earliest phases of galactic evolution, especially with regard to early CNO element enrichment: the ejecta of low-metallicity very massive stars, in fact, are expected to be very rich in C and O \citep{gosw21} and underabundant in N \citep{salv19}.

Single massive stars with initial masses between about 10 and 120~M$_\odot$ undergo all stages of nuclear burning to the point of Fe collapse. In the case of a successful conversion of the collapse into an explosion, the mantle experiences shock heating and explosive burning. In spherically symmetric 1D simulations, the transition from collapse to explosion is usually forced, e.g., by locating a piston\footnote{Other authors use kinetic \citep[e.g.,][]{limo18} or thermal bombs \citep[e.g.,][]{tomi07}.} somewhere in between the Fe core edge and the base of the O shell and setting the kinetic energy of the ejecta at infinity to a value inferred from observations: the piston is first moved inward, then rebounced generating a shock wave \citep[see, e.g.,][]{woos95,woos07}. In the last decade, three-dimensional (3D) radiation hydrodynamic simulations of the collapse have succeeded in producing powerful explosions without any artifice, starting from 3D progenitor models. Spherical symmetry breaking, neutrino heating behind a stalled shock, and neutrino-driven turbulence\footnote{The turbulence arising in the multidimensional initial models is a key factor in providing the seed turbulence that rejuvenates the shock \citep{vart22}.} are the key drivers of the explosion of most CCSNe \citep{burr21}, while the thermonuclear \citep[e.g.,][]{jone19} and magnetically-driven \citep[e.g.,][]{burr07,eise22} channels underlie a minor fraction of the global events, respectively, at the low- and high-mass ends.

\paragraph{Yields from massive stars}

Following \cite{maed92}, the net yield of element $j$ of a massive star of initial mass $m$ consists of two contributions:
\begin{equation}
m_j^{\rm{new}} = m p_j = m p_j^{\rm{wind}} + m p_j^{\rm{SN}}.
\label{eq:3}  
\end{equation}
The first term on the right-hand side of Eq.~(\ref{eq:3}) refers to the contribution from stellar winds,
\begin{equation}
m p_j^{\rm{wind}} = \int_0^{\tau(m)}[X_j(t) - X_j^{\rm{init}}] \, \dot{M}(t) \, {\rm d} t,
\label{eq:4}
\end{equation}
while the second one generally accounts for the effects of explosive nucleosynthesis,
\begin{equation}
m p_j^{\rm{SN}} = \int_{m_{\rm{remn}}}^{\widetilde{m}(\tau)}[X_j(m') - X_j^{\rm{init}}] \,{\rm d} m'.
\label{eq:5}
\end{equation}
In the above formulae, $\tau(m)$, $\dot{M}(t)$, $X_j^{\rm{init}}$, $X_j(t)$, $X_j(m')$, $m_{\rm{remn}}$, and $\widetilde{m}(\tau)$ are, respectively, the stellar lifetime, the assumed mass loss rate, the initial abundance of element $j$, the abundance of element $j$ at the surface at any time, the abundance of element $j$ at the Lagrangian mass coordinate $m'$, the mass of the remnant, and the remaining mass at age $\tau$. Again, to obtain the total amount of mass ejected in the form of element $j$ one needs to add to $m_j^{\rm{new}}$ the mass originally in the form of element $j$ that is injected in the ISM without having suffered any nuclear processing inside the star [see Eq.~(\ref{eq:2})].

Quantitative estimates of the yields from numerical simulations of massive star evolution performed in systematic studies that span a large range of initial masses and chemical compositions are exactly what chemical evolution modellers need. Several teams, adopting different (1D) stellar evolutionary codes and different input microphysics, have provided homogeneous grids of massive star yields for use in GCE models. All these works have important limitations. In the following, we comment on a few that have been widely used in GCE studies.

Back in the nineties, \citet{maed92} investigated the effects of metallicity on the yields for two different initial chemical compositions, $Z =$~0.001 and 0.02. His models include mass loss, but the evolutionary tracks of massive stars stop at the end of central C-burning. Moreover, only a few representative elements are considered. \citet{woos95} computed the evolution and explosion  of massive stars with initial metallicities $Z/Z_\odot =$ 0, 0.0001, 0.01, 0.1, and 1 and masses in the range 11--40~M$_\odot$ and provided detailed yield tables for several isotopes. However, their nuclear reaction network is truncated at Zn, convective core overshooting is not taken into account and, more important for the purposes of the present discussion, mass loss is not considered. \citet{thie96} used a larger reaction network, but investigated only solar-metallicity stars, had a lower number of mass points in their grid and, again, did not consider the effects of mass loss during the pre-SN evolution. \citet{limo06,limo12} computed yields for solar-metallicity and metal-free stars with initial masses between 11 and 120~M$_\odot$ (a slightly reduced mass grid was employed for zero-metallicity stars). They took mass loss into account \citep[see][for the adopted mass loss prescriptions as well as other details of the code]{limo06}. In the meantime, the Geneva group published the first grids of yields including the effects of both mass loss and rotation \citep{meyb02,hirs05,hirs07,ekst08}. Their models are limited to the pre-SN evolution, but this should not be a serious issue for galactic CNO evolution studies (see below). More recently, \citet{nomo13} presented a compilation of extant, homogeneous yields for CCSNe and hypernovae\footnote{These are highly energetic explosions, more luminous than any known SN \citep{pacz98}, occurring in stars more massive than 20--25~M$_\odot$.} (HNe) from the Japanese team, to which they added unpublished, consistent yields for super-solar metallicity stars \citep[see][their Sect.~4.2]{nomo13}. \citeauthor{nomo13}'s models include the effects of mass loss, but not those of rotation. The upper mass limit is set to 40~M$_\odot$, apart from the grid of zero-metallicity models that extends to 100~M$_\odot$. Finally, \cite{limo18} presented new yields from a grid of models with initial metallicities [Fe/H]~= $-3$, $-2$, $-1$, and 0 (corresponding to $Z = 3.236 \times 10^{-5}$, $3.236 \times 10^{-4}$, $3.236 \times 10^{-3}$, and $1.345 \times 10^{-2}$), initial masses $m =$ 13, 15, 20, 25, 30, 40, 60, 80, and 120~M$_\odot$, and equatorial velocities at the beginning of the main sequence $\upsilon =$ 0, 150, and 300~km~s$^{-1}$; the model stars are susceptible to mass loss and some may explode as PISNe.

\begin{figure}
\centering
\includegraphics[width=0.95\textwidth]{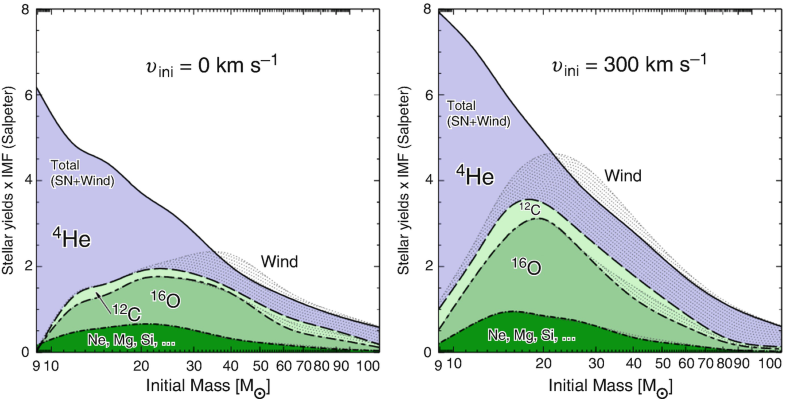}
\caption{ IMF-weighted stellar yields (see text) of different elements for non-rotating \emph{(left-handed panel)} and rotating \emph{(right-handed panel)} solar-metallicity stellar models as a function of the initial mass, piled up on top of each other. The dotted areas highlight the contribution of the wind. Note that $^{16}$O has negative SN yields for $m >$ 50~M$_\odot$ ($m >$ 30~M$_\odot$) in the non-rotating (rotating) case. Image from \cite{hirs05}, reproduced with permission \copyright \, ESO.}
\label{fig:hirs}
\end{figure}

At this point, it is worth recalling that CNO elements in massive stars are synthesized mostly during hydrostatic burning and seldom suffer modification of their yields during explosive burning \citep[e.g.,][]{woos07}. Therefore, the yields depend crucially on the pre-SN evolution and, hence, on the assumed nuclear reaction rates, convection treatment, mixing processes, and mass loss rates, while having weak or no dependence on other uncertain parameters of the models, such as the mass cut\footnote{This is the point that separates the ejecta from the collapsing core.}. At high metallicities, in the presence of mass loss large amounts of He and C are vented out and escape further processing to oxygen and its descendants; as a consequence, high-metallicity high-mass stars have large C, but low O yields \citep{maed92}. In hypernovae, oxygen burning takes place at higher temperatures in more extended regions, which results in lower C and O abundances in the ejecta \citep{umed02}. On the other hand, rotation may significantly increase the yield of oxygen. Figure~\ref{fig:hirs} shows the stellar yields [computed according to Eq.~(\ref{eq:3})] from the solar-metallicity, non-rotating \emph{(left-handed panel)} and rotating \emph{(right-handed panel)} models of \cite{hirs05}, weighted by an extrapolated \citeauthor{sala55}'s (\citeyear{sala55}) IMF and multiplied by an arbitrary constant. Rotation clearly boosts the production of oxygen and, to a lesser extent, that of carbon, a result qualitatively confirmed by \cite{limo18}. At low metallicities, mass loss plays a minor role, but rotation-induced instabilities trigger the interplay between the core He-burning and the H-burning shell that results in a significant primary production of $^{13}$C, $^{14}$N, $^{17}$O, and $^{18}$O \cite[][see also \citealt{hirs07}]{limo18}. This, as we will see in Sect.~\ref{sec:models}, has a dramatic impact on the predictions of chemical evolution models.

In closing this section, it is worth mentioning the possibility that a large fraction of massive stars with primordial, or nearly primordial, metallicity explode with lower explosion energies, i.e., as faint SNe. In this case, higher [(C + N)/Fe] and [(C + N)/Mg] ratios, corresponding to smaller ejected Fe masses and to larger compact remnant masses, have to be expected \citep[e.g.,][]{hege10,tomi14} with important measurable consequences for the chemical composition of the stars at extremely low metallicities. More details on this issue will be given in Sect.~\ref{sec:dsphufd}.

\subsection{Binary systems}
\label{sec:bin}

So far, I have discussed the evolution of stars that spend their lives in isolation. However, most stars form in clusters and associations \citep{lada03}, where interactions are frequent. In this section, I focus on binary systems in which close interaction may boost CNO element production. A thorough review of binary classes and their pathways can be found in \cite{dema17}.

Most young Galactic stars with $m \ge$ 15~M$_\odot$ interact with a companion before exploding as CCSNe \citep{sana12}, which modifies their evolutionary paths, nucleosynthesis, SN appearance, and remnant properties \citep[see, e.g.,][]{lang12,modj19,woos19,schn21}. \citet{pods04} investigated how the presence of a close companion affects the final core structure and terminal evolution of a massive star and found, among other effects, that ECSN explosions are favoured in binary systems.

The nucleosynthetic outputs of massive binary interactions haven't been scrutinized in so much detail as to allow implementation in GCE codes yet, which is understandable. Each binary component, in fact, brings with it all the uncertainties on the treatment of convection, extra mixing, mass loss, and microphysics discussed in the previous sections. Duplicity adds further ambiguity linked, for example, to the details of mass transfer and leakage by Roche-lobe overflow or to the effects of tides on rotation velocity evolution and internal mixing \citep{song16}. The separation and the mass ratio between the primary and the secondary component of the system must also be specified. All of this makes the parameter space to expand significantly.

\begin{figure}
\centering
\includegraphics[width=0.75\textwidth]{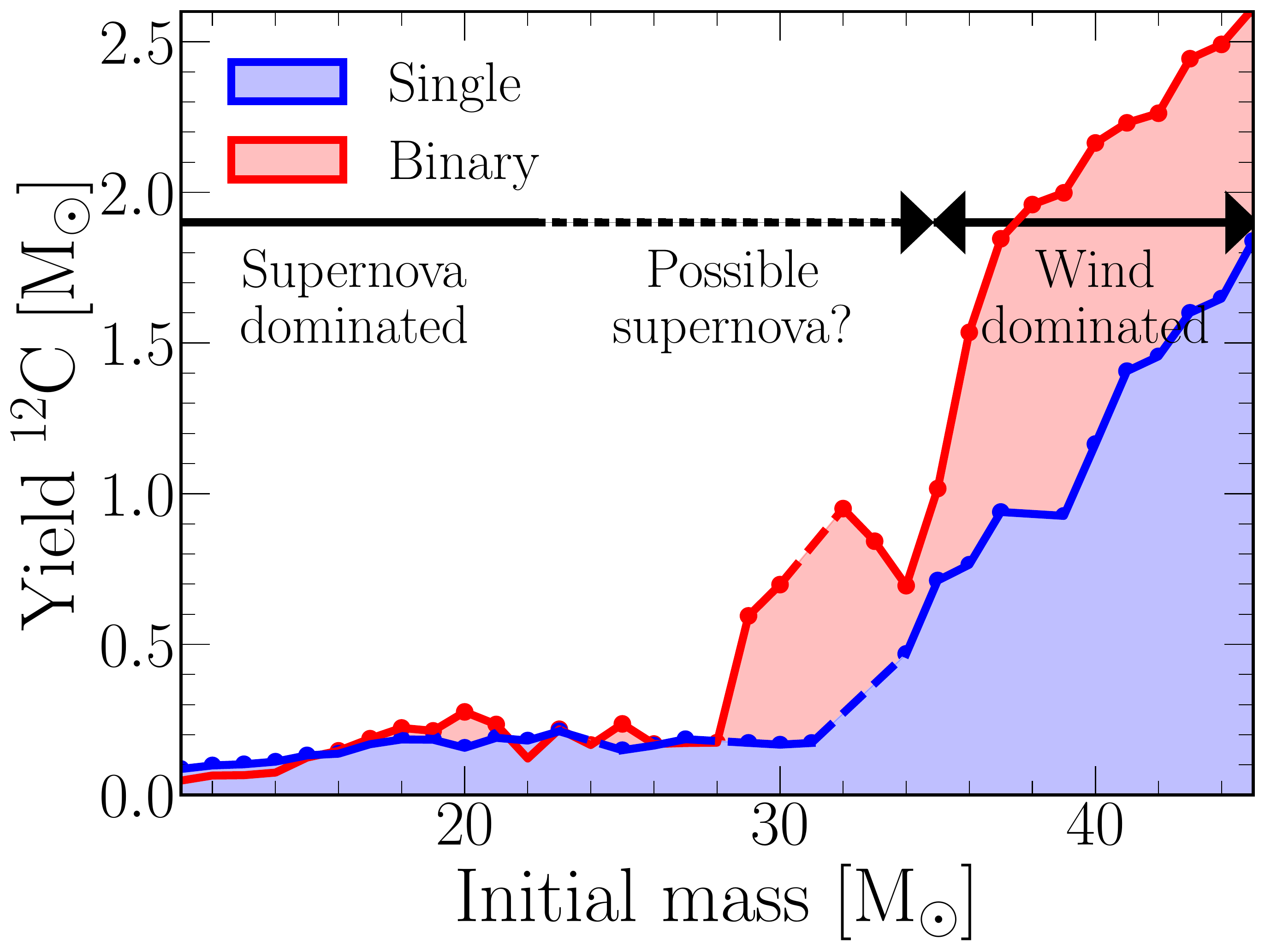}
\caption{ Net $^{12}$C yield including the contributions from stellar winds and SN explosions as a function of the initial stellar mass. Blue lines and areas denote the results from single star models, while red lines and areas represent binary models. The $^{12}$C yields are dominated by SN explosions in the range 11--22~M$_\odot$, whereas winds dominate above 35~M$_\odot$. Figure adapted from \cite{farm21}, licensed under a Creative Commons Attribution 4.0 International Licence. Reproduced by permission of the author.}
\label{fig:farm}
\end{figure}

\cite{farm21} have calculated the C yields of solar-metallicity massive stars that either evolve alone or are stripped in binary systems during H burning in shell (notice that this is only one of the many possible fates of massive stars with close companions). In binary-stripped stars the convective He-burning core does not grow (it may rather retreat) and the outermost layers disconnect forming a $^{12}$C-rich pocket that never reaches temperatures high enough for $^{12}$C burning. In single stars, instead, these layers get mixed into the growing core where $^{12}$C is readily destroyed by $\alpha$ captures first, and C burning later. As a consequence, binaries are more efficient in enriching the ISM in $^{12}$C than single stars (see Fig.~\ref{fig:farm}). However, the authors themselves caution that different choices of wind mass-loss rates could have a not negligible impact on the results. The $^{12}$C yields are, instead, remarkably robust against changes in the explosion parameters (see also Sect.~\ref{sec:high}). The main source of uncertainty (in both single and binary models) appears to be the treatment of convective boundary mixing. Yields of other species should become available in the future, which will open new avenues to our knowledge of how chemical enrichment proceeds in the cosmos taking stellar duplicity into account.

Regarding lower mass systems, it has to be emphasized that many solar-type and intermediate-mass stars have companions, but the interaction is less frequent than in the case of more massive stars \citep[see, e.g.,][]{kouw05,ragh10,moem17}. This notwithstanding, some low-mass close binaries may lead to spectacular events that have been, and still are, the subject of a great deal of theoretical work. In the remainder of this section, I focus on classical novae that prove to be efficient manufacturers of some rare CNO isotopes.

Classical nova explosions occur in close binary systems where a WD primary accretes H-rich matter from a secondary, less evolved star that overflows its Roche lobe. Once the temperature of the deepest layers accreted on top of the WD exceeds about $10^8$~K, a thermonuclear runaway is initiated \citep[see][and references therein]{gure57}. The resulting outburst does not disrupt the WD and, after the ejection of the envelope, the process is re-initiated. On average, the typical nova experiences $10^4$ outbursts during its lifetime \citep{bath78,ford78}.

\begin{figure}
\centering
\includegraphics[width=0.55\textwidth]{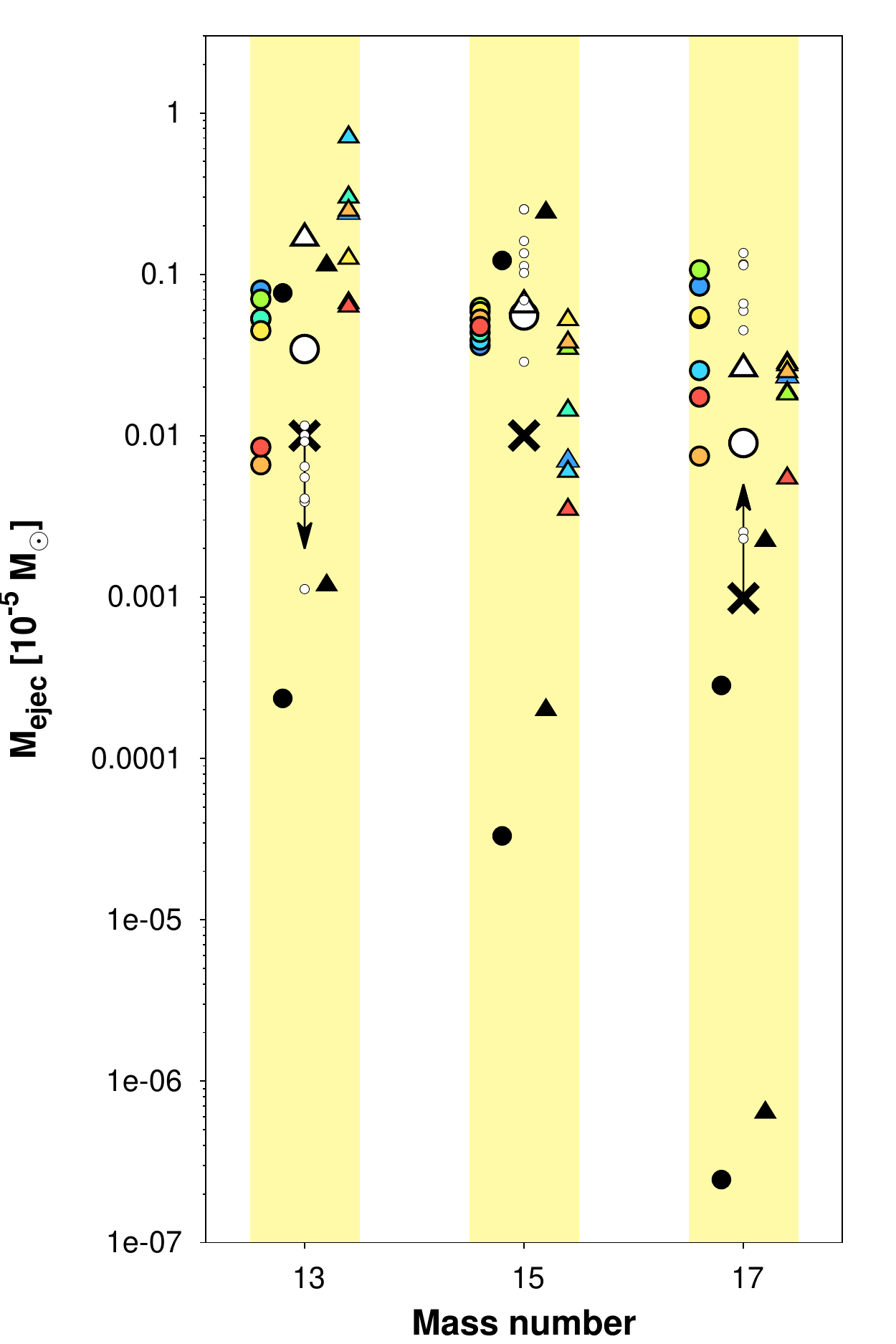}
\caption{ Yields of $^{13}$C, $^{15}$N, and $^{17}$O from hydrodynamic models of the nova outburst for CO (triangles) and ONe WDs (circles). Results from representative 1D models \citep[][coloured symbols; \citealt{yaro05}, black symbols; \citealt{starr09}, small empty circles]{jose98} are shown together with yields from simulations that switch from one to three dimensions, and viceversa \citep[123--321 models,][large empty symbols]{jose20}. In the latter suite of models, the metal enhancement of the envelope is not forced, but naturally produced. The big $\times$ signs represent the average yields adopted in chemical evolution simulations by \citet{roma17}. Figure adapted from \cite{roma17}, copyright by the authors.} 
\label{fig:roma}
\end{figure}

In their pioneering hydrodynamical studies, \citet{starr72,starr78}, \citet{pria78}, and \citet{pria86} developed models of the nova outburst and showed that for a successful outburst to develop, the envelope must be enriched in CNO elements. Moreover, only massive WDs produce outbursts: \cite{kove85} find that models with $m_{\rm{WD}} = 0.6$~M$_\odot$ do not lead to a nova outburst, but instead settle into a thermal equilibrium configuration with a luminosity typical of a red giant. In the models of \citet{starr72}, the CNO cycle is always far out of equilibrium in the outburst and the $\beta^+$-unstable nuclei $^{13}$N, $^{14}$O, $^{15}$O, and $^{17}$F become very abundant. Convection carries these unstable nuclei to the surface where they decay to $^{13}$C, $^{14}$N, $^{15}$N, and $^{17}$O and the temperatures are too low for further $p$-captures, which affects the composition of the ejecta. \citet{jose98} performed a thorough investigation of the role of classical novae as ISM polluters, adopting the 1D implicit hydrodynamic code SHIVA to follow the course of nova outbursts from the onset of accretion up to the ejection of the processed envelope. Fourteen models were computed, spanning WD masses from 0.8 to 1.35~M$_\odot$ and mixing levels between the accreted envelope and the underlying WD from 25\% to 75\%. Both CO WDs and ONe WDs were considered. To limit the parameter space, the rate of mass accretion and the initial luminosity were kept fixed, at $2 \times 10^{-10}$~M$_\odot$ yr$^{-1}$ and $10^{-2}$~L$_\odot$, respectively. From these simulations, novae turn out to be significant producers of $^{13}$C, $^{15}$N, and $^{17}$O, possibly contributing most of their Galactic abundances \citep{roma03}.

Over the last twenty years, multi-dimensional studies of mixing at the core-enve-lope interface during outbursts have demonstrated that Kelvin-Helmholtz instabilities develop, naturally resulting in an effective mixing of the outermost WD material into the accreted layers. This makes the envelope metallicity to achieve values in agreement with the observed ones, namely, $Z \sim$ 0.20--0.30 \citep[e.g.,][]{glas97,casa10,casa16,jose20}, which ensures powerful nova outbursts. Simulations implementing -- or guided by -- these results have started to produce more coherent and self-consistent nova yields \citep{jose20,starr20}.

Figure~\ref{fig:roma} compares the masses ejected in the form of $^{13}$C, $^{15}$N, and $^{17}$O in a single outburst from different nova models. We display both results from 1D simulations \citep{jose98,yaro05,starr09} and results from models that combine 1D and 3D hydrodynamics \citep[123--321 models,][]{jose20}. Yields computed with different codes and specifics may differ by up to four orders of magnitude. The $\times$ signs denote the values adopted in the GCE models of \citet{roma17}. A detailed discussion of nova nucleosynthesis implementation in GCE models is deferred to Sect.~\ref{sec:mw}.

\section{Modeling the evolution of CNO elements in galaxies}
\label{sec:models}

From all the above it is clear that the stellar yields bring with them considerable uncertainties. Nonetheless, they are fundamental ingredients of GCE models \citep[actually, their most important inputs; see][]{roma10,cote17}.

GCE models follow the variation of the chemical composition of the ISM in space and time in galaxies taking into account various processes, such as gas inflows and outflows, star formation, feedback from stars, and radial motions of gas and stars \citep[see][and references therein]{matt21}. They rely on empirical formulae and semi-analytical, yet physically-motivated laws to describe complex processes. This makes the computational cost affordable even in the case of a full screening of the parameter space. This kind of models can thus be used, among other things, to efficiently test different grids of stellar yields, providing useful hints to more elaborate simulations -- it would be rather unpleasant to run expensive chemodynamical simulations first, then discover that the adopted stellar nucleosynthesis prescriptions are in need of a radical overhaul!

The theoretical abundances and abundance ratios likewise depend on the adopted galaxy-wide stellar IMF (gwIMF), which is used to weigh the yields. A broad consensus has emerged over the years that the first stars -- called, somewhat counterintuitively, Pop~III stars -- formed with a top-heavy IMF; chemical enrichment by the first SNe then suddenly turned the star formation mode from high-mass to low-mass dominated \citep[e.g.,][]{karl13}. While such claims have solid theoretical foundations, they still lack an unambiguous observational confirmation. In fact, it is found that low-mass stars can form at very low metallicity \citep[$Z < 7 \times 10^{-7}$;][]{cafb11}. Also, the average abundance patterns of Milky Way halo stars -- as well as those of some weird objects -- do not necessitate a peculiar gwIMF to be reproduced \citep[e.g.,][]{limo03}. From the point of view of chemical evolution, Pop~III stars, if ever existed, produced negligible effects in the earliest phases of Galactic evolution \citep{ball06}. Possible systematic variations of the gwIMF in dependence of a galaxy's metallicity and star formation rate have been proposed \citep[][and references therein]{jera18,yanz20,yanz21} that deserve further investigation\footnote{At present, there are still contrasting views concerning possible systematic variations of the IMF with the environment \citep[see, e.g.,][]{bast10,hopk18,smit20}.}.

In particular, it has been suggested that the abundance ratios of some specific CNO isotopes may provide an effective litmus test of possible gwIMF variations in starburst galaxies, once reliable measurements are analysed by means of GCE models \citep[][see Sect.~\ref{sec:beyond}]{henk93,papa14,roma17,zhan18}. This procedure, however, requires an a-priori careful calibration of the models against local CNO abundance data. This is the subject of the following section.

\subsection{The local universe as a benchmark}
\label{sec:bench}

In this section we deal with the evolution of the CNO elements and their isotopes in the Milky Way and other selected galaxies in and beyond the Local Group, up to about 20 Mpc distance from us. The latter are grouped in galaxies with (Sects.~\ref{sec:dirrbcd} and \ref{sec:andro}) and galaxies without (Sect.~\ref{sec:dsphufd}) a detectable gas component at present.

\subsubsection{The Milky Way galaxy}
\label{sec:mw}

In a number of galaxies in the Local Group it is possible to resolve single stars. In particular, in our own Galaxy high-resolution spectra of unevolved stars can be analysed, which provides detailed, high-precision CNO abundances reflecting the chemical composition of the gas out of which the stars were born. Low-mass stars formed at different epochs and still alive today can thus be used to trace the evolution of CNO elements in the Galaxy from its earliest formation phases to the present time.

CNO abundances in stars can be determined from permitted atomic lines of C\,I and O\,I, forbidden [C\,I] and [O\,I] lines, and molecular bands of C$_2$, CH, NH, OH, CN, and CO in the UV, optical and NIR \cite[][and references therein]{barb18,rand21}. Significant efforts are ongoing to obtain accurate and precise abundances for, ideally, millions of stars in the Galaxy representing the variety of its stellar populations, from the inner bulge to the disc outskirts \citep[see][and references therein]{jofr19}. Coupling these abundance measurements with the exquisite astrometric data provided by the European Space Agency (ESA) \emph{Gaia} mission \citep{perr01,gaia16} and with precise stellar ages \citep{gall19,migl21} allows an unprecedented multidimensional mapping of stars within about 10~kpc from the Sun. These high-precision positions, motions, abundances, and ages can be effectively used to constrain galaxy formation and evolution scenarios as well as stellar evolution and nucleosynthesis theory \citep{free02,niss18}.

In the last fifteen years, large ground-based spectroscopic surveys, such as the \emph{Gaia-}ESO Survey \citep[GES,][]{rand13,rand22,gilm22}, the Apache Point Galactic Evolution Experiment \citep[APOGEE,][]{maje17}, the Large sky Area Multi-Object fiber Spectroscopic Telescope (LAMOST) survey \citep{zhao12}, and the GALactic Archaeology with HERMES survey \citep[GALAH,][]{desi15}, have measured detailed properties of several hundred thousand stars in the Milky Way discs, finding that the stars populate two distinct sequences in the [$\alpha$/Fe]--[Fe/H] space \citep{miko14,nide14,reci14,hayd15,yu21}. The observed dichotomy confirms and extends to a larger portion of the disc original findings by \citet{fuhr98}, who found the thin- and thick-disc components to be chemically disjunct from the analysis of a sample of solar neighbourhood stars \citep[see also][]{grat96}.

Pure chemical evolution models for the Milky Way disc have been developed assuming either a sequential or a parallel formation of its components \citep[for a thorough review, see Sect.~5 of][]{matt21}. The two-infall model of \citet{chia97,chia01}, for instance, assumes two main episodes of gas infall, out of which the thick and thin discs form in a sequence. \cite{spit19} fine-tuned this model by leveraging recent constraints on stellar ages from asteroseismology to reproduce the [$\alpha$/Fe] versus [Fe/H] relationship at different Galactic epochs, the age-metallicity relation, and the metallicity and age distributions of the stars populating the high-$\alpha$ and low-$\alpha$ sequences. A crucial assumption of their revised two-infall model is the longer time delay imposed for the start of the second gas infall event, i.e., $\sim$4 Gyr as opposed to 1~Gyr in the original version. The parallel approach developed by \cite{gris17,gris18} envisages instead the formation of the thin and thick discs out of two distinct accretion episodes on different timescales. Also in this case, the parallel sequences in the [$\alpha$/Fe]--[Fe/H] plane and the stellar metallicity distributions of the thin- and thick-disc populations can be reproduced. Chemically distinct discs are also predicted in the context of chemodynamical simulations of the formation and evolution of Milky Way-sized galaxies. In some models the two components emerge from separate evolutionary pathways \citep[e.g.,][]{gran18,mack18,ager21,khop21}, whereas others propose a continuous star formation history and stellar migration to generate the double sequence \citep[][and references therein]{shar21}. To fine tune the free parameters relevant to the chemical enrichment process, the stellar metallicity distributions and [$\alpha$/Fe]--[Fe/H] trends are commonly used. The evolution of the CNO elements and their isotopes are rarely investigated. This is partly due to the inherent difficulties in obtaining trustworthy CNO element abundances that reflect the ISM enrichment processes.

\paragraph{The Sun}

It is sobering, in fact, that even the solar photospheric abundances of the CNO elements have been revised several times over the years. \cite{lamb78} suggested $A$(O)$_\odot = 8.92 \pm 0.10$\footnote{On the usual abundance scale, $A$(X)~$\equiv \log(N_\mathrm{X}/N_\mathrm{H}) + 12$, where $N_\mathrm{X}$ and $N_\mathrm{H}$ are the abundances by number of element X and hydrogen, respectively.}, $A$(C)$_\odot = 8.67 \pm 0.10$, and $A$(N)$_\odot = 7.99 \pm 0.10$, using as primary abundance indicators the [C\,I]\,$\lambda$\,8727~$\AA$ line, the CH\,A-X and C$_2$ Swan systems, the N\,I lines and the forbidden atomic oxygen lines at 6300 and 6363~$\AA$. In their prominent review, \citet{ande89} recommended a value of $A$(O)$_\odot = 8.93 \pm 0.04$. The solar O abundance has undergone major downwards revisions since, culminating with the low value of $A$(O)$_\odot = 8.66 \pm 0.05$ suggested by \citet{aspl04}. Later on, \cite{ayre06} pointed again to a high value, $A$(O)$_\odot = 8.84 \pm 0.06$, which was revised downwards to $A$(O)$_\odot = 8.76 \pm 0.07$ by \cite{caff08}. Recommended literature values for the solar photospheric C abundance similarly oscillate from high to low figures -- for instance, $A$(C)$_\odot = 8.592 \pm 0.108$ in \cite{holw01}, $A$(C)$_\odot = 8.39 \pm 0.05$ in \cite{aspl05}, and $A$(C)$_\odot = 8.50 \pm 0.06$ in \cite{caff10}. The determination of the solar abundance of nitrogen also suffered from important downwards revisions: \cite{ande89} list $A$(N)$_\odot = 8.05 \pm 0.04$, \cite{holw01} recommends $A$(N)$_\odot = 7.93 \pm 0.11$, while \cite{caff09} favour $A$(N)$_\odot = 7.86 \pm 0.12$, to cite only a few relevant references. Incorrect or incomplete modelling of the different lines considered in different studies offers a likely explanation for the discrepancies \citep[see, e.g.,][]{caff13}.

The chemical portrait of the Sun by \cite{aspl21}, based on 3D spectral line formation calculations also taking into account departures from local thermodynamic equilibrium (non-LTE), ends up with a set of present-day photospheric abundances, $A$(O)$_\odot = 8.69 \pm 0.04$, $A$(C)$_\odot = 8.46 \pm 0.04$, and $A$(N)$_\odot = 7.83 \pm 0.07$, that nicely agree within the mutual errors with previous 3D-based studies \citep[][see the above-quoted values]{caff08,caff09,caff10}. In their most recent work, \cite{magg22} recommend $A$(O)$_\odot = 8.77 \pm 0.04$, $A$(C)$_\odot = 8.56 \pm 0.05$, and $A$(N)$_\odot = 7.98 \pm 0.08$, again in excellent agreement with the results by Caffau and collaborators. The concordance of these latest studies is very encouraging.

Concerning the minor isotopes, values of $91.4 \pm 1.3$, $2738 \pm 118$, and $511 \pm 10$ (at 1~$\sigma$) are inferred for the $^{12}$C/$^{13}$C, $^{16}$O/$^{17}$O, and $^{16}$O/$^{18}$O ratios, respectively, from IR (2--6~$\mu$m) rovibrational bands of CO in the solar spectrum\footnote{The authors caution that the observed $^{12}$C$^{17}$O features are very weak.} using 3D convection models \citep{ayre13}. The nitrogen isotopic ratio presents outstanding variability on the Solar System scale, which can be explained by invoking fractionation processes \citep{furi15}. For the protosolar nebula, a low ratio, $^{14}$N/$^{15}$N~$\simeq 441 \pm 6$, is suggested by \citet{mart11}.

\paragraph{Other stars}

Modern 3D line formation models including non-LTE effects provide solar photospheric abundances consistent with helioseismic constraints \citep[][see also \citealt{pins09,cafa11}]{magg22}. Consideration of 3D non-LTE effects is even more crucial in the low-metallicity regime. \cite{amar19} derived homogeneous 3D non-LTE carbon, oxygen, and iron abundances for a sample of 187~F and G~dwarfs belonging to the Galactic discs and halo. The inferred abundance trends are tighter than -- and, at low metallicities, deviate markedly from -- those obtained in the classical 1D LTE approximation. Thick-disc stars display [C/Fe] and [O/Fe] ratios higher than those of their thin-disc counterparts, while the opposite is true in the [C/O]--[O/H] plane \citep[see also][but see \citealt{bens06}, for different results]{niss14}. Notably, the upturn of [C/O] for metallicities below [O/H]~$\simeq -$1 spotted out in 1D LTE analyses \citep{aker04} is not recovered by 3D non-LTE line formation models \citep[see][their Fig.~13]{amar19}. This has profound implications for the characterization of the stellar factories responsible for the synthesis of C and O at early times.

\begin{figure}
\centering
\includegraphics[width=0.75\textwidth]{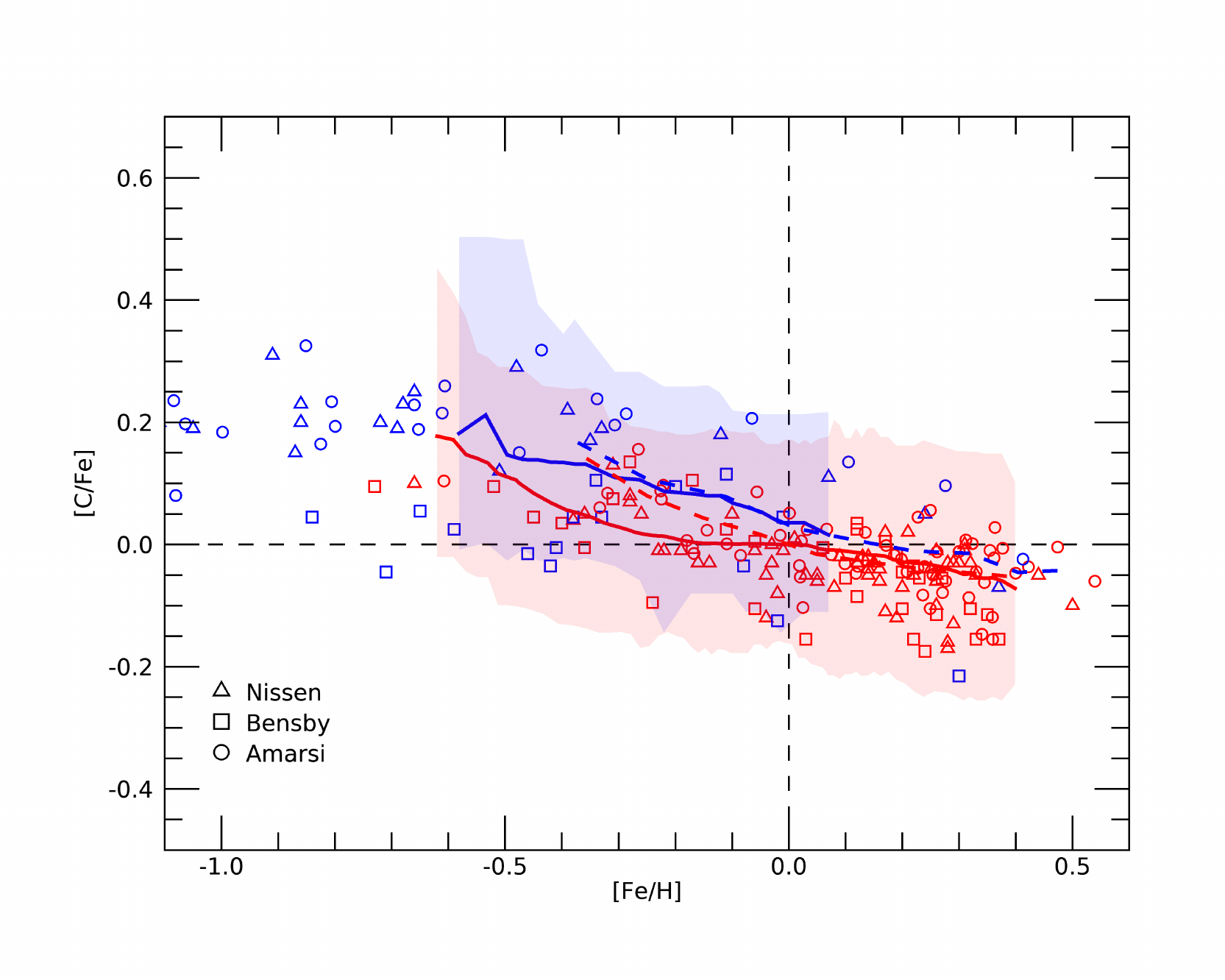}
\includegraphics[width=0.75\textwidth]{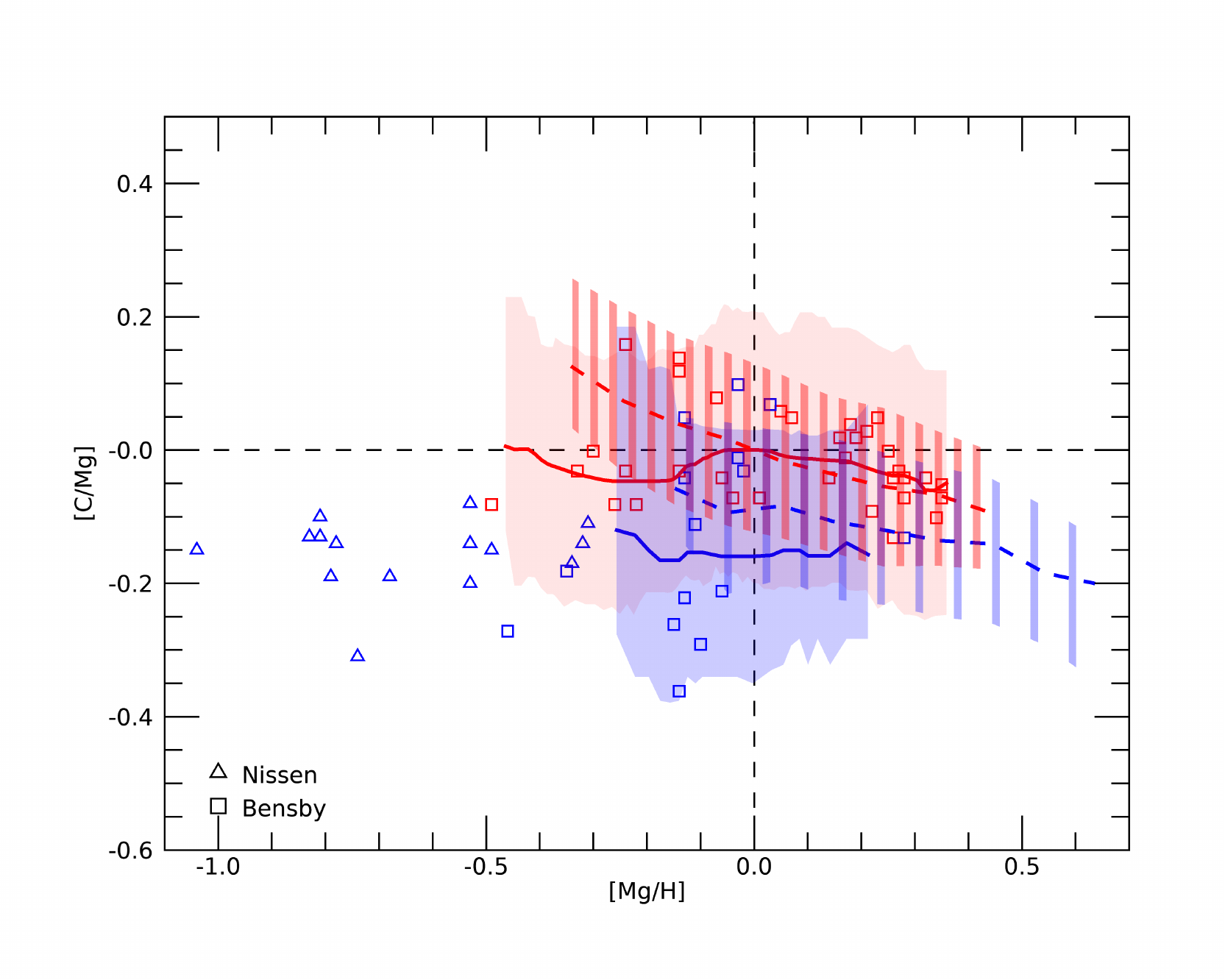}
\caption{ Carbon-to-iron \emph{(top panel)} and carbon-to-magnesium \emph{(bottom panel)} abundance ratios as functions of [Fe/H] and [Mg/H], respectively. Red and blue symbols are abundance estimates for thin- and thick-disc unevolved stars (triangles: \citealt{niss10,niss14}; squares: \citealt{bens06,bens14}; circles: \citealt{amar19}). Solid red and blue lines are abundance ratio trends of, respectively, $\sim$1\,300 thin- and $\sim$100 thick-disc dwarf stars selected on the basis of chemical criteria from GES iDR5 \citep{fran20}. Dashed red and blue lines are abundance ratio trends of more than 12\,000 high-Ia and low-Ia stars, respectively, from GALAH DR2 \citep{grif19}. First-order zero-point offsets (see text) are applied to the data. Pale red and blue areas show the spread in thin- and thick-disc GES iDR5 abundance data within boundaries corresponding to the 10$^\mathrm{th}$ and 90$^\mathrm{th}$ percentiles. Red and blue dashed areas show the analogous spread in the GALAH DR2 thin- and thick-disc datasets (\emph{bottom panel} only). Figure from \cite{roma20}, reproduced with permission \copyright \, ESO.} 
\label{fig:coohSN}
\end{figure}

Apart from the inclusion or omission of 3D and/or non-LTE effects in the derivation of the abundances, other systematics may affect the data, impacting the comparison between observations and theoretical predictions and, hence, the conclusions of GCE models. As an example, in Fig.~\ref{fig:coohSN} we display different datasets in the [C/Fe]--[Fe/H] and [C/Mg]--[Mg/H] planes (\emph{top} and \emph{bottom panels,} respectively). The red and blue symbols refer to abundance measurements from high-resolution, high signal-to-noise ratio (SNR) spectra of single thin- and thick-disc stars, respectively, presented in \citet{bens06,bens14,niss10,niss14,amar19}. The solid red and blue lines represent the average trends derived by \cite{fran20} from, respectively, about 1\,300 thin-disc and nearly 100 thick-disc dwarf stars from the fifth internal data release of GES (GES iDR5). The dashed red and blue lines have similar meanings, but refer to the analysis of more than 12\,000 stars from the second data release of GALAH (GALAH DR2), with the stars chemically divided into high-Ia and low-Ia sequences based on the higher/lower Fe contribution from SNeIa in the thin-/thick-disc phases \citep{grif19}. Only unevolved stars with C abundances not altered by evolutionary processes are considered. In order to remove at least partly systematic offsets, zero-point corrections are applied to each dataset, so that thin-disc stars with [X/H]~$\simeq 0$ also have [C/X]~$\simeq 0$ (where X stands for either Fe or Mg).

From Fig.~\ref{fig:coohSN}, \emph{top panel,} it appears that, in general, the agreement between the offset [C/Fe] versus [Fe/H] trends traced by GES and GALAH data is very good. Moreover, there is an overall agreement with the area covered by individual data points from the other high-resolution spectroscopic studies, especially if the spread in the survey data (here only for the GES dataset -- pale red and blue areas for thin- and thick-disc stars, respectively) is taken into account. The rather large extension of the two scatter areas, $\pm 0.2$~dex, is likely due to a combination of uncertainties in the derived abundance ratios -- some of the GES spectra have relatively low SNR -- and cosmic scatter. The fact that the individual points from the selected high-resolution studies, which are affected by much smaller uncertainties, span similar wide areas indicates that the cosmic scatter is significant for both thin- and thick-disc stars. Figure~\ref{fig:coohSN}, \emph{bottom panel,} with Mg as the metallicity tracer, points to a less favourable figure: the GES [C/Mg] trends are flatter than the GALAH ones for both discs and agree better with the loci of the individual observations, especially for the thick-disc component. The spreads in GES and GALAH data are comparable in size (when also taking the different number of stars in the two samples into account) and resemble that of the individual points, thus supporting the presence of a cosmic scatter also in [C/Mg]. In this respect, it is worth noting that examination of [O/Fe] abundance ratios of Galactic disc stars with $H$-band spectra from APOGEE reveals a star-to-star cosmic variance at a given metallicity of about 0.03--0.04 dex in both the thin- and thick-disc components that is tentatively associated with the wide range of galactocentric distances spanned by APOGEE targets \citep{bert16}.

It is, therefore, interesting to analyse the behaviour of stars that span a narrower distance range. \cite{delg21} have derived homogeneous C abundances for a volume-limited sample (most targets are within 60~pc distance from us) of 757 FGK dwarfs with $T_{\mathrm{eff}} > 5200$~K and $-1.2 <$~[Fe/H]~$< 0.6$. The stars have been observed at very high resolution ($R \sim$~115\,000) with the High Accuracy Radial velocity Planet Searcher (HARPS) spectrograph \citep{mayo03} mounted on the ESO La Silla 3.6~m telescope. The sample does not contain any fast rotators and active stars. Two well-known atomic C lines (C\,I\,$\lambda$\,5052~$\AA$ and C\,I\,$\lambda$\,5380~$\AA$) have been subjected to a standard LTE analysis. The two lines provide consistent C abundances in the considered $T_{\mathrm{eff}}$ range \citep[see Fig.~1 of][]{delg21}. Oxygen abundances have also been derived using either the O\,I\,$\lambda$\,6158~$\AA$ line or the [O\,I]\,$\lambda$\,6300~$\AA$ line. The usage of the first O indicator results in a tighter [C/O] versus [Fe/H] relation, suggesting that the O abundances derived from the permitted line at 6158~$\AA$ are more reliable. For the hot dwarf stars analysed by \cite{delg21}, the forbidden line at 6300~$\AA$ provides systematically higher carbon-to-oxygen ratios.

\begin{figure}
\centering
\includegraphics[width=0.75\textwidth]{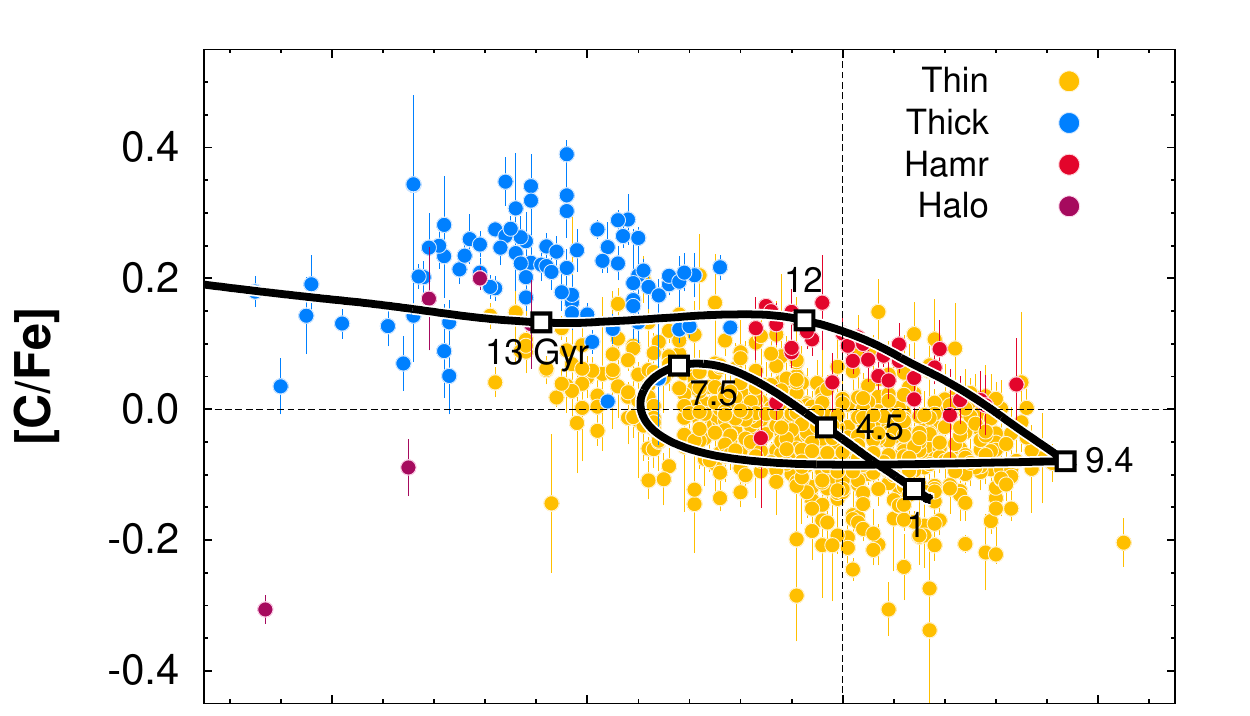}
\includegraphics[width=0.75\textwidth]{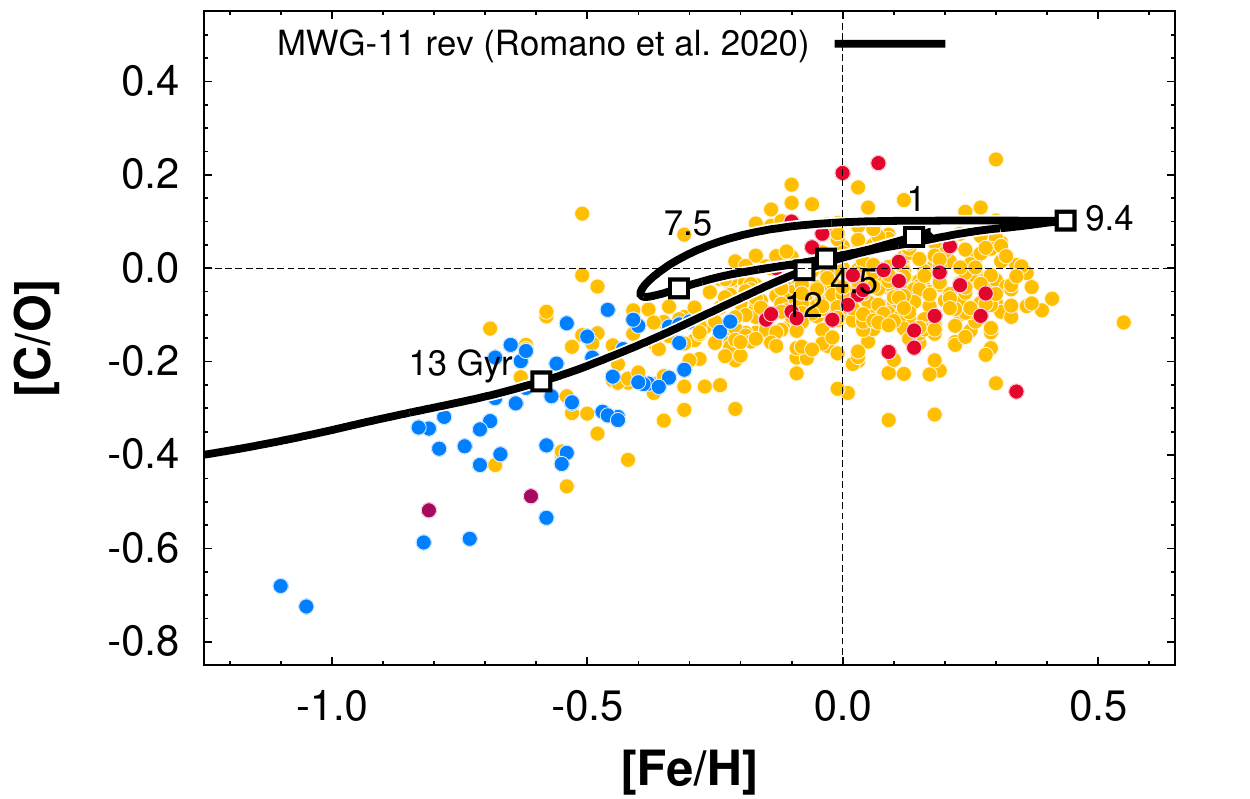}
\caption{ Circles: carbon-to-iron \emph{(top panel)} and carbon-to-oxygen ratios \emph{(bottom panel)} as functions of metallicity for neighboring stars ($d \le$ 60~pc) from \cite{delg21}. Carbon abundances are from two well-known atomic lines; oxygen abundances have been derived from the permitted O\,I\,$\lambda$\,6158~$\AA$ line (see text). Different colours indicate different Galactic populations (legend in the \emph{top panel}). The observational sample comprises only stars for which the C and O abundances are measured reliably, i.e., dwarf stars hotter than 5200~K. Solid lines are theoretical predictions for the solar vicinity from \citet{roma20}, their model MWG-11\,rev. The empty squares along the theoretical tracks highlight the abundance ratios at the specified ages. Stars older than 9.4~Gyr belong to the thick-disc component \citep[see][]{spit19}. Data and models are normalized to solar abundances from \cite{delg21}.} 
\label{fig:mwg11revC}
\end{figure}

In Fig.~\ref{fig:mwg11revC}, the dataset of \cite{delg21} is compared to the predictions of one of the GCE models for the solar vicinity discussed in \citet{roma20}, i.e., their model MWG-11\,rev. The observed stars are assigned to different groups -- thin-disc, thick-disc, halo, and high-$\alpha$ metal-rich stars -- based on chemical criteria \citep[see][and references therein]{delg21}. The model displayed in the figure is a two-infall model, assuming that the two disc components form in a sequence. In particular, it assumes a delay of 4.3~Gyr for the beginning of the second accretion event that originates the thin disc, based on recent constraints on the stellar ages from asteroseismology \citep{spit19}. The adopted stellar yields\footnote{For low- and intermediate-mass stars ($1 \le m/{\mathrm M}_\odot < 9$), the adopted yields are from \cite{vent13,vent14,vent18,vent20} and account for the evolution of the most massive intermediate-mass stars as super-AGB stars. The yields of CCSN progenitors are taken from \cite{limo18}. In particular, for [Fe/H]~$< -1$~dex their yield set R with initial rotational velocities $\vel_{\mathrm{rot}} =$~300 km~s$^{-1}$ and mass limit for full collapse to black holes of 30~M$_\odot$ is adopted. For higher metallicities, the yields are those for non-rotating stars, with 60~M$_\odot$ as the limiting mass for full collapse to black holes \citep[see][]{roma20}.} guarantee a satisfactory fit to the data. From Fig.~\ref{fig:mwg11revC}, \emph{top panel,} it is seen that the model reproduces very well the separation between thin- and thick-disc stars in the [C/Fe]--[Fe/H] plane, the decreasing [C/Fe] ratio at super-solar metallicities, and the subsolar [C/Fe] ratios for solar-metallicity stars \citep[also pointed out by other authors; see, e.g.,][]{bote20,fran20}. Furthermore, the model naturally explains the [C/Fe] versus [Fe/H] pattern of the high-$\alpha$ metal-rich stars, in the hypothesis that they represent the metal-rich tail of the thick disc as proposed by \cite{bens14}. A metal-rich tail of thick-disc stars (according to a kinematic classification) has been also found in other high-resolution studies \citep[e.g.,][]{amar19}. Some shortcomings are present though: model MWG-11\,rev does not reproduce the observed [C/Fe] bump for thick-disc stars in the $-0.8 <$~[Fe/H]~$< -0.4$ range, it predicts [C/O] ratios higher than observed (by 0.10~dex at high metallicities and up to 0.4~dex in the metal-poor regime), and it does not reproduce the flattening\footnote{The analysis based on the [O\,I]\,$\lambda$\,6300~$\AA$ line returns rather an increasing trend \citep[see also][]{fran21}.} of the carbon-to-oxygen ratio at super-solar metallicities (Fig.~\ref{fig:mwg11revC}, \emph{bottom panel}). In this respect, we note that the adopted massive star yields \citep{limo18} are computed for four initial metallicity values, [Fe/H]~$= -$3, $-$2, $-$1, and 0. Therefore, when the ISM metallicity computed by the GCE code exceeds the solar value, one has to extrapolate arbitrarily the solar-metallicity grid. The inclusion of yields tailored to super-solar metallicity stars could improve the model performances. In particular, the implementation of suitable mass loss rates is crucial in determining the final outcome of the models, including the yields (see Sect.~\ref{sec:single}). Our results may imply that a milder mass loss is needed at super-solar metallicities, resulting in lower (higher) yields for carbon (oxygen). In the metal-poor domain, the inclusion of HNe (\citealt{limo18} consider only normal CCSNe) could improve the agreement with the data. Another intriguing possibility would be to take into account the existence of very massive stars exploding as PISNe in the early Galaxy. Such a possibility is explored by \citet{gosw21} for a number of elements comprising oxygen, but not for carbon.

In the parallel Galaxy formation scenario \citep{gris17,gris18}, thin- and thick-disc stars are predicted to populate distinct sequences in the [C/Fe]--[Fe/H] and [C/O]--[O/H] spaces \citep[see][their Fig.~4]{roma20} but, at variance with the sequential framework discussed above, they may have a common range of ages. While precise stellar ages for larger samples of stars and an exact quantification of the effects of radial migration are required to allow discriminating among different evolutionary pathways, pure GCE models can still be used to provide some indications of the robustness of the adopted stellar yields. In fact, when considering the run of abundance ratios with metallicity the effects of changes in several free parameters of the GCE models tend to cancel out, while those of different nucleosynthesis prescriptions are maximized \citep{tosi88}.

To that end, it is important to notice that different C diagnostics produce different trends \citep[e.g.,][]{bens06,suar17}. It has become clear by now that it is not a good idea to mix different kinds of data and try to fit the grossly averaged trend; rather, efforts must be made to understand any systematics affecting the data, in order to define meaningful trends and better constrain the models \citep[see, e.g., discussions in][]{jofr19,roma20,delg21}.

The above considerations apply, of course, also to other elements. Nitrogen abundances in both the [N/Fe] versus [Fe/H] and [N/O] versus [O/H] planes are characterized by a large scatter at low metallicities, even if only unmixed stars\footnote{These are giant stars that have not suffered any significant mixing process yet, as witnessed by the moderately diluted $^7$Li in their atmospheres \citep{spit05}.} are considered \citep{spit05}. The dispersion could be due to inhomogeneous enrichment during the early Galaxy assembly, either from massive stars with different initial rotational velocities \citep{matt86,chia06,pran18,rom19b} or from AGB stars \citep[][and references therein]{koba22}. \citet{spit05} further notice that N determinations from the NH band and from the CN band at 388.8~nm in giants show systematic differences of 0.4~dex that they attribute to physical parameters of the NH band. The scatter in N measurements sensibly reduces at disc metallicities, where reliable N abundances are derived homogeneously for statistically significant samples of dwarf stars from the NH molecular band at 336~nm \citep[e.g.,][]{suar16}. We note that, at present, high-resolution spectra of dwarf stars including the NH feature can be obtained, basically, only with one instrument, the Ultraviolet and Visual Echelle Spectrograph \citep[UVES,][]{dekk00} mounted on the ESO Very Large Telescope (VLT).

Atomic lines suitable to O abundance derivation are scarce in late-type stars. Moreover, each of them presents specific challenges. The strong O\,I IR triplet lines at $\lambda$\,777~nm are easily accessible and free of blends, but highly sensitive to 3D non-LTE effects \citep[e.g.,][]{cava97,caff08,amar19}. Other features (the O\,I\,$\lambda$\,6158~$\AA$ and [O\,I]\,$\lambda$\,6363~$\AA$ lines) are safe from non-LTE effects, but are weak and difficult to measure. High-resolution, high SNR spectra permit the derivation of trustworthy abundances, though \citep{bert15,delg21}. The forbidden [O\,I]\,$\lambda$\,6300~$\AA$ line is very weak at low metallicities; it can also be contaminated by telluric features and severely blended by two isotopic Ni\,I lines \citep{lamb78}. The two forbidden [O\,I] lines also contain weak CN blends \citep[e.g.,][]{skul15}. The picture is no better as regards the numerous OH bands in the near UV and NIR. Although they are strong enough to be measured in main-sequence stars down to metallicities [Fe/H]~$\simeq -$3 \citep{isra98,isra01,boes99}, the extraction of the abundances is not that straightforward, because of poorly-understood physical processes affecting the lines.

\begin{figure}
\centering
\includegraphics[width=0.75\textwidth]{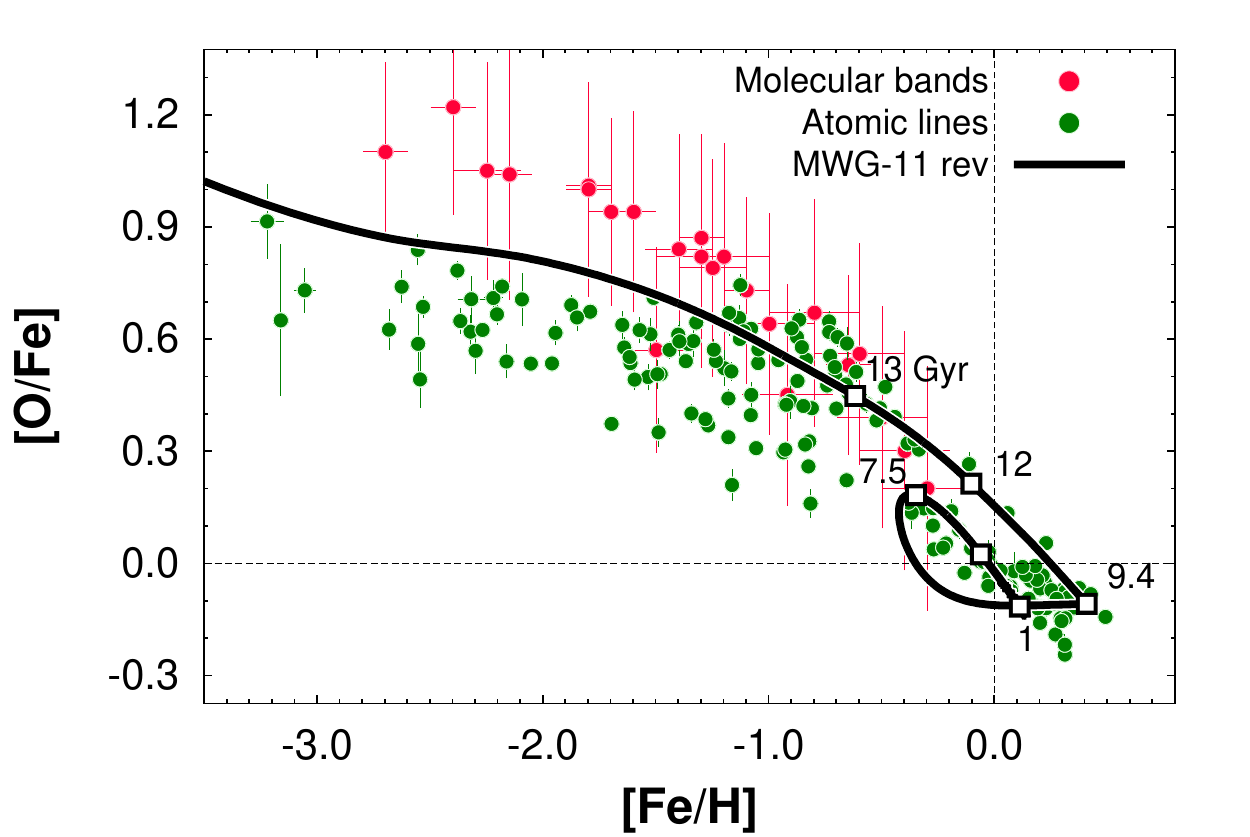}
\caption{ Oxygen-to-iron ratio as a function of metallicity in the solar vicinity predicted by model MWG-11\,rev (solid black line) and inferred from observations of atomic lines (green circles) and molecular bands (red circles) in high-resolution stellar spectra. The empty squares along the theoretical track highlight the abundance ratios at the labeled ages (13, 12, 9.4, 7.5, 4,5, and 1~Gyr). Data and models are normalized to solar abundances from \cite{amar19}.} 
\label{fig:mwg11revO}
\end{figure}

This notwithstanding, it has been claimed that GCE models can reproduce fairly well the main features of the [O/Fe] versus [Fe/H] diagram of solar neighbourhood stars \citep[e.g.,][]{roma10,pran18,koba20}. In Fig.~\ref{fig:mwg11revO}, we show the run of [O/Fe] as a function of [Fe/H] predicted by model MWG-11\,rev of \cite{roma20}. The theoretical trend is compared to O abundance determinations for two samples of unevolved stars, resting on a set of atomic oxygen lines the one \citep{amar19} and on the UV molecular OH band the other \citep{isra98}. The atomic lines have been subjected to a full 3D, non-LTE analysis. While at relatively high metallicities, [Fe/H]~$> -1.0$, the two datasets almost overlap and the theoretical predictions are in reasonable agreement with the observations, in the low-metallicity regime the linear increase of the [O/Fe] ratio with decreasing [Fe/H] drawn from the OH data sharply contrasts with the flattening of the ratio traced by the 3D, non-LTE analysis of the atomic lines. The model predictions during the earliest phases of Galactic evolution (it takes only $\sim$350 Myr to reach [Fe/H]~$= -1$ in the local disc) are marginally consistent with both the flat trend and the steep increase of [O/Fe] with decreasing [Fe/H] suggested by \cite{amar19} and \cite{isra98}, respectively. Actually, if we were to consider the data all together, we would be led to happily conclude that the gross [O/Fe] versus [Fe/H] trend is well reproduced; on closer inspection, however, the inconsistency of the two datasets and the shortcomings of the model for [Fe/H]~$< -1.5$~dex become apparent. Accounting for the existence of HNe and/or PISNe (both not included in the MWG-11\,rev model shown in Fig.~\ref{fig:mwg11revO}) would decrease the predicted [O/Fe] ratio at low metallicities (see Fig.~8 of \citealt{roma10} and Fig.~11 of \citealt{gosw21}, respectively, for the effects of HNe and PISNe on the model predictions). Moreover, some low-metallicity stars might be not rotating fast (our reference model in Fig.~\ref{fig:mwg11revO} assumes that all massive stars below [Fe/H]~$= -1$ are spinstars), which would again bring down the predicted [O/Fe], because of the lower O production expected from non-rotating massive star models \citep{limo18,pran18}.

Things get even more complicated when it comes to the minor isotopes. The local enrichment history of the secondary CNO isotopes can be unlocked only by measuring their abundances in unevolved stars. In fact, since the abundances of the minor CNO isotopes are severely altered in the advanced phases of stellar evolution (see Sect.~\ref{sec:nuc}), measurements in giant stars do not reflect the original composition of the ISM out of which the stars were born. The problem is that measuring CNO isotopic ratios in dwarf stars is extremely challenging, if not impossible. \cite{spit06} determined the $^{12}$C/$^{13}$C ratio of a sample of 35 extremely metal-poor stars on the lower RGB and found it to be very close to 30 in a subsample of stars that likely suffered minimal mixing processes. \cite{chia08} adopted a GCE model especially deviced to the Galactic halo including gas infall and outflow and demonstrated that, if fast rotators are common in the early universe, $^{12}$C/$^{13}$C ratios between 30 and 300 are obtained for [Fe/H]~$< -3$ dex. This result implies that some stellar mixing is needed to explain \citeauthor{spit06}'s (\citeyear{spit06}) data, but by far less extreme than what would be necessary if low-metallicity massive stars were not rotating at all. Stellar rotation, in fact, triggers the production of primary $^{13}$C in the first generations of massive stars, significantly lowering the C isotopic ratio in the ISM of the early Galaxy \citep[see][their Fig.~1]{chia08}. Measurements of $^{12}$C/$^{13}$C ratios are available also for carbon-enhanced metal-poor (CEMP) stars at the turnoff \citep[e.g.,][]{siva06}, but in CEMPs these ratios may mirror the chemical composition of material transferred from AGB companions \citep[e.g.,][]{aoki06}, or localized enrichment from faint SNe or spinstars \citep[e.g.,][]{hans16}. The first and still unique measurement of the carbon isotopic ratio in an ancient, metal-poor, C-normal star that just passed the turnoff rests on the analysis of exquisite spectra obtained with ESPaDOnS at the Canada-France-Hawaii Telescope (CFHT), ESPRESSO at the VLT, and HARPS at the ESO 3.6~m telescope and returns $^{12}$C/$^{13}$C~= $33^{+12}_{-6}$ \citep{spit21}, very close to previous estimates for unmixed giants \citep{spit06}.

\begin{figure}
\centering
\includegraphics[width=0.75\textwidth]{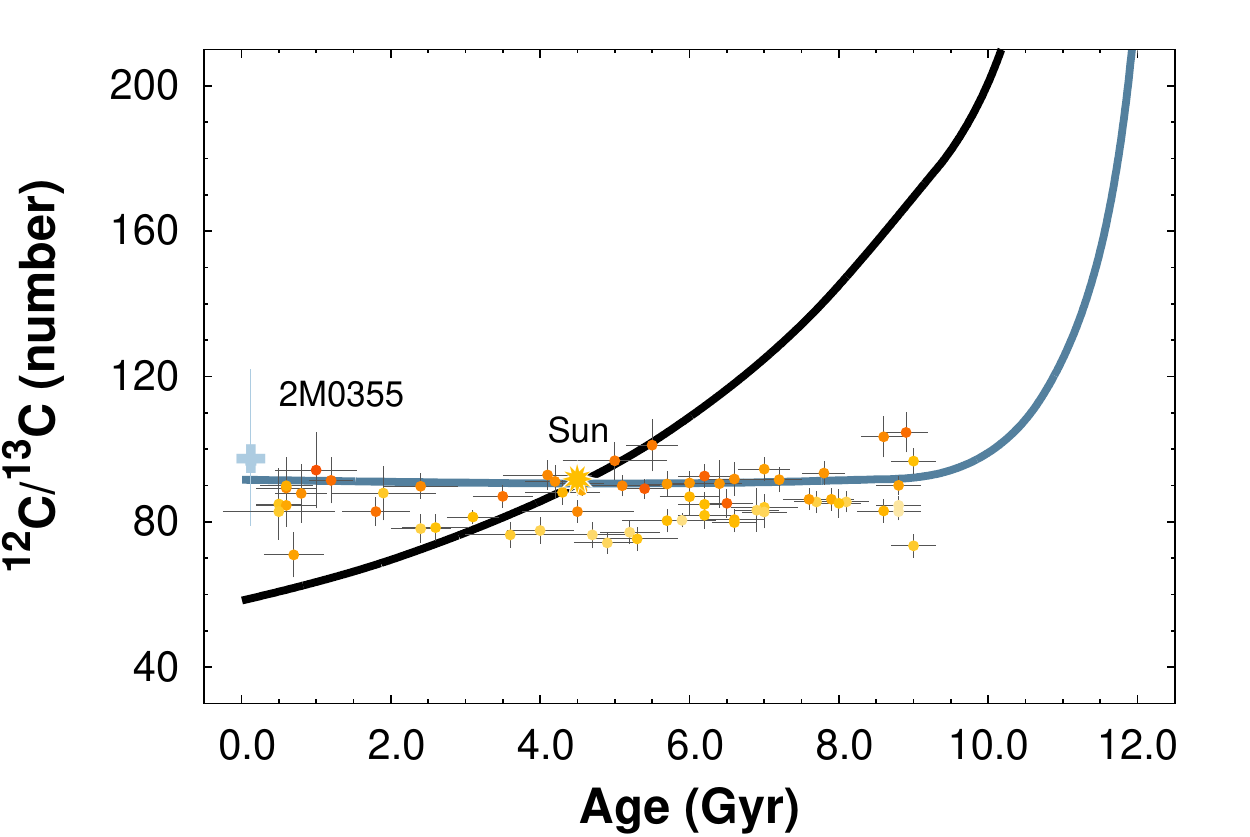}
\includegraphics[width=0.75\textwidth]{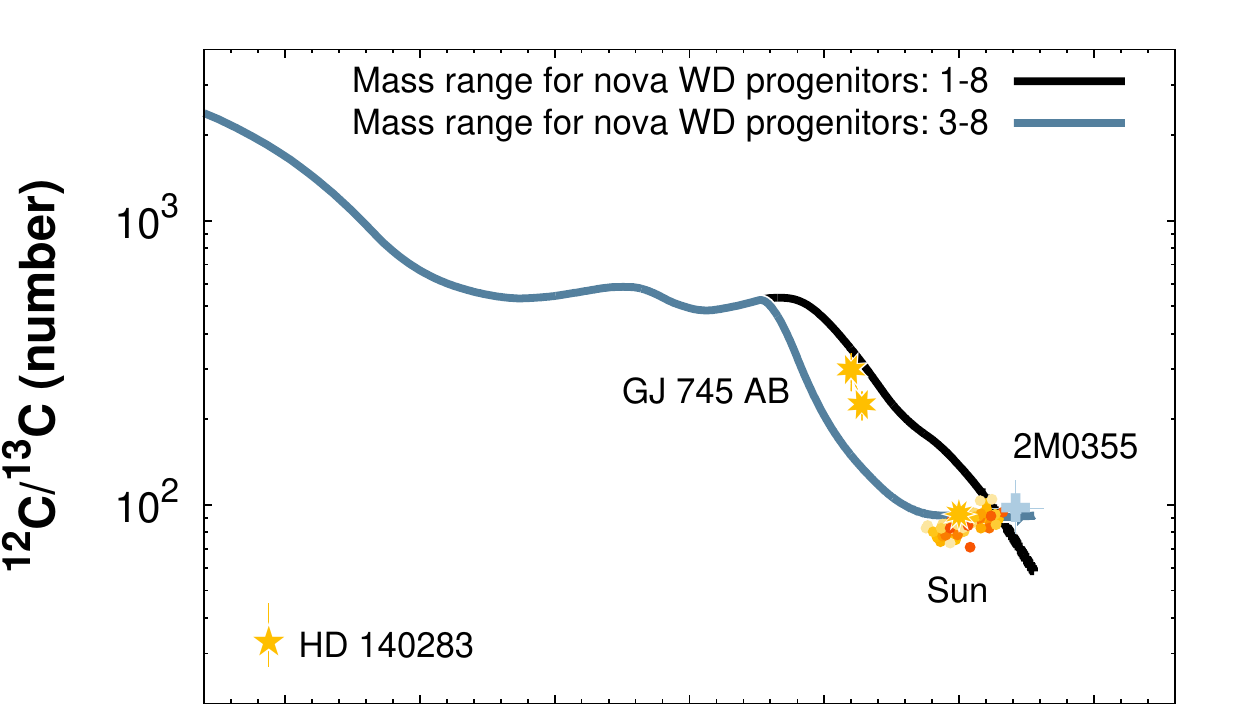}
\includegraphics[width=0.75\textwidth]{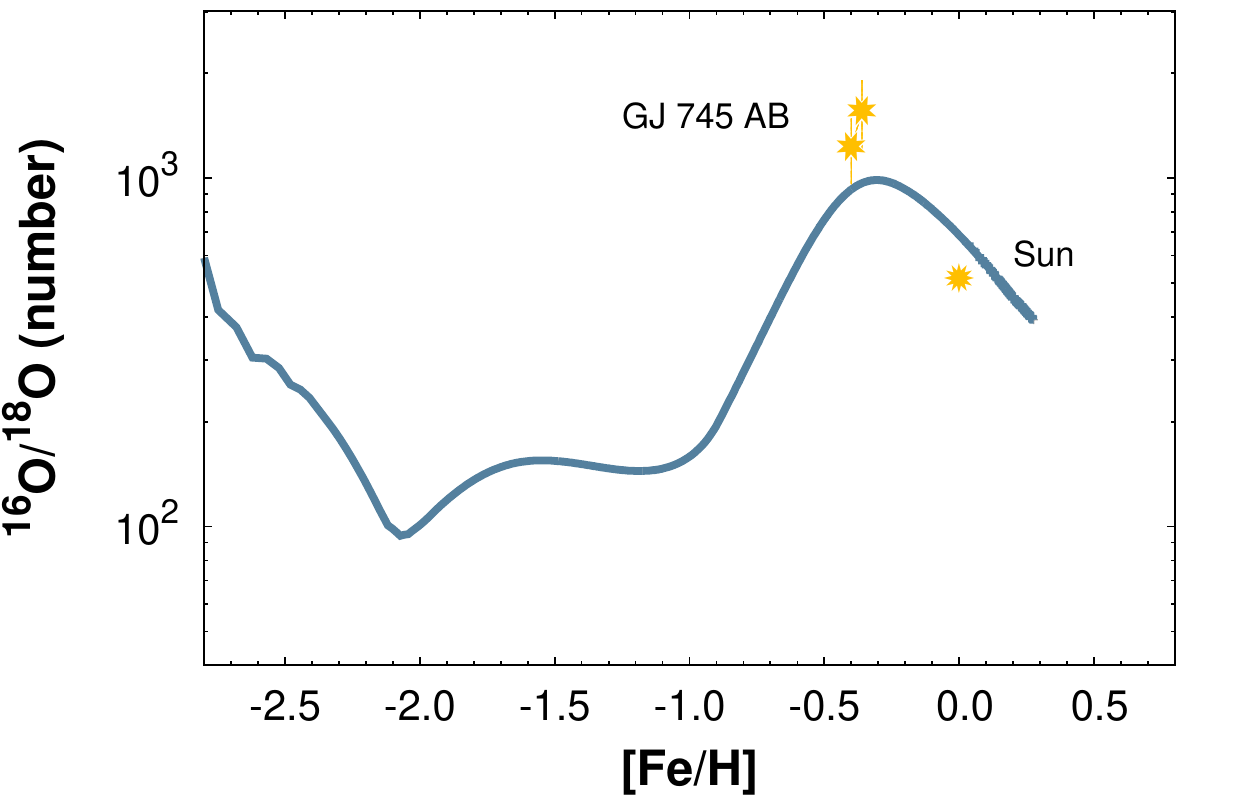}
\caption{ Expected $^{12}$C/$^{13}$C ratio as a function of age \emph{(top panel)} and metallicity \emph{(middle panel)} according to the GCE models for the solar vicinity of \cite{roma21}. The models implement nova nucleosynthesis assuming that the nova WD progenitors are either in the range 1--8~M$_\odot$ (solid black lines) or in the range 3--8~M$_\odot$ (solid blue lines). The evolution of the $^{16}$O/$^{18}$O ratio is also shown \emph{(bottom panel)}. The entry model is a parallel model. Here we show only the evolution of the thin-disc component. Data are from \citet[][HD\,140283]{spit21}, \citet[][GJ\,745\,AB]{cross19}, \citet[][63 solar twins, colour-coded from light-yellow to dark-orange going from metal-poorer/older to metal-richer/younger objects]{bote20}, \citet[][the Sun]{ayre13}, and \citet[][2M0355, light blue cross]{zhan21}.} 
\label{fig:isoSN}
\end{figure}

Useful constraints to the evolution of the C isotopic ratio in the local disc are set by \cite{bote20}, who estimate $^{12}$C/$^{13}$C ratios from CH\,A-X and CN\,B-X features in HARPS spectra of 63 solar twins spanning a wide range of isochrone ages (0.5--9 Gyr). \cite{cross19} add further information from their detailed analysis of a system of two low-mass M dwarfs orbiting in a wide binary observed at high resolution ($R =$ 70\,000) with the iSHELL spectrograph \citep{rayn16} on the NASA Infrared Telescope Facility (IRTF). The stars are fully convective, hence chemically homogeneous, hence reflective of the chemical composition of the ISM at the time of their formation. Spectroscopy of carbon monoxide (CO) at the 4.7~$\mu$m fundamental and 2.3~$\mu$m first-overtone rovibrational bandheads reveals multiple rare isotopologues. For the first component, GJ\,745\,A, $^{12}$C/$^{13}$C~=~$296 \pm 45$ and $^{16}$O/$^{18}$O~=~$1220 \pm 260$, while for the second component, GJ\,745\,B, $^{12}$C/$^{13}$C~=~$224 \pm 26$ and $^{16}$O/$^{18}$O~=~$1550 \pm 360$. While observations of the CO fundamental rovibrational band at high resolution present several advantages over previous medium-resolution observations \citep[see discussion in][]{cross19}, the huge amount of observing time per target and the scarceness of spectrographs capable of performing high-resolution, $M$-band observations make the collection of a statistical sample an unattainable task for now. Last but not the least, we mention the determination of the C isotopic ratio in a young (age~$\sim$ 125~Myr) brown dwarf: \cite{zhan21} have obtained $^{12}$CO/$^{13}$CO~=~$97^{+25}_{-18}$ for the nearby ($d = 9.1 \pm 0.1$~pc) L dwarf 2MASS J03552337+1133437 (hereinafter 2M0355) from an analysis of archival $K$-band (2.03--2.38 $\mu$m) spectra taken with the spectrograph NIRSPEC at the Keck~II 10~m telescope.

In Fig.~\ref{fig:isoSN}, we compare the stellar measurements discussed in previous paragraphs to the predictions of recent GCE models including nova nucleosynthesis \citep[][and Romano et al., in prep.]{roma21}. Nova systems are expected to inject non-negligible amounts of $^{13}$C, $^{15}$N, and $^{17}$O in the ISM, with yields varying widely in dependence of the assumptions underlying any specific model of the outburst (see Sect.~\ref{sec:bin} and, especially, Fig.~\ref{fig:roma}). The current nova rate in the Galaxy is also largely uncertain. Recent studies report rates of $43.7^{+19.5}_{-8.7}$ yr$^{-1}$ \citep{de21}, $26 \pm 5$ yr$^{-1}$ \citep{kawa22}, and $28^{+5}_{-4}$ yr$^{-1}$ \citep{rect22}. The past Galactic nova rate is basically unknown, though some population synthesis studies suggest the existence of an anticorrelation between metallicity and the number of novae produced: for instance, according to \cite{kemp22}, the number of novae at $Z = 0.03$ is roughly half that at $Z = 10^{-4}$. Clearly, there is a high degeneracy between the nova rate and the nova yields assumed by any GCE model. Moreover, the models discussed here only accommodate average yields (see, again, Sect.~\ref{sec:bin}). From all the above, it is clear that we are treading on thin ice: extreme caution is required when discussing the impact of novae on the Galactic enrichment.

This notwithstanding, it is common wisdom that a large fraction of the meteoritic lithium ($^7$Li) abundance comes from low-mass stars, with novae being the favored candidates \citep[see][and discussions therein]{roma99,roma21}. Resting on the analysis of a sample of open cluster (OC) and field stars with $^7$Li abundance measurements from GES iDR6, \cite{roma21} have suggested that a better match between theoretical predictions about $^7$Li evolution and observations is possible if the lower mass limit of primary stars entering the formation of nova systems is increased from $\sim$1~M$_\odot$ \citep[a standard assumption of GCE models including novae; see][]{dant91,roma03} to 2--3~M$_\odot$. This is to say that systems hosting small WDs are not effective in developing the outbursts \citep[see also][]{kove85}. Interestingly, while the standard assumption leads to a continuous, mild increase of the predicted nova rate in time, increasing the mass threshold for WD progenitors that potentially lead to nova outbursts results in a nova rate that peaks at early times and declines afterwards, in agreement with more general findings that the fraction of close binary systems decreases with increasing [Fe/H] \citep[e.g.,][]{gao14}. In \citeauthor{roma21}'s \citeyear{roma21} model, this happens because the more massive the allowed WD progenitors, the more closely the theoretical nova rate follows the star formation rate (which is overall decreasing in time in the Milky Way), due to the shorter time lag introduced by the lifetimes of more massive stars \citep[although the peak in the nova rate is always delayed relative to that in the star formation rate, because of the time required for the WDs to cool at a level that ensures strong enough nova outbursts; see][and references therein]{roma21}.

A model assuming a mass range for WD progenitors in nova systems intermediate between the two extreme cases discussed here would agree better with the data in the $^{12}$C/$^{13}$C versus age diagram (see Fig.~\ref{fig:isoSN}, \emph{top panel}). However, the increase of the C isotopic ratio with increasing metallicity spotted out in \citeauthor{bote20}'s (\citeyear{bote20}) sample (Fig.~\ref{fig:isoSN}, \emph{middle panel}) remains difficult to decipher. A viable explanation could be recent dilution of the local ISM by poorly-processed gas, which would lower the metallicity whilst the $^{12}$C/$^{13}$C ratio would remain almost constant or slightly decrease in time. The possibility that some of the solar twins are intruders, i.e., stars that were born at inner radii and migrated to their current positions, should be investigated as well. The chemical properties of any intruders, in fact, should not be compared to the predictions of models for the solar neighbourhood \citep[e.g.,][]{guig19,roma21}. Also shown in Fig.~\ref{fig:isoSN} is the run of the $^{16}$O/$^{18}$O ratio with [Fe/H] \emph{(bottom panel)}. Both $^{16}$O and $^{18}$O are produced mostly by massive stars, without any significant contribution from novae, so the two models described above overlap in the O isotope plot. The $^{16}$O/$^{18}$O ratio first decreases because of primary $^{18}$O production triggered by stellar rotation \citep{limo18}. Then, the trend is reversed when the yields from non-rotating stars are adopted. Finally, at higher metallicities secondary $^{18}$O production becomes dominant and the isotopic ratio decreases again. The model predictions are marginally consistent with \citeauthor{cross19}'s (\citeyear{cross19}) data as well as with the solar value.

\paragraph{Planet hosts}

Modeling planet formation and evolution requires detailed information about the chemical composition of the host stars \citep[e.g.,][and references therein]{sant17}. In particular, it is important to ascertain the role of CNO elements in the planet formation process, with special regards to C and O, which enter the composition of complex organic molecules that are the precursors of biogenic compounds \citep{case12}. A thorough recap of this and related issues would require a dedicated review and is beyond the scope of the current discussion. However, for the sake of completeness we report below a few facts about the determination of CNO abundances in planet-harboring solar-type stars.

In their pioneering work, \citet{mele09} used high-resolution abundances for a dozen solar twins and the Sun itself and highlighted that our star has a low ratio between refractory and volatile elements\footnote{Refractory elements encompass Na, Mg, Al, Si, Ca, Mn, and Ni, among others. Volatile elements include C, N, O, S, and Zn.} relative to the majority of the solar twins. They propose the sequestration of refractory elements in terrestrial planets as a possible explanation, further suggesting that the planets may imprint the signature of their existence in the [X/Fe] versus elemental condensation temperature relation. However, the refractory-to-volatile element ratios may be related to the stellar ages and birth radii: \cite{adib14} point out that old stars and stars with an inner disc origin display lower refractory-to-volatile element ratios.

The tendency of stars with giant planets to have a metallicity excess with respect to stars without planets was first spotted out by \citet{gonz97} and has been confimed by many subsequent studies \citep[see, e.g., the review by][]{gonz06}. However, this correlation seems to disappear if the planets have masses lower (higher) than 0.05 (4) times the mass of Jupiter \citep[e.g.,][]{buch12,sant17}. As regards the CNO element abundances, we stress again that detailed abundances of CNO elements are difficult to determine in stars. \cite{delg21} find tentative evidence from high-resolution, high-quality spectra that stars hosting low-mass planets have [C/Fe] ratios higher than their counterparts without planets for [Fe/H]~$< -0.2$. For higher metallicities, there is no C-enhancement associated with the presence of planets. Oxygen seems to follow a similar trend, but given the larger errors affecting O abundance determinations, no strong conclusion can be drawn \citep[see also][]{niss14}. Homogeneous nitrogen determinations for 74 solar-type stars by \cite{suar16} show that planet hosts (42 stars, 57\% of the sample) are N-rich with respect to single stars. A likely explanation resides in the fact that their planet hosts are metal-rich and, according to GCE models, higher [N/Fe] ratios are expected at higher [Fe/H].

We conclude that a good understanding of the evolution of the CNO elements in different Galactic components is mandatory in order to improve our knowledge of exoplanet formation and evolution. High-accuracy, high-precision abundances of statistically significant control samples of stars and increasingly refined GCE models are necessary tools to reach this goal.

\begin{figure}
\centering
\includegraphics[width=0.75\textwidth]{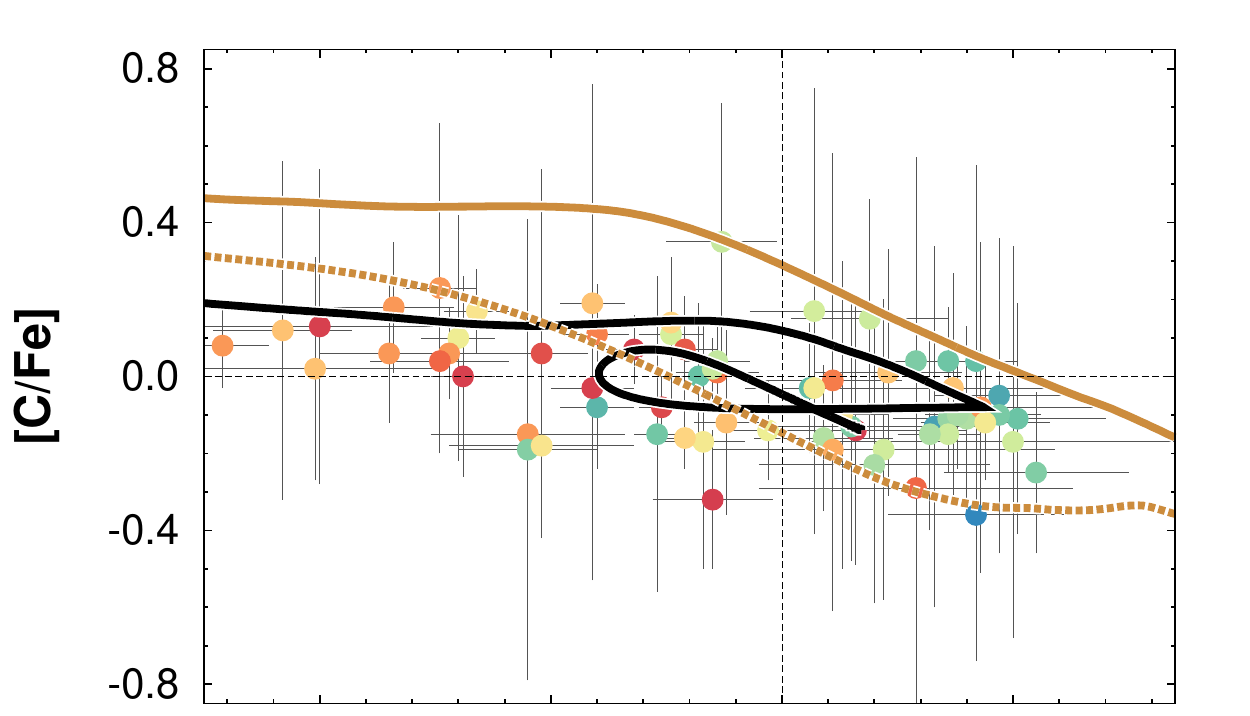}
\includegraphics[width=0.75\textwidth]{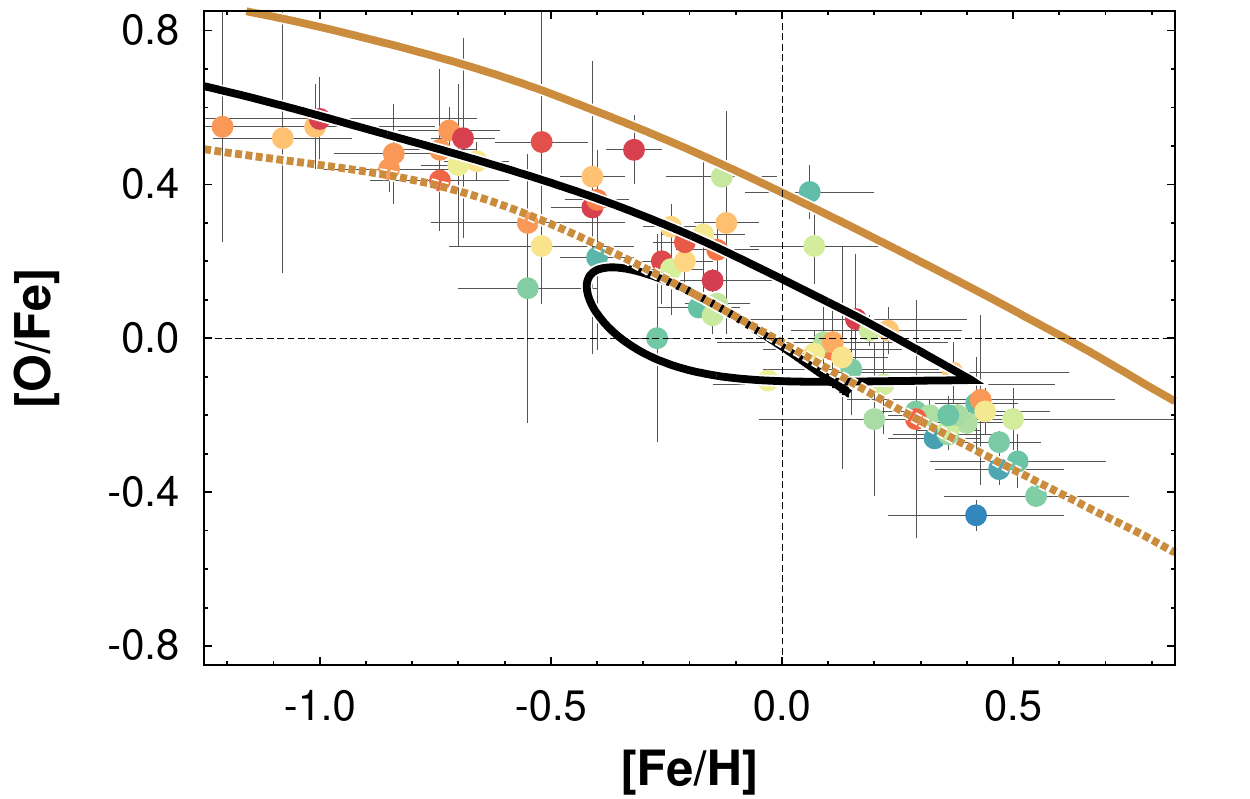}
\caption{ Carbon-to-iron \emph{(top panel)} and oxygen-to-iron \emph{(bottom panel)} ratios as functions of [Fe/H] for microlensed unevolved stars in the bulge \citep[][circles]{bens21}. Red, orange, yellow, green, and blue colours indicate stars from older to younger. Predictions from models for the solar neighbourhood (solid black curves) and for the Galactic bulge (solid and dotted gold curves) from \cite{roma20} are also shown. The solar neighbourhood model assumes a gwIMF slope $x = 1.7$ for massive stars, while the bulge models assume the Salpeter slope, $x = 1.35$. The solar neighbourhood model assumes the yields by \cite{limo18} for fast-rotating massive stars below [Fe/H]~$= -1$, while above this threshold the ones for non-rotating stars are used. The models for the bulge are computed assuming either that all massive stars are fast rotators (solid curve), or that all massive stars do not rotate (dotted curve), independently of metallicity.} 
\label{fig:buA}
\end{figure}

\begin{figure}
\centering
\includegraphics[width=0.75\textwidth]{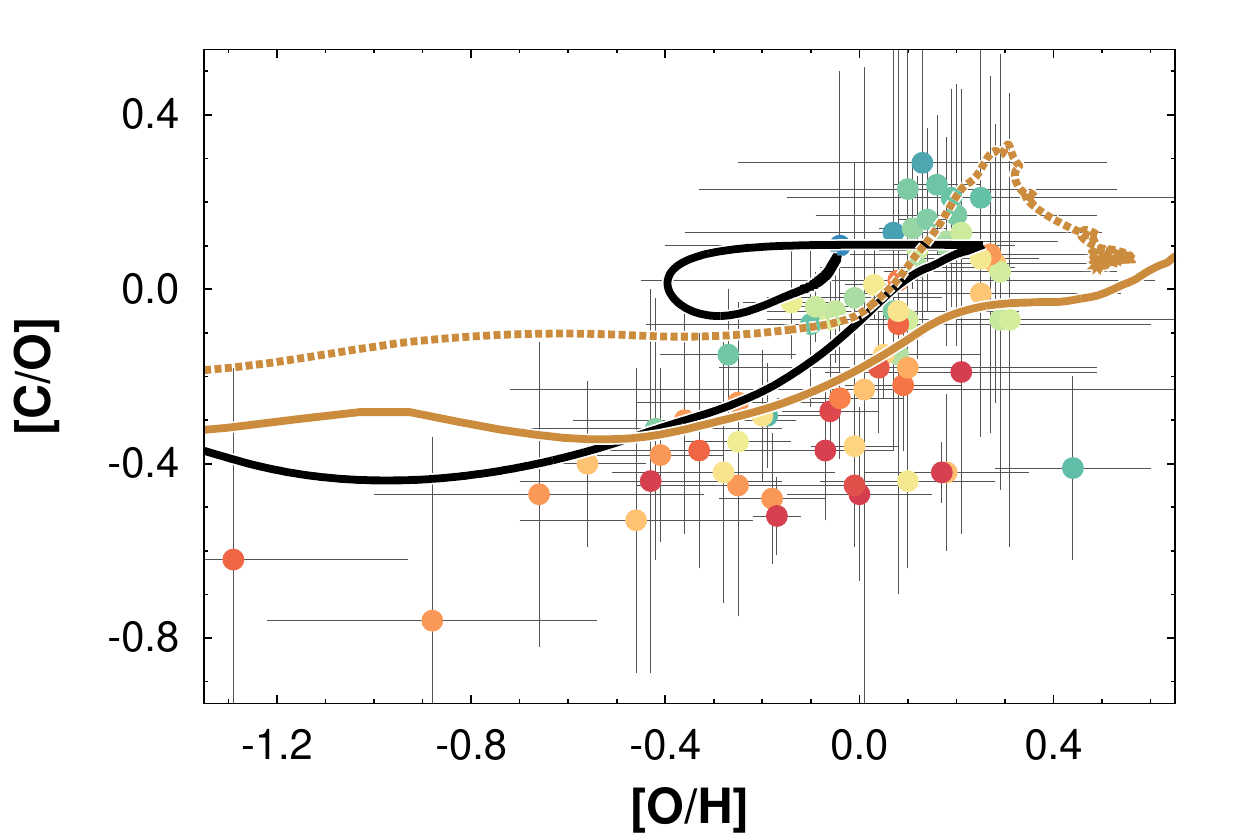}
\caption{ Same as Fig.~\ref{fig:buA}, but for [C/O] versus [O/H].} 
\label{fig:buB}
\end{figure}

\paragraph{The Galactic bulge}

\cite{bens21} have determined the abundances of C and O of 70 microlensed dwarf, turnoff, and subgiant stars found in the Galactic bulge today, thereby providing the first statistically significant sample of bulge stars with C abundance measurements free from the effects of stellar evolution. To the best of our knowledge, before their study C abundances of unevolved bulge stars were available only for three objects \citep{john08,cohe09}, making it hard to sensibly constrain carbon enrichment scenarios for the inner Galaxy \citep[][but see \citealt{cesc09}]{roma20}. To measure carbon, \cite{bens21} rely on spectral synthesis of six C\,I lines around 910~nm, while the O\,I triplet at 777~nm is used to derive oxygen abundances. The abundances are corrected for non-LTE effects.

The trends of [C/Fe] and [O/Fe] as functions of [Fe/H] obtained by \cite{bens21} are similar to the corresponding trends observed in the local thin and thick discs, with bulge stars older than 8~Gyr behaving as thick-disc members and stars younger than 8~Gyr following the thin-disc trend \citep{bens21}. In Fig.~\ref{fig:buA}, \citeauthor{bens21}'s data are compared to the predictions of model MWG-11\,rev for the solar vicinity\footnote{Models specifically tailored to inner galactocentric distances ($R_\mathrm{GC} \simeq$~4--6~kpc) would agree better with the observations, because they would produce theoretical trends shifted to higher metallicities.} and to the outputs of two models tailored to the bulge from \cite{roma20}. In the high-mass domain, a gwIMF flatter than the one used for the solar neighbourhood is assumed for the bulge, in order to fit better the stellar metallicity distribution function and the run of [$\alpha$/Fe] with [Fe/H] observed for bulge stars \citep[][and references therein]{matt19}. The models for the bulge assume two extreme prescriptions about stellar rotation -- massive stars are all rotating fast in one case, or do not rotate at all in the other, independently of metallicity. The new data seem consistent with a scenario in which the bulk of the bulge stars owes its origin to secular evolution of the disc component and with the existence of a metallicity threshold for stellar rotation. The latter hypothesis allows to explain also the N abundances of a sample of giant bulge stars selected from Data Release 16 (DR16) of the APOGEE-2 survey \citep{kisk21}, after some corrections for stellar evolutionary effects are applied \citep[see][]{gris21}.

The [C/O] ratio in the bulge is found to steeply increase with increasing [O/H] (Fig.~\ref{fig:buB}). This contrasts with the almost flat behaviour of [C/O] at super-solar metallicities found by \cite{delg21} for nearby stars. The different abundance diagnostics used in the two studies may be at the origin of the discrepancy.

As a final word of caution, it is worth reminding that the inner Milky Way region is very complex. It hosts different stellar populations that coexist close to the Galactic midplane \citep{quei21}: an inner thin- and thick-disc component, a bar, a population of counter-rotating stars, and a spheroidal component. The latter experienced an early rapid, vigorous star formation, in agreement with the predictions of classical GCE models \citep{matt19}. When modeling the inner Galactic regions, the coexistence of these distinct components has to be kept in mind \citep[see, e.g.,][]{hayw18}.

\paragraph{The Galactic gradient}

\begin{figure}
\centering
\includegraphics[width=0.75\textwidth]{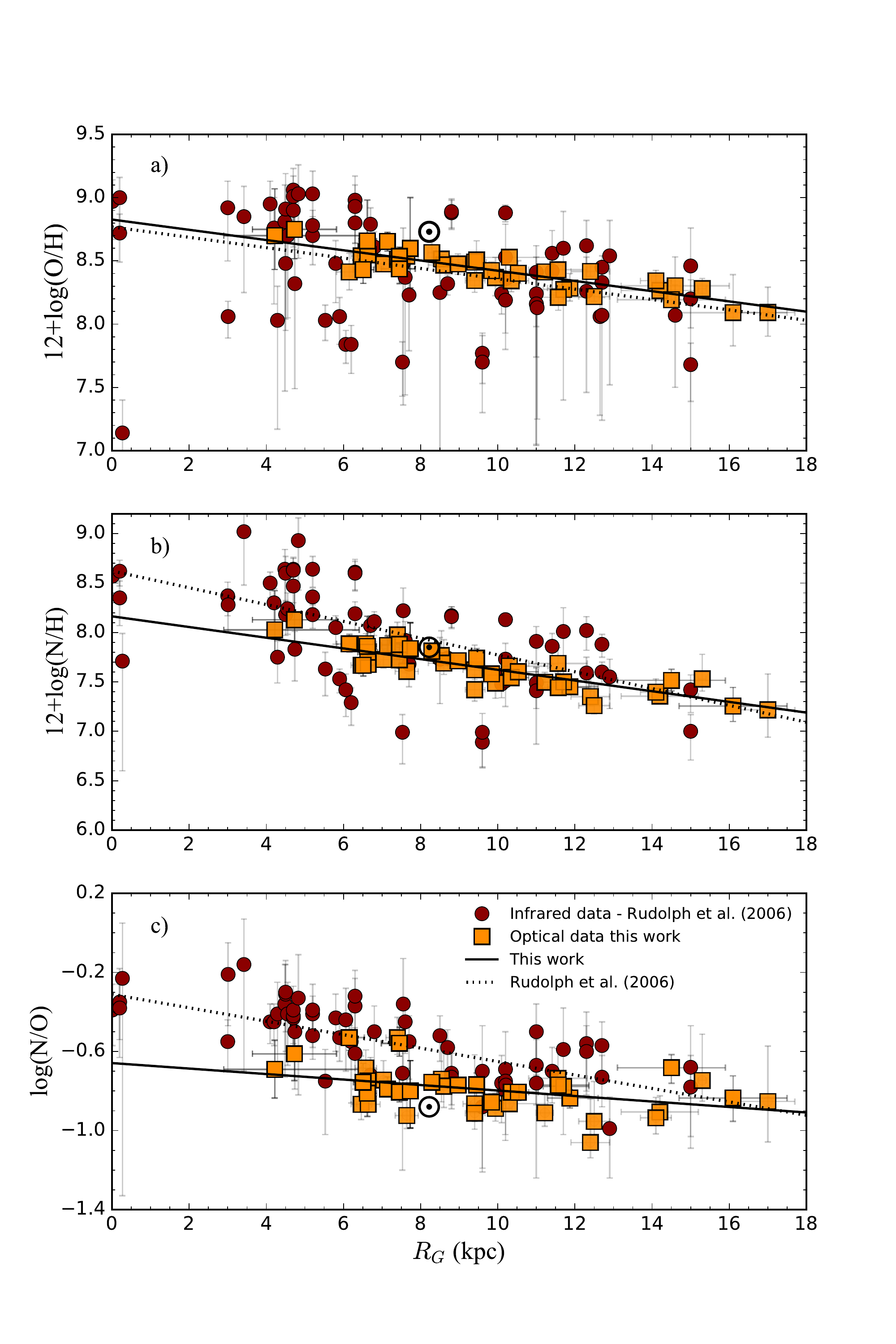}
\caption{ Gradients of oxygen \emph{(top panel)}, nitrogen \emph{(middle panel)} and N/O \emph{(bottom panel)} at the present time from H\,II regions. Orange squares are results from optical emission lines \citep{arel20,arel21}. Burgundy circles are results from FIR observations \citep{rudo06}. Solid and dotted lines are the fits to the two datasets. The Sun symbols show the solar photospheric values after \cite{lodd09}. Figure reproduced from \cite{arel21}, copyright by the authors.} 
\label{fig:arel}
\end{figure}

Nowadays, samples of stars (both in the field and in OCs) from large surveys spanning wide ranges of distance and metallicity can be used to undisclose the full chemical enrichment histories of different Galactic components, revealing the details of the Milky Way assembly process over an unprecedented Galactic volume \citep[e.g.,][]{hayd15,xian19,gaia22, rand22}. Additional information on the dynamical processes that regulate the distribution of chemical species along the disc comes from the abundance gradients.

Recent observations of H\,II regions, classical Cepheids, B-type stars, and OCs provide sound estimates of present-day gradients of CNO elements along the disc. \citet{arel20,arel21} put forth a homogeneous analysis of deep spectra of 42 H\,II regions with galactocentric distances $R_{\mathrm{GC}} \simeq$~4--17 kpc. All the nebulae have direct determinations of the electron temperatures, $T_{\mathrm{e}}$, and revised distances based on \emph{Gaia} parallaxes of associated stars. Moreover, the most appropriate ionization correction factor (ICF) scheme is chosen on an element-by-element basis. Resting on their careful analysis, \citet{arel20,arel21} conclude that the present-day ISM is fairly well mixed, at least in the disc quadrant covered by their observations. Fig.~\ref{fig:arel} illustrates the improvement obtained with respect to previous investigations. The new determinations are consistent with linear radial abundance gradients for oxygen and nitrogen with slopes of $-0.042 \pm 0.009$~dex kpc$^{-1}$ and $-0.057 \pm 0.011$~dex kpc$^{-1}$, respectively \citep{arel21}. A flattening of the O gradient in the inner disc \citep[e.g.,][]{este18} is ruled out. The N/O ratio gradient is flatter, with a small negative slope of $-0.015 \pm 0.007$~dex kpc$^{-1}$ \citep{arel21}. For C/H and C/O, \citet{arel20} report slopes of $-0.072 \pm 0.018$~dex kpc$^{-1}$ and $-0.037 \pm 0.016$~dex kpc$^{-1}$, respectively\footnote{The findings of \citet{arel20,arel21} are confirmed by a reanalysis of the same data by \cite{mend22}.}.

\cite{brag19} find a steeper gradient of $-0.07$ dex kpc$^{-1}$ for oxygen in the outer disc ($8.4 \le R_{\mathrm{GC}} \le 15.6$) from a sample of 28 young OB stars with high-resolution spectra obtained with the spectrograph MIKE on the Magellan Clay 6.5-m telescope in Las Campanas. \citet{stan18} use planetary nebulae (PNe) as tracers of the radial O gradient. By separating their sample in PNe with progenitors younger (older) than 1~Gyr (7.5~Gyr), they find that young objects define an O gradient, $-0.015$~dex kpc$^{-1}$, shallower than that derived from older objects, $-0.027$~dex kpc$^{-1}$. \citet{magr18} present N and O abundances in a sample of 17 GES OCs spanning wide ranges of galactocentric distances (4.5--15~kpc) and ages (0.1--7~Gyr), while \citet{dono20} discuss the O abundance gradient based on a high-quality sample of 71 OCs with IR data from APOGEE DR16; both find general agreement with data published in the literature. Finally, the gradient from Cepheids that is derived by \cite{luck18} based upon the analysis of 1127 spectra for 435 stars and adopting Bayesian distance estimates based on \emph{Gaia} data from \citet{bail18} has slopes of $-0.0665 \pm 0.0038$~dex kpc$^{-1}$, $-0.0470 \pm 0.0040$~dex kpc$^{-1}$, and ~$-0.0429 \pm 0.0023$~dex kpc$^{-1}$ for C, N, and O, respectively. It is particularly reassuring that updated gradients from gaseous and stellar diagnostics resting on homogeneous analyses of large samples of objects agree very well with one another within the errors (see Table~\ref{tab:grad}).

\begin{table}
\begin{center}
\caption{Slopes (in dex kpc$^{-1}$) of current X/H abundance gradients (X = C, N, O) from recent observations of H\,II regions and Cepheids. Theoretical predictions from selected GCE models are listed in the last four columns for comparison.}
\label{tab:grad}
\begin{tabular}{c@{\hskip 22pt}cc@{\hskip 22pt}cccc}
\hline
Element & \multicolumn{2}{c}{Observations} & \multicolumn{4}{c}{Models$^c$}\\
        & H\,II regions$^a$ & Cepheids$^b$ & MW\,B & MW\,F & MW\,D & MW\,E\\
\hline
C & $-0.072 \pm 0.018$ & $-0.0665 \pm 0.0038$ & -- & -- & -- & -- \\
N & $-0.057 \pm 0.011$ & $-0.0470 \pm 0.0040$ & $-0.0344$ & $-0.0713$ & $-0.0931$ & $-0.0646$ \\
O & $-0.042 \pm 0.009$ & $-0.0429 \pm 0.0023$ & $-0.0305$ & $-0.0734$ & $-0.0961$ & $-0.0688$ \\
\hline
\end{tabular}
\end{center}
\vspace{2pt}
\footnotesize{
$^a$From \cite{arel20,arel21}.
$^b$From \cite{luck18}.
$^c$From \cite{pall20}.
}
\end{table}

\begin{figure}
\centering
\includegraphics[width=\textwidth]{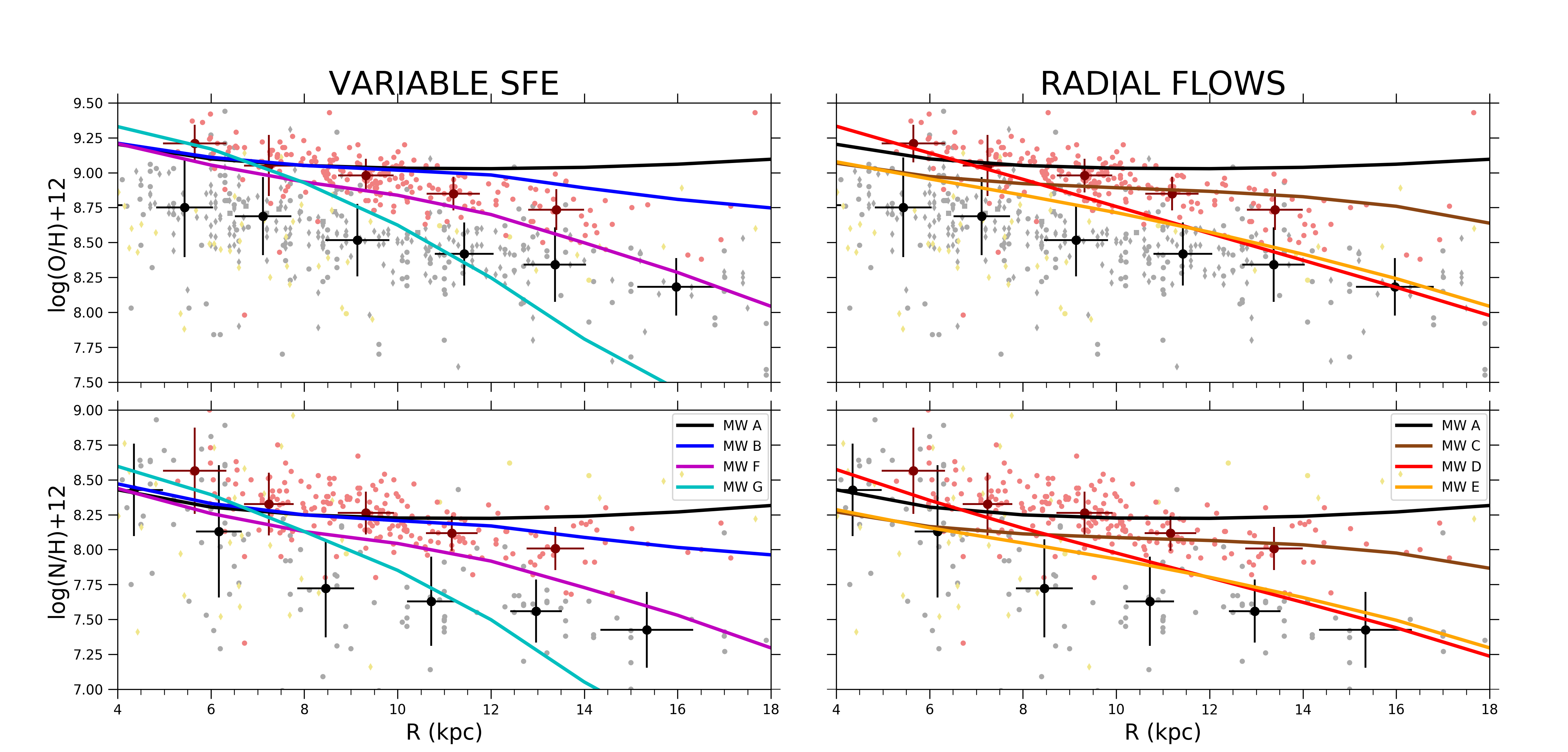}
\caption{ Oxygen \emph{(top panels)} and nitrogen \emph{(bottom panels)} gradients predicted by GCE models. The models on the right adopt a constant star formation efficiency and different prescriptions about the intensity of radial gas flows along the disc. The models on the left adopt a star formation efficiency that decreases along the disc without (blue lines) or with (magenta and cyan lines) radial gas flows. The black curves represent the reference model with constant star formation efficiency and no gas flows. Different symbols denote different datasets for H\,II regions, Cepheids, young OCs, and PNe \citep[see][for references]{pall20}. Figure adapted from \cite{pall20}, copyright by the authors.}
\label{fig:pall}
\end{figure}

The effects of different model assumptions on the predicted current radial abundance gradients are nicely summarized in \cite{pall20}. We report their results for O and N in Fig.~\ref{fig:pall} and Table~\ref{tab:grad}. Besides an inside-out formation of the thin disc \citep{lars76,matt89}, a variable star formation efficiency \citep[see][their Sect.~2.1.3]{bois99} and radial gas flows \citep{lace85,port00,spit11} must be considered to reproduce the observed gradient. In the early work of \citet{lace85}, radial gas flows arise as a consequence of the viscosity of the gas, the angular momentum difference between the gas accreted by infall and that already present in the disc, and the presence of non-axisymmetric density patterns that modify the angular momentum of the gas. Nowadays, high-resolution, magnetohydrodynamical cosmological simulations of the formation of Milky Way-like galaxies can be used to study the radial transport of material within the disc planes \citep[e.g.,][]{okal21}, which provides scaling laws that can be usefully employed to parametrize radial gas flows between concentric rings in GCE models. The tight relations between abundance and position along the disc obtained through improved data analyses \citep[e.g.,][see Table~\ref{tab:grad}]{luck18,arel20,arel21} have the potential to significantly constrain the models and, hence, the relative importance of the physical mechanisms underlying the emergence of Galactic gradients. Finally, PN and OC samples can be separated in subgroups of different ages, thus enabling the study of the temporal evolution of the gradient. This is especially important, since it allows discriminating between models that predict a steepening \citep[e.g.,][]{chia01} or a flattening of the gradients in time \citep[e.g.,][]{pran00,vin18a}.

\begin{figure}
\centering
\includegraphics[width=0.49\textwidth]{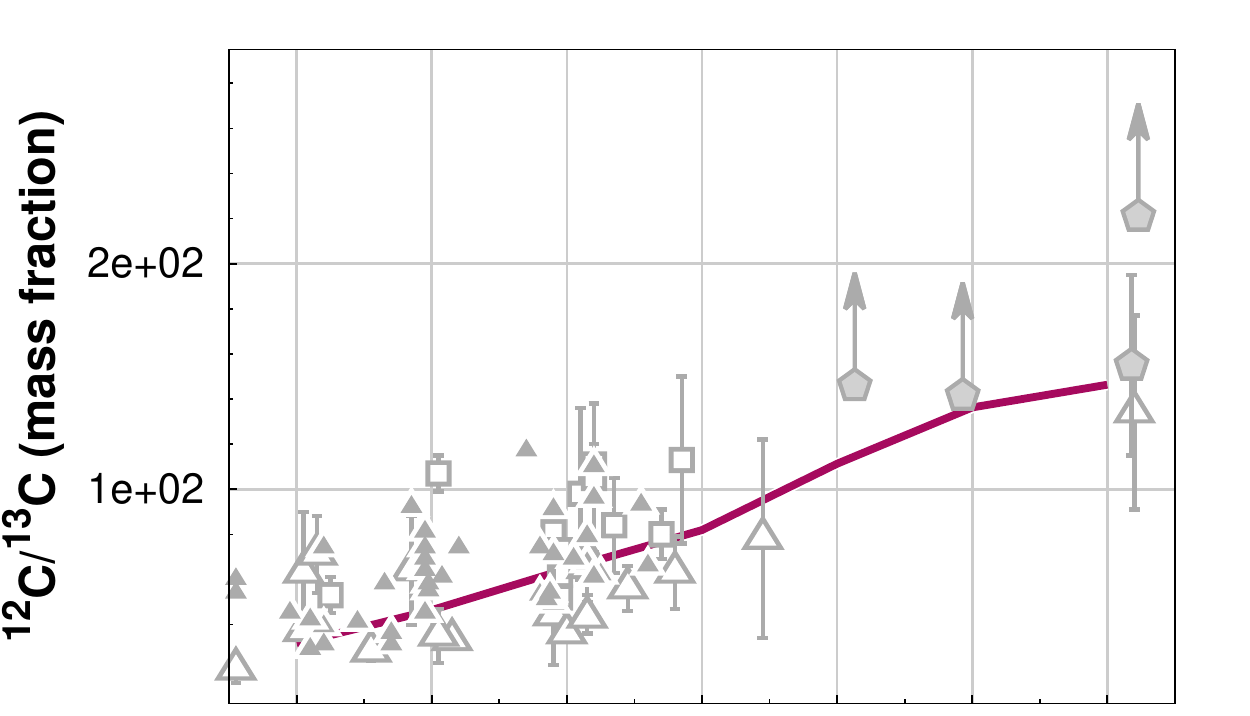}
\includegraphics[width=0.49\textwidth]{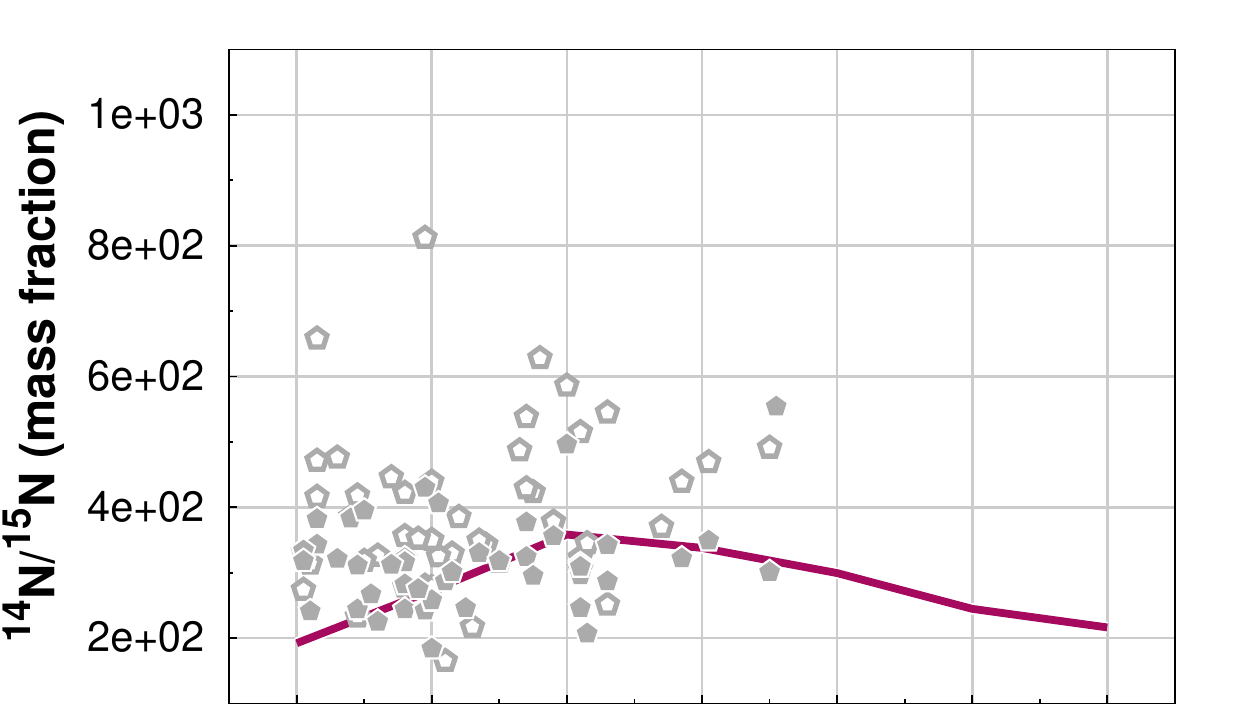}
\includegraphics[width=0.49\textwidth]{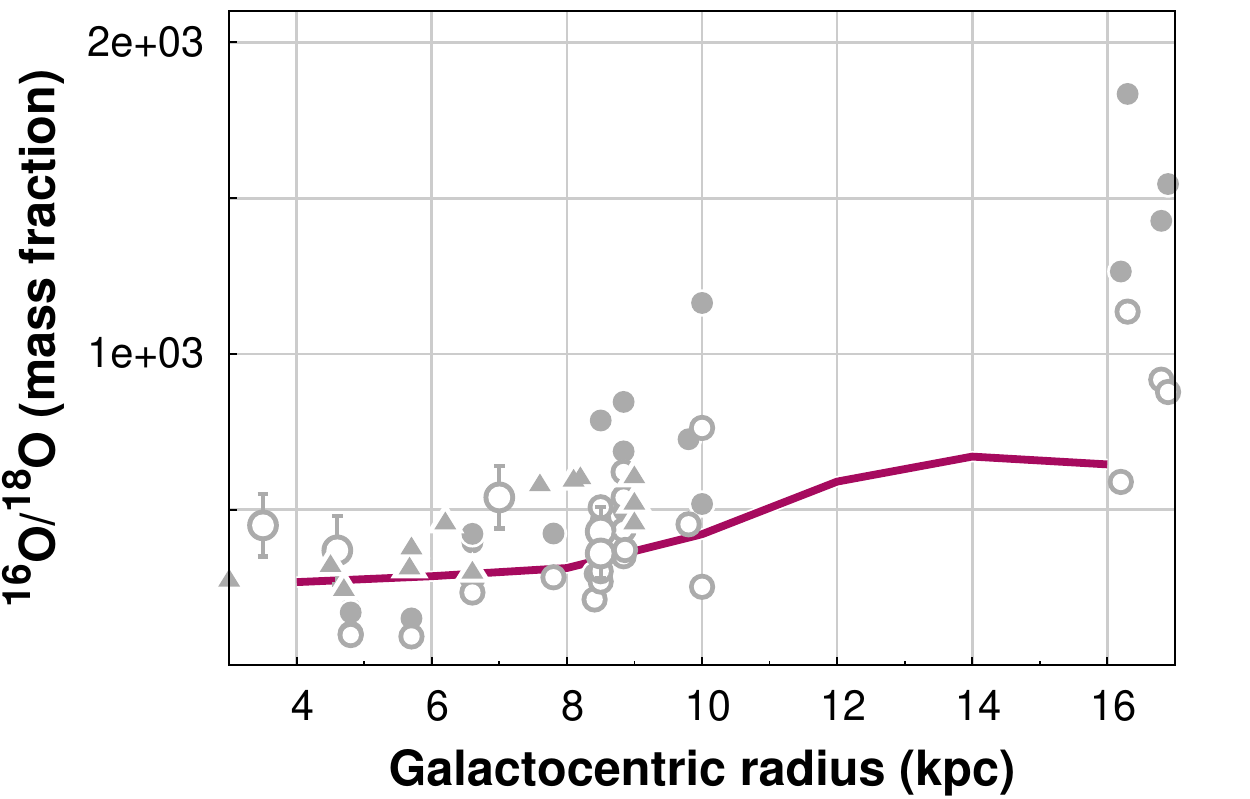}
\includegraphics[width=0.49\textwidth]{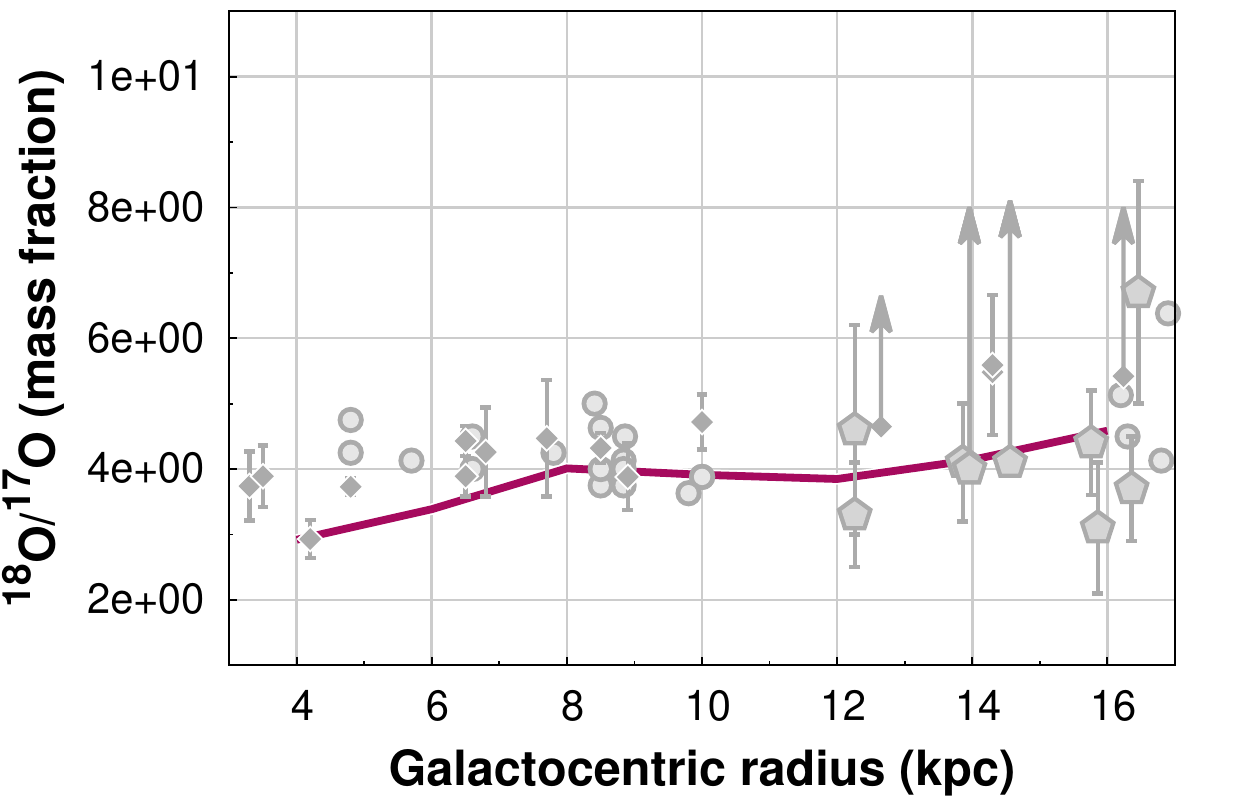}
\caption{ \emph{Clockwise from top left:} present-time gradients of C, N, and O isotope ratios across the Milky Way disc. The solid lines are the predictions of model MWG-11 of \citet{rom19b}, including primary production of $^{13}$C, $^{14}$N, $^{17}$O, and $^{18}$O from massive fast rotators at low metallicity and $^{13}$C, $^{15}$N, and $^{17}$O production from novae. Data are from \citet[][filled triangles]{wils94}, \citet[][empty triangles]{mila05}, \citet[][large empty circles]{pole05}, \citet[][small circles]{wout08}, \citet[][diamonds]{li2016}, \citet[][small pentagons]{colz18}, \citet[][empty squares]{boog00}, and a preliminary analysis of proprietary data \citep[][large pentagons]{rom19b}. Figure adapted from \citet{rom19b}, copyright by the authors.} 
\label{fig:isograd}
\end{figure}

Derivation of CNO isotopic ratio gradients along the disc involves measurements of $^{12}$C/$^{13}$C, $^{14}$N/$^{15}$N, $^{16}$O/$^{18}$O, and $^{18}$O/$^{17}$O from various CNO-bearing molecules, such as CO, CN, CH$^+$, H$_2$CO, HCN, HNC, OH, and their isotopologues. A summary of observations can be found in \citet{roma17,rom19b}. Nowadays, the increased sensitivity and high spectral resolution of millimeter telescopes allow to reach far outer Galactic disc regions and probe their low-metallicity ISM. This provides unique information on the synthesis of the minor CNO isotopes in metal-poor environments \citep[see][]{roma17,rom19b,zhan20,font22}. Spectroscopy of unevolved stars, in fact, provides only fragments of the local enrichment histories of $^{12}$C/$^{13}$C and $^{16}$O/$^{18}$O (see Fig.~\ref{fig:isoSN}), with basically no hope to get any information about $^{14}$N/$^{15}$N. Observations of the $^{17}$O/$^{18}$O ratio in stars are limited to a few giants, where the ratio is known to be altered by stellar evolutionary processes \citep[e.g.,][]{denu17}. Therefore, molecular data are a perfect complement to the scarce stellar data. It is necessary to bear in mind, though, that switching from a given isotopologue abundance ratio to its corresponding isotope abundance ratio is potentially complicated by local chemical fractionation effects \citep[e.g.,][]{roue15}, like isotope-selective photodissociation \citep[e.g.,][]{caso92}. Neglecting local fractionation processes might result, for instance, in highly scattered diagrams for $^{14}$N/$^{15}$N \citep{colz19,berg20,evan22}.

Figure~\ref{fig:isograd} shows the radial gradients of several CNO isotopic ratios predicted by a specific GCE model \citep[model~MWG-11 of][]{rom19b} compared to the observations. The current gradients of $^{12}$C/$^{13}$C and $^{16}$O/$^{18}$O (Fig.~\ref{fig:isograd}, \emph{left-handed panels}) are in satisfactory agreement with the observed trends, although the latter are characterized by a significant dispersion. As for nitrogen, the model predicts an increase of the isotopic ratio when moving from the inner Galaxy to the solar position and a decrease afterwards, when moving from solar to outer radii (Fig.~\ref{fig:isograd}, \emph{top right-handed panel}). Such a prediction still waits to be proven, or disproven, by the observations. Finally, the theoretical $^{18}$O/$^{17}$O gradient presents a moderate variation from a value of approximately 3 at $R_{\mathrm{GC}} =$~4 kpc to $^{18}$O/$^{17}$O~$< 5$ at $R_{\mathrm{GC}} =$~16 kpc (Fig.~\ref{fig:isograd}, \emph{bottom right-handed panel}), which fits only in part the data.

We close this section with a word of caution regarding the model predictions discussed here for the inner Galaxy: the adopted yields for massive stars do not extend to super-solar metallicities, meaning that arbitrary extrapolations are needed when the computed metallicity exceeds solar. It is hard to foresee how this may affect the GCE model predictions, but secondary elements such as $^{18}$O are likely to be the most affected.

\subsubsection{Dwarf spheroidal and ultrafaint dwarf galaxies}
\label{sec:dsphufd}

The local universe contains a large number of dwarf galaxies, the faintest of which are still being found lurking in and around the haloes of the Milky Way and Andromeda, after the first discovery of an UFD by \citet{will05}. The properties of stellar populations in local dwarf galaxies are summarized in excellent reviews by \citet{tols09,simo19,anni22}, to which we address the interested reader.

Being concerned here with the evolution of the CNO elements, it comes natural to divide the dwarf galaxies in two loose groups, the one comprising galaxies that still have large amounts of gas and active star formation, the other including galaxies with no detectable (or only traces of) gas at present. The first is the subject of the next section, while the second is dealt with in the following paragraphs.

Many of the closest dwarf galaxies host old stellar populations and have lost their gas, hence their ability to form stars. Being placed relatively nearby, their stellar populations can be studied on a star-by-star basis with current observational facilities. Their low mean metallicity gives us the possibility to investigate how star formation and chemical enrichment proceeded in the early universe \citep[e.g.,][]{freb15}. High-resolution abundance measurements, however, are possible only for giant stars. A fair comparison between the predictions of chemical evolution models and the observations thus implies uncertain corrections to the derived abundances of C and N, which hampers the achievement of firm conclusions, at least as far as these elements are concerned \citep[e.g.,][and references therein]{lard16}.

\begin{figure}
\centering
\includegraphics[width=\textwidth]{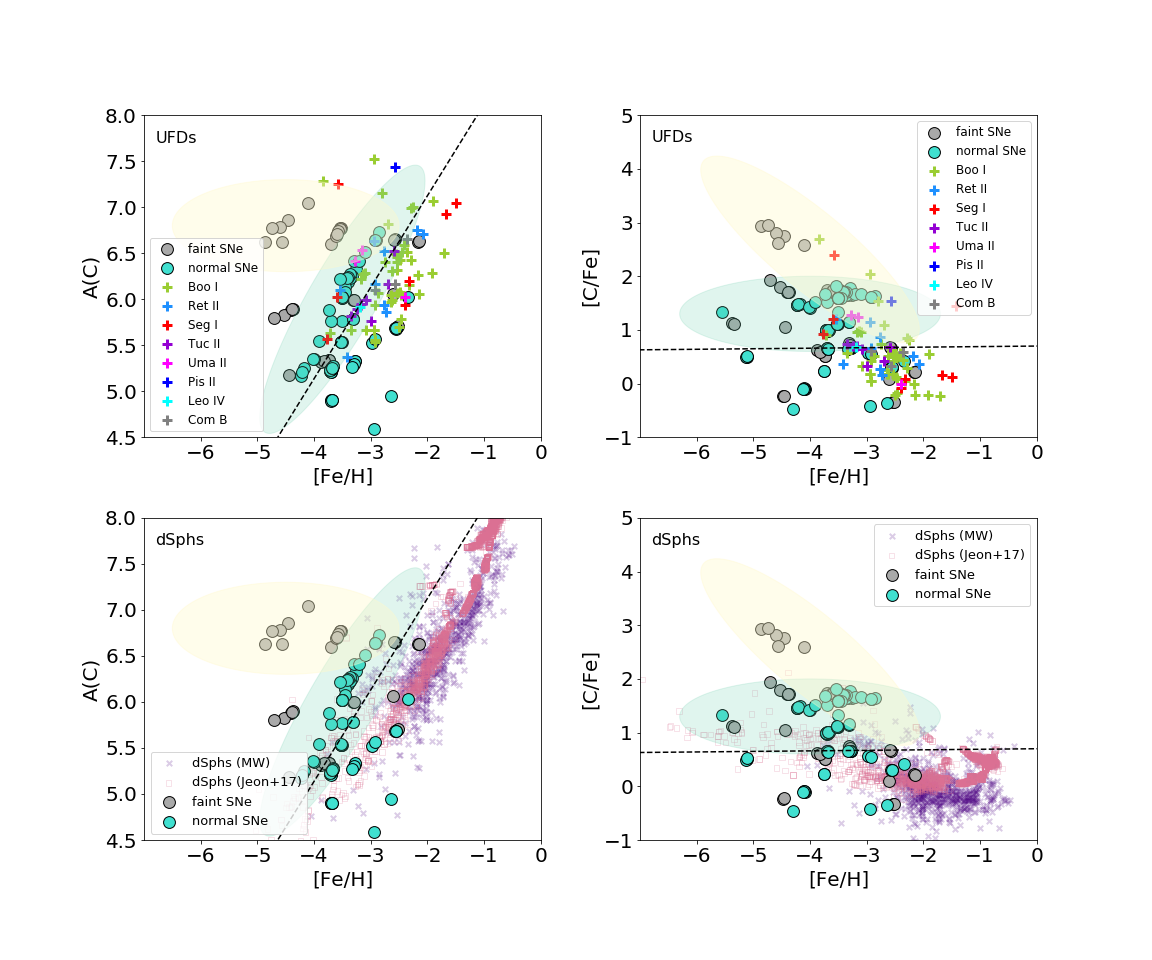}
\caption{ $A$(C) and [C/Fe] versus [Fe/H] diagrams (\emph{left-hand} and \emph{right-hand panels,} respectively) of individual stars observed in UFDs (coloured crosses, \emph{top panels}) and in dSphs (small violet $\times$ symbols, \emph{bottom panels}), compared to simulated UFD stars \citep{jeon21} assuming contributions from normal (cyan circles) and faint SNe (grey circles). Simulated dSph stars from the work of \citet{jeon17} are shown as pink squares. Reproduced with permission from \citet{jeon21}, copyright by the authors.}
\label{fig:jeon}
\end{figure}

It is, however, the existence of a non-negligible fraction of CEMP stars in these galaxies that has attracted major attention, both in an attempt to understand the origin of their anomalously high C abundances and in connection to the formation of the Galactic halo: if the dwarf galaxies orbiting the Milky Way today are the remnants of the hierarchical merging process that built up the Galactic halo many Gyr ago, the halo and the surviving satellites should contain similar fractions of CEMPs.

Metal-poor stars having [C/Fe]~$> +1$ (CEMPs)\footnote{Nowadays, the threshold value for a star being classified as a CEMP is rather set to [C/Fe]~$\ge 0.7$ dex.} are classified in subcategories taking the relative abundances of heavy neutron-capture elements into account: CEMP-$s$ and CEMP-$r$ are CEMPs characterized by $s$- and $r$-process element enhancement, respectively, CEMP-$r/s$ present both kind of enhancements, and CEMP-no no enhancement at all \citep{beer05}. For [Fe/H]~$< -1$, many stars in the Galactic halo and in dwarf satellites are CEMPs, with relative fractions with respect to C-normal stars increasing with decreasing the metallicity \citep[e.g.,][]{aoki07,yong13,salv15}, reaching 70\% below [Fe/H]~$= -4$ \citep[][]{yoon18}. For [Fe/H]~$> -3$, CEMP-$s$ stars are more numerous; their peculiar chemical composition is likely due to mass transfer from a close AGB companion \citep[e.g.,][]{aoki06}. On the other hand, CEMP-no stars with [Fe/H]~$< -2$ might either have a nucleosynthetic origin and reflect the properties of the elusive first (Pop~III) stars, or be again the result of mass transfer from an AGB star in a binary system \citep{boni15}. With their low-metallicity stellar populations, [Fe/H]~$< -2$ \citep{simo19}, UFDs are the ideal laboratory to study CEMPs and the effects of pristine chemical enrichment from Pop~III stars \citep{salv19}.

In their recent study based on cosmological hydrodynamic zoom-in simulations of isolated UFDs, \citet{jeon21} find that most CEMP-no stars in the UFDs and the Milky Way halo can be explained by normal Pop~III and Pop~II SNe (with explosion energy of the order of $10^{51}$ ergs). Faint SNe (with explosion energy of the order of $0.6 \times 10^{51}$ ergs) provide a viable path to CEMP-no stars with [C/Fe] exceeding a value of 2, though the highest C overabundances remain unexplained. Figure~\ref{fig:jeon} summarizes the results obtained by \citet{jeon21} for both UFDs and dSphs. The loci occupied by the observations are, generally, very well recovered by the models, but there are simulated stars that fall in unpopulated regions, and viceversa, densely-populated regions that are missed by the models. The same problem is faced by \citet{komi20}, using their stochastic chemical evolution model in cosmological context that implements the yields from two sets of mixing and fallback models for faint SNe. Overall, this calls for a refinement of the nucleosynthesis prescriptions underlying the simulations. Indeed, detailed analyses of high-resolution CEMP spectra providing the abundances of many elements suggest that different kinds of Pop~III polluters must be in action \citep[][see also \citealt{boni15}]{skul21}.

\subsubsection{Dwarf irregular and blue compact dwarf galaxies}
\label{sec:dirrbcd}

Local dIrrs and BCDs contain large fractions of gas and are subject to ongoing star formation. This, joint to the low metallicities, made people think of them as convenient, readily accessible counterparts of the basic galactic building blocks predicted numerous in the distant universe \citep{whit78}. However, the physical conditions within galaxies and their surroundings vary with cosmic time \citep{kewl13,shap19}, which challenges the primeval galaxy interpretation (even leaving aside the effects of internal chemical enrichment). The high spatial resolution of the \emph{Hubble Space Telescope (HST)} photometry has allowed to resolve local (up to $\sim$20~Mpc distance from us) dIrrs and BCDs in stars and to build deep colour-magnitude diagrams (CMDs) from which detailed star formation histories (SFHs) have been extracted \citep[see][and references therein]{tosi91,gall05,tols09,cign10,anni22}: local dIrrs and BCDs prove to be definitively old objects and are found to have formed stars as far back in time as the observations can reach. The disparate SFHs revealed via CMD analyses for objects that share similar mass and metallicity suggest that many different physical processes -- gas accretion, mergers, interactions with halo hosts and close companions, stellar chemical and energetic feedback -- combine in a complex way to lead to the specific configuration observed for each system.

With the notable exceptions of the Large and Small Magellanic Clouds (hereinafter, LMC and SMC, respectively), for which statistically significant samples of (giant) stars with high-resolution spectra are available \citep{pomp08,lape12,nide20}, chemical abundances in dIrrs and BCDs are mainly measured from optical and NIR spectroscopy of bright H\,II regions \citep[see][and references therein]{anni22} and UV nebular emission lines \citep[e.g.,][]{berg19}. Thus, for these galaxies we do not have a complete fossil record of the past enrichment history, but only the current snapshots \citep[see, e.g.,][]{chia03}.

\begin{figure}
\centering
\includegraphics[width=\textwidth]{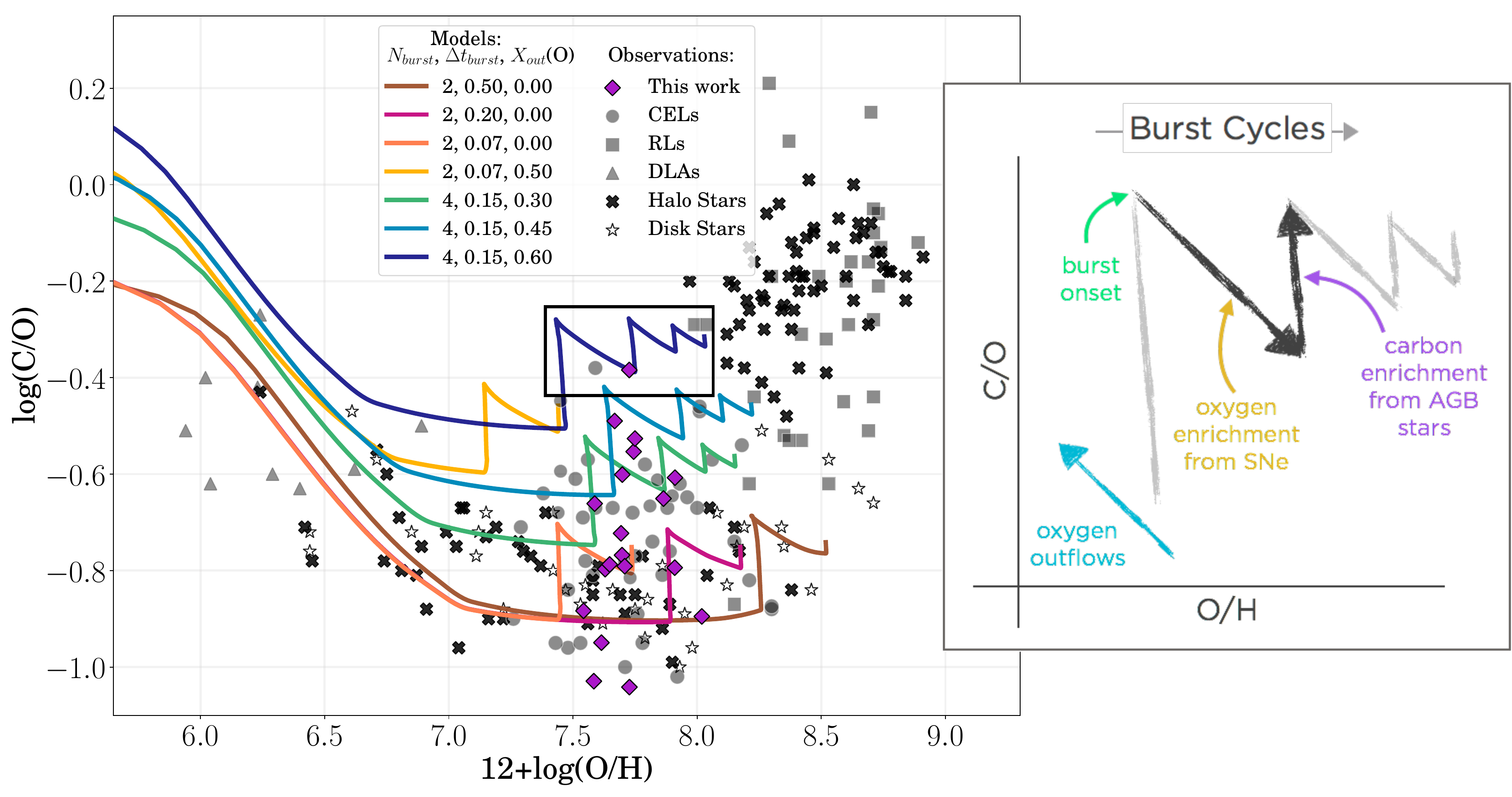}
\caption{ Theoretical tracks in a log(C/O) versus log(O/H)\,+\,12 plane (coloured curves) predicted by chemical evolution models assuming a series of star formation bursts of different duration as well as various fractions of oxygen expelled by galactic winds, as given in the legend \citep{berg19}. Different symbols are data for Damped Lyman $\alpha$ systems, Milky Way halo and disc stars, and H\,II regions in dwarf galaxies \citep[references in][]{berg19}. The panel to the right illustrates the physical processes responsible for the sawtooth behaviour of the theoretical tracks (see text). Figure from \citet{berg19}, licensed under a Creative Commons Attribution 4.0 International Licence.}
\label{fig:berg}
\end{figure}

In Fig.~\ref{fig:berg}, taken from \citet{berg19}, determinations of the C/O ratio in different systems are compared to the predictions of generic chemical evolution models. The sawtooth behaviour \citep[see][]{pily92} seen in the theoretical curves arises from the assumption of a bursty star formation: each burst makes the metallicity increase and the C/O ratio decrease due to the short time delay with which oxygen, produced by short-lived massive stars, is injected into the ISM relative to carbon, which owes a large fraction of its production to low- and intermediate-mass stars acting on longer timescales \citep{tins79}. Differential galactic outflows, venting out of the system mainly the products of CCSN nucleosynthesis, contrast both the increase of O/H and the decrease of C/O. Nitrogen follows a qualitatively similar path. Interestingly, from their analysis \cite{berg19} conclude that the C/N ratio of the low-metallicity sample is constant with metallicity, but has a significant scatter: they suggest that this may hint to dominant production of C from stars in a different mass range relative to that responsible for most of N. The broad variations in the observations can be reproduced by heuristic chemical evolution models that vary solely the efficiency of star formation, its duration, and the fraction of ejected oxygen \citep[see][their Fig.~12]{berg19}. However, as pointed out by \cite{roma06}, if a name is given to the generic point in the C/O, N/O, C/N diagrams, and a present-day gas fraction, a stellar mass, and a detailed SFH can be associated to it, it may become difficult to reproduce the observed abundances. This is because a better-constrained GCE model often reveals all the limitations of its simple approach.

While a galactic outflow that expels preferentially metal-rich matter processed by massive stars is a convenient way of reducing the effective yield of a stellar population, the very low star formation rates observed in some metal-poor and extremely metal-poor dwarf galaxies would actually point to a mild outflow effect. In fact, star formation rates lower than 1~M$_\odot$ yr$^{-1}$ imply a deficit of massive clusters able to generate massive stars \citep{yanz17}, hence a less effective SN feedback (both in terms of metal and energy production). \citet{yanz20} apply the integrated galactic IMF theory that allows to compute the gwIMF as a function of a galaxy star formation rate and metallicity, and find a mildly bottom- and top-light gwIMF for the UFD Bo\"otes~I, which leads to a good agreement of their GCE model predictions with the data. We note that a gwIMF in which oxygen production from massive stars is depressed, whilst nitrogen production from intermediate-mass stars is enhanced, could provide a compelling explanation for the high N/O ratios observed in some extremely metal-poor, gas-rich dwarf galaxies, such as Leo~P, DDO~68, and Leoncino (AGC\,198691) \citep[][]{skil13,ann19a,aver22}, which lie significantly above the very narrow low-metallicity plateau at log(N/O)~= $-1.60 \pm 0.02$ suggested by \citet{izot99}.

Another ingredient with which GCE models can play a bit to reduce the ISM metallicity of a galaxy is the rate of infall of primordial gas. The effect of infall is maximum on O/H, but tends to cancel out when considering the abundance ratios (except for a second order effect on the yields, due to their metallicity dependence). In a recent study by \citet{pasc22}, $N$-body hydrodynamical simulations are used to investigate whether the extremely low metallicity of DDO\,68 \citep{ann19a} can be explained by infall of gas of primordial chemical composition. The models, tailored to reproduce the main observed kinematic and structural properties of this galaxy \citep{anni16,ann19b}, show that the infalling mass necessary to dilute the oxygen that is produced according to the CMD-inferred SFH and a canonical gwIMF is inconsistent with any possible dynamical solution, unless the measured metallicity refers to the accreted material and not to that of the main galaxy.

The roles of all the different physical mechanisms mentioned above in setting the current chemical properties of dwarf galaxies have been recognized and discussed long ago \citep{matt83}. Nowadays, also thanks to the parallelization of algorithms that boosts the computational power of supercomputers, hydrodynamical simulations can be coupled to chemical evolution models to better explain the observations of gas-rich small galaxies and increase the predictive power of both \citep[e.g.,][among others]{recc01,recc07,emer20}. Because of the dramatic increase of the necessary computational resources when introducing a new chemical species, however, usually a choice has to be made and only a few elements can be selected. We stress that it would be important to include at least the main CNO elements in the simulations.

Turning to the minor CNO isotopes, abundance estimates of CNO isotopes in dwarf galaxies are available only for the Magellanic Clouds. From a spectral line survey of the prominent star-forming region N\,113 in the LMC covering parts of the frequency range from 85~GHz to 357~GHz, ratios of $^{12}$C/$^{13}$C~$\simeq 49 \pm 5$, $^{16}$O/$^{18}$O~$\simeq 2000 \pm 250$, and $^{18}$O/$^{17}$O~$\simeq 1.7 \pm 0.2$ are estimated by \cite{wang09}. Atacama Large Millimeter/submillimeter Array (ALMA) observations in the $^{12}$CO($J$ = $2-1$), $^{13}$CO($J$ = $2-1$), C$^{18}$O($J$ = $2-1$), $^{12}$CO($J$ = $3-2$), and $^{13}$CO($J$ = $3-2$) lines toward the active star-forming region N83C in the SMC are presented by \citet{mura17}, from which the $^{12}$C/$^{13}$C and $^{16}$O/$^{18}$O isotopic ratios can be derived. We will comment on this in Sect.~\ref{sec:beyond}.

\subsubsection{The Andromeda galaxy}
\label{sec:andro}

We conclude this roundup of local galaxies with Andromeda (M\,31), a giant spiral galaxy that is also the most massive member of the Local Group. At variance with the disc of the Milky Way, that suffered a last major merger about 10~Gyr ago, leading to the formation of the thick disc \citep{helm18}, and evolved peacefully since, M\,31 is representative of a lively history of mergers, with clear signs of recent encounters \citep{mcco09}. \citet{bhat22} provide oxygen abundance measurements from direct detection of the [O\,III]\,$\lambda$\,4363~$\AA$ line for a magnitude-limited sample of 205 PNe in the disc of M\,31. The sample is divided in high- and low-extinction PNe that are associated, respectively, with ages~$\le 2.5$~Gyr and $\ge 4.5$~Gyr, namely, with thin- and thick-disc populations. The sample as a whole defines a flat gradient, $0.001 \pm 0.003$~dex kpc$^{-1}$, consistent with zero as also found by previous studies based on PNe \citep[e.g.,][]{pena19}. However, it is also seen that older PNe are metal-poorer and trace a mild positive gradient of $0.006 \pm 0.003$~dex~kpc$^{-1}$, while younger PNe are metal-richer and define a steeper gradient of $-0.013 \pm 0.006$~dex kpc$^{-1}$. The latter is marginally consistent with the O abundance gradient measured from other young indicators, H\,II regions \citep[$-0.023 \pm 0.002$~dex kpc$^{-1}$,][]{zuri12}.

$N$-body simulations of galaxy interactions show that mergers tend to flatten the radial metallicity gradient \cite[e.g.,][]{rupk10}, while classic GCE models emphasize the role of the gas threshold and efficiency of the star formation process in steepening or flattening the gradient \citep{marc10}. Ad hoc simulations specifically tailored to reproduce the observed characteristics of Andromeda, like the ones presented by \citet{yin09} and \citet{robl14}, are necessary to quantify the relative importance of all the processes at play.

\subsubsection{Other systems}

\begin{figure}
\centering
\includegraphics[width=0.75\textwidth]{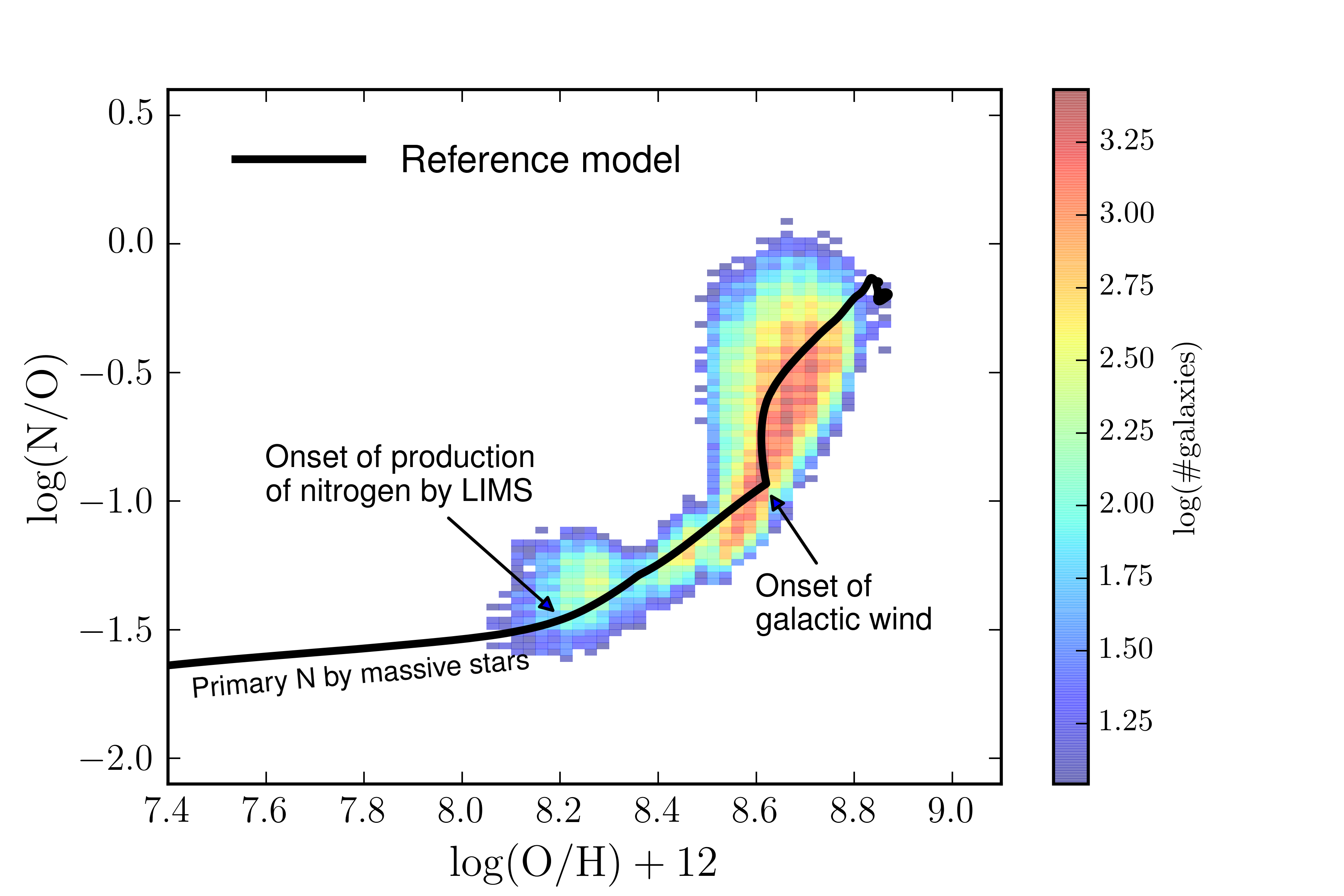}
\caption{ Heuristic chemical evolution model from \citep[][black solid line]{vinc16} reproducing the main features of local galaxies drawn from the SDSS DR7 in the log(N/O)--log(O/H)\,+\,12 plane. The observed distribution is shown as a 2D histogram, with a bin size along both axes of 0.025 dex. Different colours represent the number of galaxies in each bin. The changes in slope of the predicted relation are linked to different physical properties of the model, as indicated. Image reproduced from \citet{vinc16}, copyright by the authors.}
\label{fig:vince}
\end{figure}

For local elliptical galaxies, indications about the abundances of CNO elements come essentially from absorption features in integrated starlight, while emission lines in UV and optical spectra are used for gas-rich star-forming galaxies. A comprehensive review of these topics can be found in \cite{maio19}. In the following, we report a couple of examples focusing on the theoretical implications.

\begin{table}
\begin{center}
\caption{$^{12}$CO/$^{13}$CO, $^{12}$CO/C$^{18}$O, and $^{13}$CO/C$^{18}$O abundance ratios for external galaxies from the literature \citep[adapted from Table~1 of][copyright by the authors]{roma17}.}
\label{tab:ext}
\begin{tabular}{lcccccl}
\hline
Name & Type & Redshift & $^{12}$CO/$^{13}$CO & $^{12}$CO/C$^{18}$O & $^{13}$CO/C$^{18}$O & Refs$^a$\\
\hline
SPT stacking & SMG & 3.0 & 100--200 & $>$100--200 & $>1$ ($3\sigma$) & 1\\
Cloverleaf & QSO & 2.5579 & 300--10\,000 & -- & -- & 2, 3\\
MA2.53 & DLA & 2.525 & $>$40 & -- & -- & 4\\
Eyelash & SMG & 2.3 & 100 & 100 & 0.8 & 5\\
MA1.15 & DLA & 1.15 & $>$53 ($3\sigma$) & -- & -- & 6\\
MA0.89 & Spiral & 0.89 & $27\pm 2$ & $52\pm 4$ & 1.9 & 7\\
MA0.68 & Spiral & 0.68 & $38\pm 5$ & $\sim$80 & $\sim$2 & 8\\
Mrk\,231 & ULIRG & 0.042170 & 100 & 100 & 1.0 & 9\\
NGC\,6240 & ULIRG & 0.0245 & 300--500 & -- & 1.6 & 10, 11\\
Arp\,193 & ULIRG & 0.023299 & $\sim$150 & -- & -- & 10\\
VV\,114 & LIRG & 0.020067 & 229 & -- & -- & 12\\
Arp\,220 & ULIRG & 0.018126 & -- & 70--130 & 1.0 & 13\\
Arp\,220 & ULIRG & 0.018126 & -- & $>$80--100 & 1.0 & 14\\
NGC\,1614 & LIRG & 0.015938 & 130 & -- & $>$6.6 & 15\\
LMC & Dwarf & 0.000927 & 49 &  2000 & 27$\pm$9 & 16, 17\\
NGC\,253 & LIRG & 0.000811 & $>56$ & $145\pm 36$ & 2.6 & 18, 9,  19\\
M\,82 & LIRG & 0.000677 & $>138$ &  $>350$ & 2.2--3.7 & 18, 20, 21\\
NGC\,1068 & LIRG & 0.003793 & -- & -- & 3.33 & 22\\
NGC\,1614 & LIRG & 0.015938 & $>$36.5 & $>80$ & $>2.2$ & 23\\
NGC\,4945 & LIRG & 0.001878 & $>$17 & $>61$ & 3.6 & 24\\
Cen\,A & Radio & 0.001825 & 17 & $>56$ & $>3$ & 25, 26\\
NGC\,2903 & Normal & 0.001834 & $>10$ & -- & $>7$ ($3\sigma$) & 27\\
IC\,860 & LIRG & 0.011164 & $>20$ & $>20$ & 1 & 28\\
NGC\,3079 & Normal & 0.003723 & $>17$ & $>91.5$ & 5.4 & 28\\
NGC\,4194 & LIRG & 0.008342 & $>18.7$ & $>47$ & $>2.5$ & 28\\
NGC\,7469 & LIRG & 0.016317 & $>21$ & $>142$ & 6.8 & 28\\
NGC\,7771 & LIRG & 0.014267 & $>14$ & $>65$ & 4.7 & 28\\
NGC\,660 & Normal & 0.002835 & $>17$ & $>46$ & 6.9 & 28\\
NGC\,3556 & Normal & 0.002332 & $>12.5$ & $>161$ & 12.9 & 28\\
NGC\,7674 & LIRG & 0.028924 & $>14.5$ & $>40$ & 2.8 & 28\\
UGC\,2866 & Radio & 0.004110 & $>21$ & $>120$ & 5.8 & 28\\
Circinus & Normal & 0.001448 & $>13$ & $>54$ & 5.1 & 29, 30, 31\\
IC\,10 & Normal & 0.001161 & $>7$ & $>100$ ($3\sigma$) & $>15$ & 32, 33\\
IC\,342 & Normal & 0.000103 & $>10$ & $>70$ & 3-7 & 34\\ 
M\,51 & Normal & 0.002000  & $>10$ & $>46$ & 4.5 & 35\\
Maffei\,2 & Normal & 0.000057 & $>10$ & $>43$ & 4.3 & 36\\
NGC\,1808 & Normal & 0.003319 & $>17$ & $>49$ & 3 & 37\\ 
NGC\,3256 & LIRG & 0.009354 & $>33$ & $>135$ & 4 & 37\\ 
NGC\,7552 & LIRG & 0.005365 & $>14$ & $>35$ & 3 & 37\\ 
NGC\,4826 & Normal & 0.001361 & $>8$ & $>21$ & 4 & 37\\ 
NGC\,2146 & LIRG & 0.002979 & $>12.5$ & $>36$ & 3 & 37\\ 
NGC\,4418 & LIRG & 0.007268 & -- & -- & 8.3 & 38\\
IRAS\,04296+2923 & Normal & 0.007062 & $>21$ & $>94$ & 3.7 & 39\\
(starburst) &  &  &  &  &  & \\
IRAS\,04296+2923 & Normal & 0.007062 & $>16$ & $>45$ & 1.7 & 39\\
(CNZ) &  &  &  &  &  & \\
NGC\,6946 & Normal & 0.000133 & $>$13.3 & $>21$ & 2.5 & 40\\
\hline
\end{tabular}
\begin{flushleft}
\footnotesize{$^a$1:  \cite{spil14}; 2:  \cite{henk10}; 3:  \cite{lutz07}; 4:  \cite{note17}; 5:  \cite{dani13}; 6:  \cite{levs06}; 7:  \cite{mull06}; 8:  \cite{wall16}; 9:  \cite{henk14}; 10: \cite{papa14}; 11: \cite{pasq04}; 12: \cite{sliw13}; 13: \cite{gonz12}; 14: \cite{mar11b}; 15: \cite{sliw14}; 16: \cite{wang09}; 17: \cite{heik98}; 18: \cite{mart10}; 19: \cite{harr99}; 20: \cite{mao00}; 21: \cite{tan11}; 22: \cite{papa99}; 23: \cite{koen16}; 24: \cite{curr01}; 25: \cite{espa10}; 26: \cite{salo16}; 27: \cite{mura16}; 28: \cite{cost11}; 29: \cite{zhan14}; 30: \cite{davi14}; 31: \cite{for12}; 32: \cite{nish16}; 33: \cite{yin10}; 34: \cite{meie01}; 35: \cite{wata14}; 36: \cite{meie08}; 37: \cite{aalt95}; 38: \cite{cost15}; 39: \cite{meie14}; 40: \cite{meie04}. 
}
\end{flushleft}
\end{center}
\end{table}

\cite{conr14} model stacked spectra of early-type galaxies from the SDSS \citep[see also][]{joha12} as a function of velocity dispersion ($\sigma$). The abundances of C, N, and O measured from the integrated light reveal that [O/Fe] increases from about 0 for $\sigma =$ 90~km s$^{-1}$ to 0.25 for $\sigma =$ 300~km s$^{-1}$. In the same velocity dispersion range, C and N track O and their ratios relative to Fe exceed 0.2 at the high $\sigma$ end. The observed ratios are better explained if the gwIMF of elliptical galaxy progenitors is skewed in favour of massive stars \citep[see][see also \citealt{yanz21}]{roma20}. The idea of a flatter IMF slope in the high stellar mass range for elliptical galaxies relative to that assumed for the solar neighbourhood is not new: it has been proposed long ago to fit the [Mg/Fe] abundance ratios of ellipticals derived from Lick indices \citep{matt94}.

The log(N/O) versus log(O/H)\,+\,12 diagram of star-forming galaxies in the local universe has been studied by \citet{vinc16}. These authors use [O\,II]~$\lambda \lambda$~3726, 28, [O\,III]~$\lambda$~5007, H\,$\beta$, H\,$\alpha$, [N\,II]~$\lambda$~6584, and [S\,II]~$\lambda \lambda$~6717,\,31 line fluxes in emission to infer gas-phase O and N abundances of galaxies selected from SDSS Data Release 7 spanning a wide metallicity range. The effects of dust depletion are taken into account. The main characteristics of the diagram are well reproduced by the heuristic GCE model sketched in Fig.~\ref{fig:vince} that includes: (i) primary N production from fast-rotating massive stars at low metallicities, (ii) a major contribution to N production (mainly of secondary origin) from intermediate-mass stars at higher metallicities, and (iii) the effects of galactic outflows. The outflows are assumed to eject preferentially the products of massive star nucleosynthesis (oxygen), which explains the steepening of the relation relative to the flatter slope obtained at previous times, due solely to the late injection of N from low- and intermediate-mass stars.

Local gas-rich galaxies can also be targeted for isotopic CO transition measurements. While the limits of past observational capabilities severely hampered the investigations, which were restricted to metal-rich galaxies \citep[see][]{encr79}, ALMA now allows to reach much fainter objects. In Table~\ref{tab:ext}, that is a slightly modified version of Table~1 of \citet{roma17}, determinations of $^{12}$CO/$^{13}$CO, $^{12}$CO/C$^{18}$O, and $^{13}$CO/C$^{18}$O in a number of local and high-redshift objects (see Sect. \ref{sec:highz}) are reported. The abundance ratios of isotopes that originate from stars in different mass intervals are seen to vary widely. In particular, in starburst galaxies the ratio of two generic isotopes L and M, where M is produced mainly by massive stars on short timescales and L is produced mainly by low- and intermediate-mass stars on longer timescales (or has a mixed origin) is lower than in secularly evolving objects \citep{henk93,roma17,zhan18,brow19}. We will come back to this in Sect.~\ref{sec:highz}.

\subsection{Moving beyond the local universe}
\label{sec:beyond}

\subsubsection{The high-redshift universe}
\label{sec:highz}

\begin{figure}
\centering
\includegraphics[width=0.65\textwidth]{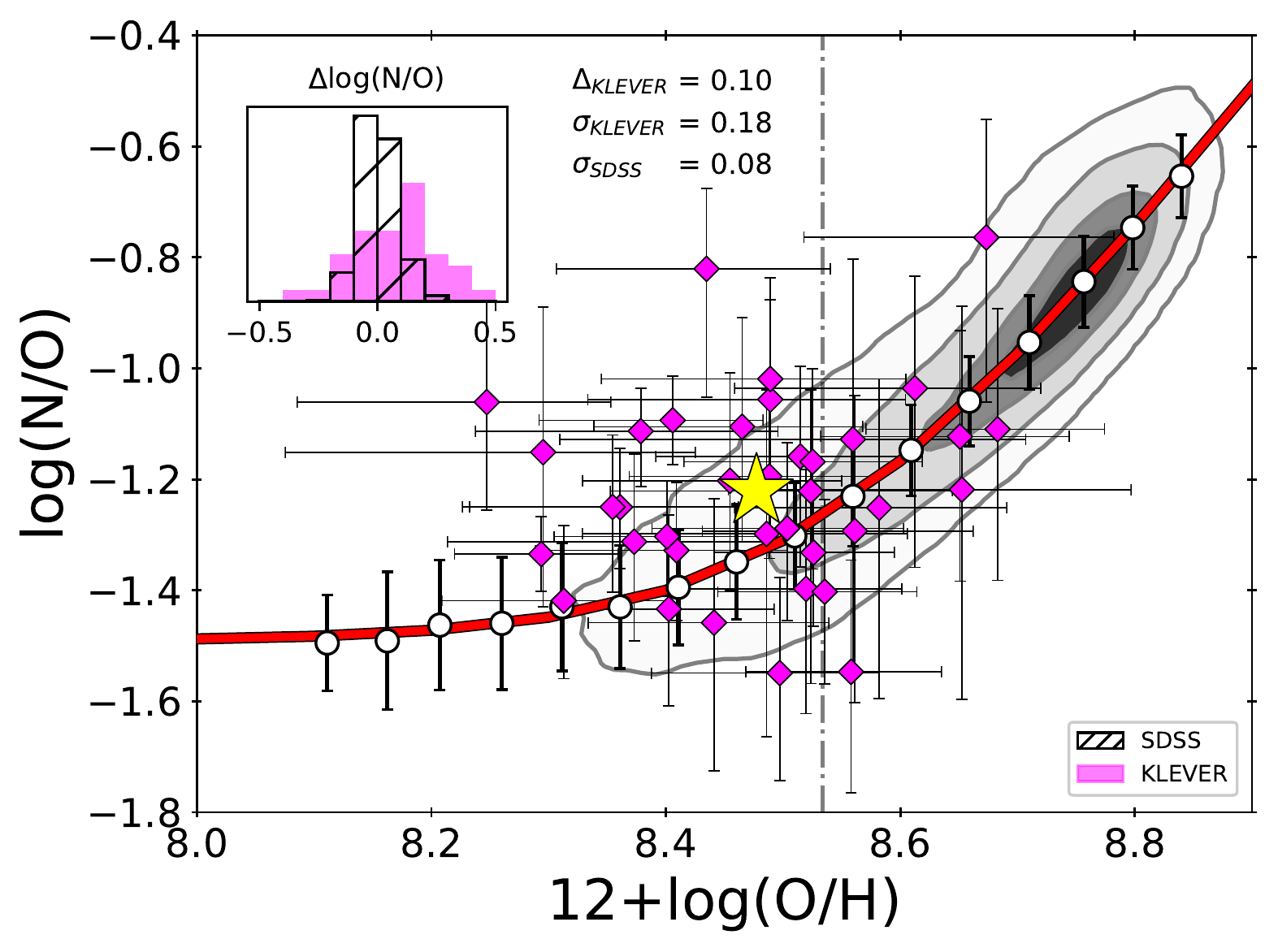}
\includegraphics[width=0.65\textwidth]{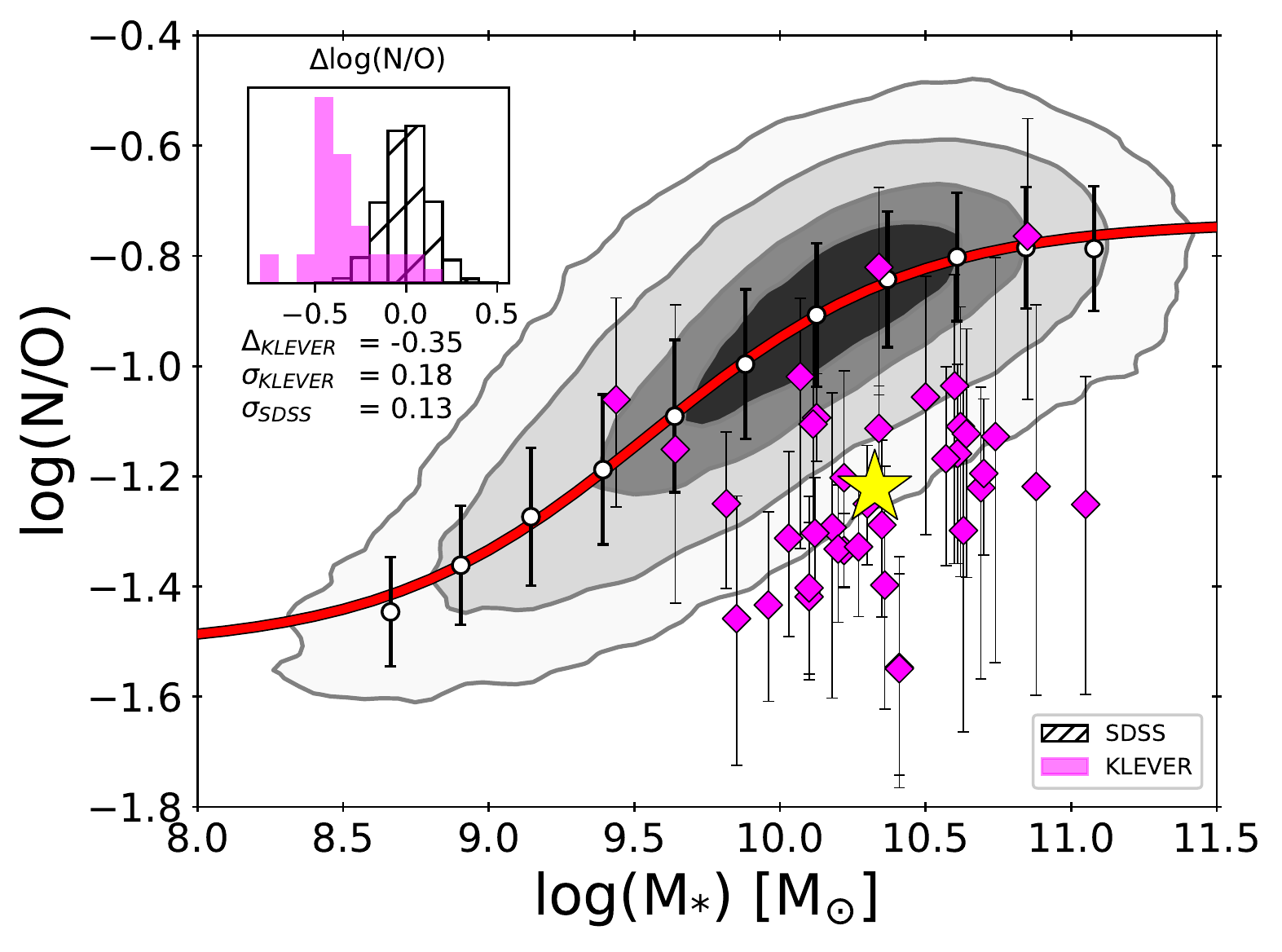}
\caption{ \emph{Top panel:} nitrogen-to-oxygen abundance ratio versus gas-phase metallicity (as traced by oxygen) for the SDSS (grey contours) and KLEVER (magenta points) galaxy samples discussed by \citet{hayd22}. The white points are the median binned SDSS data with corresponding standard deviations. The solid red curve is the best fit to local data. The yellow star refers to the average position of the KLEVER galaxies. The dot-dashed vertical line indicates the metallicity above which secondary N production from low- and intermediate-mass stars becomes dominant. The histogram in the top-left corner shows the deviations of the SDSS sample and the KLEVER sample from the best-fitting line to the local data. \emph{Bottom panel:} nitrogen abundance as a function of the stellar mass for the SDSS and KLEVER samples. Contours, line and symbols have the same meanings as in the top panel. Figures reproduced from \citet{hayd22}, copyright by the authors.}
\label{fig:hayd}
\end{figure}

\begin{figure}
\centering
\includegraphics[width=0.75\textwidth]{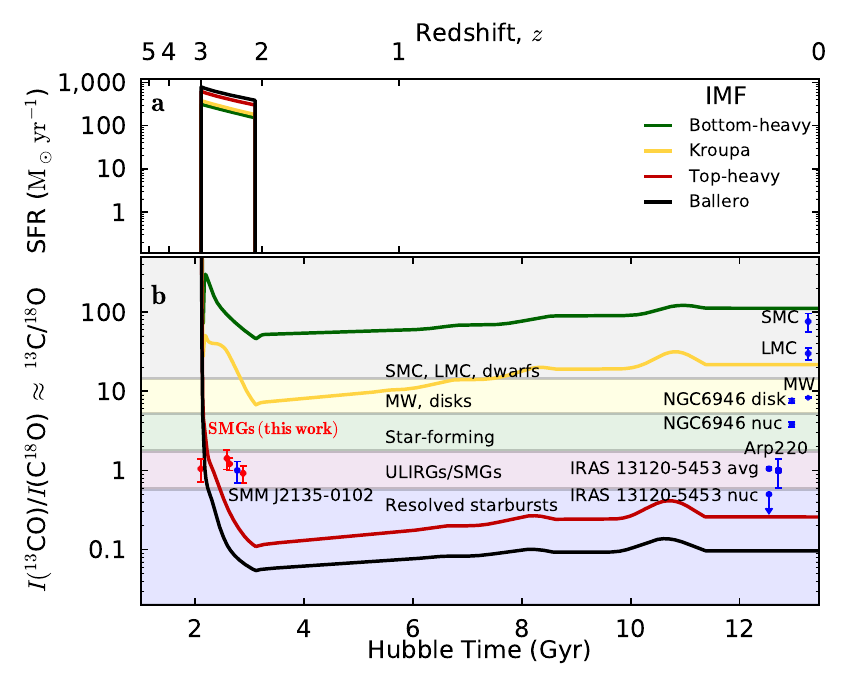}
\caption{ Lines: run with time of the theoretical $^{13}$C/$^{18}$O isotopic ratio in the ISM \emph{(bottom panel)} for a reference galaxy that forms stars in an early burst lasting 1~Gyr \emph{(top panel)}. The different colours refer to different choices of the gwIMF. The IMF slopes for $m >$ 1~M$_\odot$ are $x = 1.7$ for the green and yellow curves (changing the prescriptions for the low-mass stars), $x = 1.1$ for the red curve, and $x = 0.95$ for the black one, where $x = 1.35$ is the Salpeter slope. Red symbols indicate the SMGs in \citeauthor{zhan18}'s (\citeyear{zhan18}) sample. Blue symbols refer to the $^{13}$CO/C$^{18}$O ratios measured in a few representative galaxies. Image reproduced from \cite{zhan18}, copyright by the authors.}
\label{fig:zhang}
\end{figure}

The properties of local star-forming galaxies can be usefully compared to those of their higher-redshift counterparts, providing important clues on the evolution of the CNO elements, in particular the N/O ratio. \citet{hayd22} compare the trends of log(N/O) as functions of metallicity and stellar mass emerging from a sample of 37 $z \sim$ 2 galaxies drawn from the KMOS Lensed Emission Lines and VElocity Review (KLEVER) Survey to the corresponding patterns of a sample of local star-forming galaxies selected from the SDSS (see Fig.~\ref{fig:hayd}). Despite the large scatter in the data \citep[see also][]{stro18}, a modest enhancement of $+0.1$~dex is apparent when comparing the KLEVER sample to local galaxies in the log(N/O)--log(O/H)\,+\,12 plot, whereas at a fixed stellar mass a depletion is observed, at the level of $-0.35$~dex. For local galaxies, a strong anticorrelation is found between N/O and star formation rate in the N/O--stellar mass plane, resembling that between O/H and star formation rate in the mass-metallicity relation. Such anticorrelation is used to construct a fundamental N relation analogous to the fundamental metallicity relation. The KLEVER galaxies are consistent with both. Therefore, the depletion of N/O in the KLEVER sample at a fixed stellar mass can be interpreted as an effect of the redshift evolution of the mass-metallicity relation coupled to a redshift-invariant N/O--O/H relation \citep{hayd22}, a scenario consistent with the models discussed in \cite{vinc16}.

Moving to even higher redshifts, $z$ up to 8--9, the [C\,II] and [O\,III] gas emission can be detected in normal galaxies thanks to ALMA superb sensitivity \citep[][]{capa15,comb18,hodg20}. Moreover, the most powerful starbursts, thought to be the progenitors of the current population of quenched, passively-evolving, massive early-type galaxies \citep{toft14}, can be observed through their enshrouding dust curtains. \citet{zhan18} have observed the rotational transitions of the $^{13}$CO and C$^{18}$O isotopologues in the cold molecular gas of four submillimeter galaxies (SMGs) at $z \sim$~2--3 with ALMA and determined their $^{13}$C/$^{18}$O abundance ratio. Importantly, the determination of this ratio is immune to the effects of dust \citep{zhan18}.

Figure~\ref{fig:zhang} shows the predictions of reference chemical evolution models in comparison to \citeauthor{zhan18}'s dataset as well as other data from the literature. The effects of different assumptions about the gwIMF on the predicted $^{13}$C/$^{18}$O ratio are explored. It is seen that, assuming star formation rates that oscillate between 100 and 1000 M$_\odot$~yr$^{-1}$ for the observed SMGs, the high-redshift measurements are best explained by a gwIMF skewed towards massive stars relative to that assumed to fit the Milky Way properties. This is due to the fact that $^{13}$C and $^{18}$O come mainly from intermediate- and high-mass stars, respectively, so a $^{13}$C/$^{18}$O ratio as low as $\sim$1, as observed, requires a higher fraction of massive stars. Although the requirement of a top-heavy IMF in \citet{zhan18} is based on a set of yields that do not take the effects of stellar rotation into account, it has been demonstrated that by adopting the new yields for fast rotators from \cite{limo18} the conclusions are qualitatively the same \citep{rom19b}. This is better understood since rotation produces its largest effects at low-metallicity regimes, while SMGs reach soon solar metallicities because of the vigorous star formation regimes they experience.

\section{Building the bridge from ``here and now" to ``there and then"}
\label{sec:bridge}

In the Local Group, reliable CNO element abundances can be derived from high-resolution stellar spectra in the Milky Way and its closest satellites, with recent large surveys performed with multifiber spectrographs having led to major advances in the field \citep[see][for a review]{rand21}. High-resolution spectroscopy provides also the necessary anchor to the development of new data-driven techniques that allow the determination of precise chemical abundances from low-resolution spectra \citep[see, e.g.,][]{ting18}. In this way, low-resolution ($R =$~1\,800) LAMOST spectra of millions of stars are fully exploited to recover the abundances of many elements, including the CNO ones \citep[][see also \citealt{xian19}]{whee20}.

Gaseous diagnostics are also routinely analysed to explore disc gradients and to unravel the properties of the ISM in metal-poor, gas-rich dwarf galaxies leading, generally, to a good agreement with stellar indicators \citep[e.g.,][]{bhat22,este22}. Together with estimates of the star and gas masses and distributions, the rates of star formation, the rates of CCSN and type Ia SN explosions, the nova outburst rates, and full SFHs from CMDs, these abundances provide significant constraints to GCE models of our own and other nearby galaxies \citep{matt21}.

Exploring the universe beyond our backyard add invaluable information about the temporal evolution of fundamental relations, such as the mass-metallicity relation \citep[e.g.,][]{maio08,wang22}, or invariance thereof \citep[e.g.,][]{hayd22}, and populate the galactic zoo with more exotic objects \citep[e.g.,][]{card09}, allowing us to explore different physical conditions for star formation and chemical enrichment. As we have seen in previous sections, CNO elements are fundamental to this kind of investigation. GCE models for external galaxies can be constructed, but their predictive power is weakened by the shortage of some of the fundamental observational constraints listed above. For high-redshift galaxies a minimal list would include well-determined star formation rates, stellar masses, and ISM metallicities, as well as any information on the outflows. In this respect, we note that molecular outflows, which were largely out of reach till only a few years ago, can now be studied with the most sensitive mm/sub-mm interferometers, such as ALMA and NOEMA, though large integrations are needed to unravel their signatures through the CO line emission \citep{cico18}.

Before building models for external galaxies, moreover, it would be highly desirable to have the adopted stellar nucleosynthesis prescriptions tested against observations for the Milky Way, the best observed among all the local systems. In general, all the uncertainties underlying the model ingredients (which lead to degeneracies among the possibile solutions) must be kept well in mind and, as we have seen in Sect.~\ref{sec:nuc}, the stellar yields come with large associated uncertainties.

\section{Open issues and future directions}
\label{sec:next}

The above discussion leads to a list of desiderata that I translate and split in two lists of theoretical and observational challenges for the years to come in the remaider of this section.

\subsection{Theoretical challenges}
\label{sec:thch}

There are several improvements to the theory that would be highly welcome. In the following, a minimal personal selection of the most crucial items is given.
\begin{itemize}
\item The secular evolution of stars is still out of reach of 3D hydrodynamic simulations. At present, grids of yields covering a range of initial masses and metallicities adequate for chemical evolution studies are obtainable only via 1D models that rely on empirical prescriptions and parametric recipes to implement complex physical phenomena, such as convective motions and mass loss. However, small portions of the stars as well as small fractions of their lifetimes can be studied through 3D simulations. HPC resources should be devoted to these studies, since they provide fundamental hints to 1D modeling.
\item  Important physical processes leading to modification of the yields, such as stellar rotation, magnetic fields, and interactions with a companion in close binary systems, are still largely unexplored and deserve further efforts.
\item GCE models must include the effects of stellar duplicity on nucleosynthesis. Besides type Ia SNe and novae, which originate from low- and intermediate-mass stars, massive binaries must be implemented. Stellar duplicity is particularly relevant to CNO element evolution, as we discuss in this review.
\item Full coupling of GCE models with 3D hydrodynamical simulations is not possible yet. However, it is crucial to implement the evolution of all CNO elements (and, possibly, of their minor isotopes) in the simulations.
\item Galactic outflows are critical ingredients of GCE models. High-resolution 3D hydrodynamical simulations are needed to provide information on their effectiveness not only in removing metals, but also in entraining the neutral ISM in the flow. This allows to reduce sensibly the free parameter space of GCE models, hence reducing the degeneracy of the proposed solutions.
\end{itemize}

\subsection{Observational challenges}
\label{sec:obch}

Observations provide the necessary tests to model predictions and a guidance towards the development of more refined/new theories. In the following, a list of observations that may potentially have a huge impact on the implementation of GCE models is given.
\begin{itemize}
\item Related to the second item in the previous list, more time should be spent to improve the statistics of the binary fraction, period, and mass ratio distributions. Particular attention should be devoted to high-mass stars. In fact, massive stars often interact with companions that influence their lives at every stage, but more so during the giant phases in which the stars lose copious mass.
\item Although considerable progress has been made in recent years, further observational efforts are needed to understand which CNO diagnostics are the best to use in stars in dependence of their physical parameters. It is also important to keep in mind the usage that has to be made of the abundance determinations. For instance, \citet{delg21} point out that, since the O\,I\,$\lambda$\,6158~$\AA$ line has an high excitation potential, as the C\,I lines, in relatively warm dwarf stars in the disc, it is better to use O\,I\,$\lambda$\,6158~$\AA$ lines to evaluate the C/O ratio in those stars. Since using different indicators may give different trends, the information and physical justification provided by \citet{delg21} is very useful to GCE modelers.
\item The N\,I atomic line at 7468~$\AA$ is very weak also in solar-type stars, which forces to study the very crowded molecular band at 3360~$\AA$, where continuum placement is challenging. High resolution, of the order of $R =$ 80\,000, and SNR~= 150 are necessary for useful NH analysis in stars. This should be considered in building the next generation of instruments that will probe the UV band (e.g., CUBES, HRMOS).
\item Better characterization of the external regions of the Galactic disc is needed to improve our understanding on how CNO element enrichment proceeds at low metallicities. The project CHEMOUT \citep{font22}, now in its infancy, will address this point. Other surveys dedicated to the outer Galaxy, exploring different quadrants, would be highly welcome.
\item PISNe and faint SNe seem necessary to improve the predictions of GCE model on CNO evolution at low metallicities: JWST/WFIRST promises to provide important clues on the nature of these objects at high redshifts.
\end{itemize}

\begin{acknowledgements}
  First, I would like to thank the Board of Editors of {\it The Astronomy and Astrophysics Review} for their invitation and their patience in waiting for this review. In particular, I am grateful to the Editor-in-Chief, Francesca Matteucci, for her guidance and encouragement. I am obliged to several colleagues and collaborators for sharing their data in advance of publication and/or for the enlightening conversations, suggestions, and advice provided over the years on topics pertinent to this review -- a non-exhaustive list includes E.~Caffau, I.~J.~M.~Crossfield, F.~Matteucci, A.~Mucciarelli, P.~P.~Papadopoulos, E.~Spitoni, M.~Tosi, P.~Ventura, and Z.-Y.~Zhang. Special thanks are due to the referees for their constructive comments and suggestions that helped to improve the quality of this paper. Lastly, I want to acknowledge the International Space Science Institute (ISSI) in Bern, Switzerland, and the International Space Science Institute Beijing (ISSI-BJ) in Beijing, China, the warm hospitality and support provided to the Team~444 {\it ``Chemical abundances in the ISM: the litmus test of stellar IMF variations in galaxies across cosmic time''} (PIs: D.~Romano and Z.-Y.~Zhang).
\end{acknowledgements}
\phantomsection
\addcontentsline{toc}{section}{References}
\bibliographystyle{spbasic-FS}      
\bibliography{DRomano_BIB}

\end{document}